\documentclass[manuscript,11pt]{aastex}
\usepackage{epsfig}
\usepackage{soul}
\usepackage{color}

\usepackage{physymb}
\usepackage{mathrsfs}

\usepackage{lineno}

\def\hei{He\,{\sc i}}

\def\fexxiv{Fe\,{\sc xxiv}}
\def\fexxv{Fe\,{\sc xxv}}
\def\fexxvi{Fe\,{\sc xxvi}}
\def\fexxiii{Fe\,{\sc xxiii}}

\def\arxviii{Ar\,{\sc xviii}}

\def\sxvi{S\,{\sc xvi}}
\def\sxv{S\,{\sc xv}}
\def\sixiv{Si\,{\sc xiv}}
\def\sixiii{Si\,{\sc xiii}}

\def\mgxii{Mg\,{\sc xii}}

\def\alxiii{Al\,{\sc xiii}}

\def\mgxii{Mg\,{\sc xii}}
\def\mgxi{Mg\,{\sc xi}}

\def\mathv{\textbf{\em v}}

\def\cm{\ifmmode {\rm cm}^{-1} \else cm$^{-1}$ \fi}
\def\s{\ifmmode {\rm s}^{-1} \else s$^{-1}$ \fi}
\def\cc{\ifmmode {\rm cm}^{-3} \else cm$^{-3}$ \fi}
\def\cs{\ifmmode {\rm cm}^{-2} \else cm$^{-2}$ \fi}
\def\g{\ifmmode \gamma \else $\gamma$\fi}
\def\G{\ifmmode \Gamma \else $\Gamma$\fi}
\def\Gs{\ifmmode \Gamma~ \else $\Gamma~$\fi}

\def\gc{\ifmmode \gamma_{\rm c} \else $\gamma_{\rm c}$ \fi}
\def\sw{Schwarzschild~}
\def\gsim{\mathrel{\raise.5ex\hbox{$>$}\mkern-14mu
             \lower0.6ex\hbox{$\sim$}}}
\def\lsim{\mathrel{\raise.3ex\hbox{$<$}\mkern-14mu
             \lower0.6ex\hbox{$\sim$}}}
\def\simless{\mathbin{\lower 3pt\hbox
     {$\rlap{\raise 5pt\hbox{$\char'074$}}\mathchar"7218$}}}   
\def\simmore{\mathbin{\lower 3pt\hbox
     {$\rlap{\raise 5pt\hbox{$\char'076$}}\mathchar"7218$}}}   
\def\Msun{M_\odot}                                
\def\deg{^\circ}

\newcommand{\Alfven}{Alfv$\acute{\rm e}$n~}

\def\gro1655{GRO~J1655-40}
\def\4u1630{4U1630-472}
\def\h1743{H1743-322}
\def\grs1915{GRS1915+105}

\def\windon{{\tt wind-on}}
\def\windoff{{\tt wind-off}}

\lefthead{et al.} \righthead{\4u}

\shorttitle{Disk-Wind Transition in XRBs}

\shortauthors{Fukumura et al.}

\begin{document}

\title{Modeling Magnetic Disk-Wind State Transitions in Black Hole X-ray Binaries}

\date{\today}

\author{\textsc{Keigo Fukumura}\altaffilmark{1},
\textsc{Demosthenes Kazanas}\altaffilmark{2}, \textsc{Chris Shrader}\altaffilmark{2,3},
\textsc{Francesco Tombesi}\altaffilmark{2,4,5,6},
\textsc{Constantinos Kalapotharakos}\altaffilmark{2,7} 
\textsc{and}
\textsc{Ehud Behar}\altaffilmark{8}}

\altaffiltext{1}{Department of Physics and Astronomy, James Madison University,
Harrisonburg, VA 22807; fukumukx@jmu.edu}
\altaffiltext{2}{Astrophysics Science Division, NASA/Goddard Space Flight Center,
Greenbelt, MD 20771}
\altaffiltext{3} {Catholic University of America, Washington, DC 20064}
\altaffiltext{4}{Department of Astronomy, University of Maryland, College
Park, MD20742}
\altaffiltext{5}{Department of Physics, University of Rome ``Tor
Vergata", Via della Ricerca Scientifica 1, I-00133 Rome, Italy}
\altaffiltext{6}{INAF Astronomical Observatory of Rome, Via Frascati 33, 00078 Monteporzio Catone (Rome), Italy}
\altaffiltext{7}{University of Maryland, College Park (UMCP/CRESST II), College Park, MD 20742}
\altaffiltext{8}{Department of Physics, Technion, Haifa 32000, Israel}

\begin{abstract}
\baselineskip=15pt

We analyze three prototypical black hole (BH) X-ray binaries (XRBs), \4u1630, \gro1655\ and \h1743, in an effort to systematically understand the intrinsic state transition of the observed accretion-disk winds between \windon\ and \windoff\ states by utilizing state-of-the-art {\it Chandra}/HETGS archival data from multi-epoch observations.
We apply our magnetically-driven wind models in the context of magnetohydrodynamic (MHD) calculations to constrain their (1) global density slope ($p$), (2) their density ($n_{17}$) at the foot point of the innermost launching radius and (3) the abundances of heavier elements ($A_{\rm Fe,S,Si}$). Incorporating the MHD winds into {\tt xstar} photoionization calculations in a self-consistent manner, we create a library of synthetic absorption spectra given the observed X-ray continua.
Our analysis clearly indicates a characteristic bi-modal transition of multi-ion X-ray winds; i.e.
the wind density gradient is found to steepen (from $p \sim 1.2-1.4$ to $\sim 1.4-1.5$) while its density normalization declines as the source transitions from \windon\ to \windoff\ state.
The model implies that the ionized wind {\it remains physically present} even in \windoff\ state, despite its absent appearance in the observed spectra. Super-solar abundances for heavier elements are also favored.
Our global multi-ion wind models, taking into account soft X-ray ions as well as Fe K absorbers, show that the internal wind condition plays an important role  in wind transitions besides photoionization changes.
Simulated {\it XRISM}/Resolve and {\it Athena}/X-IFU spectra are presented to demonstrate a high fidelity  of the multi-ion wind model for better understanding of these powerful ionized winds in the coming decades.

\end{abstract}



\baselineskip=15pt

\section{Introduction}

One of the most significant discoveries of \textit{ASCA}, confirmed subsequently by \textit{Chandra} and \textit{XMM-Newton} Observatories, has been the ubiquitous presence  of outflowing plasma (winds) in the X-ray spectra of active galactic nuclei (AGNs), manifested by the presence of blueshifted absorption features at various stages of ionization and velocity. These, depending on their velocities, are referred to as Warm Absorbers (WAs with $v_{\rm out} \lsim 3,000$ km/s; e.g. \citealt{ReynoldsFabian95,Reynolds97,McKernan07, Blustin05,Laha14,Laha16}) or ultra-fast outflows (UFOs), typically identified as \fexxv/\fexxvi\ transitions at $v_{\rm out} \gsim 30,000$ km/s \citep[e.g.][]{Tombesi13}. These established the presence of ionized winds, discovered earlier in the AGN UV and optical spectra, in their X-ray domain.  As noted in the earlier studies \citep[e.g.][]{CKG03}, roughly 50\% of AGNs show such absorption features in their spectra, indicating this to be a common phenomenon among  extragalactic accreting black holes (BHs).

Similar features appear also in the spectra of accreting BHs in Galactic X-ray binaries (XRBs; e.g. \citealt{King13}). Such a component, almost necessarily required for continuous accretion, can efficiently transport excessive angular momentum and energy of plasma outwards \citep[e.g.][]{BB99}.
However, the situation there is more complicated because of the different spectral (and timing) states followed by these sources \citep[e.g.][]{FenderBelloniGallo04,RM06,DoneGierlinskiKubota07,Belloni10,Kylafis12}. In most states, the absorption signatures are either absent or of very low equivalent width (EW). However, on occasion, especially in their so-called {\it high/soft} state they exhibit numerous absorption features of very high signal-to-noise (S/N) ratios \citep[e.g.][]{RM06,Miller15,Ponti12}, but at velocities significantly lower than the equivalent AGN winds ($\sim 300-2,000$ km/s).

The presence of ionized outflows in accreting BHs has led to the introduction of several models of their launching processes and properties. Their typically near-Eddington luminosities led to model winds driven by radiation pressure on UV and optical lines, which provide an effective force multiplier, in analogy to the winds of the O/B stars \citep[e.g.][]{MCGV95,PSK00,Hagino15,Hagino17,Nomura17,Mizumoto20}. However, in Galactic BH XRBs, the line force multiplier is very low (if not absent) due to their high ionization by the sources' X-rays. In these cases, thermally-driven outflows are an alternative scenario, provided that X-ray heating raises the disk thermal velocity above the local escape value \citep[e.g.][]{BMS83,Woods96,Netzer06,Luketic10,Dyda17,Higginbottom20}.
The critical (Compton) radius for launching such a wind is given by $R_c \sim 10^{11}
(M_{\rm BH}/10\Msun)/T_{\rm c8}$ cm $\sim 10^{5-6} R_g$ ($T_{\rm c8}$ is the Compton temperature in units of $10^8$ K and $R_g \equiv GM/c^2$ is the gravitational radius). Here, we ignore, for the moment, the issue of what fraction of the X-ray flux is intercepted by a thin disk at radius $R \gsim 10^5 R_g$ when emitted by a source of vertical size $\simeq 10 R_g$. 
More recently, a thermal radiative  scenario has been proposed to explain the observed disk-winds in XRBs using a post-processed Monte Carlo radiative transfer through hydrodynamic (HD) simulations \citep[e.g.][]{Tomaru19,Tomaru20,Higginbottom20}.
With the wind dynamics of these models being well formulated, however, it remains to be seen whether they can provide the observed range of absorber's observables; e.g. the hydrogen equivalent column density $N_H$, the ionization parameter\footnote[1]{Here, $L_{\rm ion}$ is ionizing luminosity and $n$ is plasma number density at distance $r$ from the BH.} $\xi= L/(n r^2)$, the line of sight (LoS) outflow velocity $v_{\rm out}$ as well as   potential correlations between these quantities with some differences between AGNs and XRBs.

On the other hand, magnetic processes (by both magnetocentrifugal and pressure gradient forces) provide another promising scenario for driving outflows in the context of magnetohydrodynamics (MHD) (e.g. \citealt{BP82}; \citealt{CL94}; \citealt{KK94}; \citealt{Ferreira97}; \citealt{Miller06a,Miller08}; \citealt{F10a}, hereafter F10; \citealt{K12}; \citealt{Miller15,Miller16a}; \citealt{Chakravorty16}; \citealt{F17}, hereafter F17; \citealt{F18}; \citealt{Kraemer18}; \citealt{Jacquemin-Ide20}). 
A large scale poloidal magnetic field\footnote[2]{While not directly observable, the presence of BH magnetosphere of large-scale fields is almost evident as it is essential for sustaining coronae and jets \citep[e.g.][]{Hirose04}.} threading the accretion disk converts efficiently the Keplerian rotation of disk plasma to poloidal motion beyond the \Alfven radius (e.g. F10; \citealt{K12}; F17; \citealt{F18}). This process is largely insensitive to the radiative state properties in XRBs.


With respect to the Galactic XRB sources, one should further note that many are transient and tend to exhibit a non-symmetric, q-shape excursions in their hardness-intensity diagram (a.k.a, ``q-diagram"); i.e. their luminosity increases from a {\it low/hard} state by way of reaching a maximum brightness to a {\it high/soft} state and returns to its {\it low/hard} state through a path. Here, the two modes are typically described as follows; 
%
(1) The {\it high/soft} state in which an accretion disk plays a dominant role in radiative output via multi-color disk (MCD) thermal radiation \citep[e.g.][]{Mitsuda84}. The disk extends down to the innermost stable circular orbit (ISCO; \citealt{Bardeen72}) at the high mass accretion rate. (2) The {\it low/hard} state where the disk is believed to recede away from the ISCO radius (i.e. become truncated) as the mass accretion rate decreases. An optically-thin, hot, compact coronal region of unknown geometrical configuration then develops and is producing the non-thermal (power-law) component in the spectrum. 

It appears that X-ray winds are preferentially (if not exclusively) observed  in {\it high/soft} states, while leaving few wind signatures during  {\it low/hard} states \citep[e.g.][]{Esin97,Ponti12}.
Based on this common hysteresis evolution, it has been suggested that the ``hard" and ``soft" states are related to the sources' luminosity as a fraction of its Eddington value\footnote[3]{Other properties, such as quasi-periodic oscillations (QPOs) and radio jets, have also been known to be associated with the state transition \citep[e.g.][]{Esin97,FenderBelloniGallo04}, issues beyond the scope of this paper.} \citep[e.g.][]{Esin97,DoneGierlinskiKubota07}. The detailed physics responsible for the q-shaped hysteresis behavior throughout the evolution are not well known despite a number of extensive studies \citep[e.g.][]{FenderBelloniGallo04,RM06,DoneGierlinskiKubota07,Belloni10,Ponti12}, although it has been suggested that it may  involve the evolution of their global magnetic field structure \citep[e.g.][]{Kylafis12}.


Most studies of XRB disk winds, due to the inherent weakness of their atomic lines,  
put emphasis on the physical properties (e.g. $v_{\rm out}$, $N_H$ and $\xi$) of the most prominent absorbers, typically those of the Fe K complex. While still informative and valuable, such a {\it single-ion approach} misses several important aspects of the observed winds: (1) Possible interconnections among ions across the broadband in X-ray that ``live" at different regimes of the ionization parameter $\xi$. (2) The possibility to forge a \textit{global} model of these transitions.

For the past several years, our group has studied the ionization structure and properties of photoionized accretion disk winds of the type enunciated by \cite{BP82} and later generalized by \cite{CL94}. Our original aim was to model the ionization of smooth outflows that can naturally accommodate the wide range of ionization states of the observed multi-ion absorbers
at grating resolution. An additional advantage of these outflow models, as also noted in \citet{K12}, is their scale invariance which affords their application to both AGNs and XRBs. 
In conventional analyses, the wide range of ($N_H, \xi, v_{\rm out}$) is constrained phenomenologically by assuming a number of mutually decoupled  components/zones that are arbitrarily chosen to fit data. However, \cite{HBK07} and \cite{B09}, in a confluence with our own views on smooth ionized outflows, identified a continuous relation between $N_H$ and $\xi$. To this end, these authors have introduced the so-called absorption measure distribution (AMD), namely a relation of the form
\begin{eqnarray}
\textmd{AMD} \equiv \frac{d N_H} {d \log \xi} = \xi \frac{d N_H}{d \xi}  \propto \xi^\alpha  \ . \label{eq:eqn1}
\end{eqnarray}
Then, by a global fit to the ensemble of transitions, they distilled the multitude of their properties to the value of a single parameter, namely $\alpha$, which was found to have a rather limited range, $0.0 \lsim \alpha \lsim 0.3$ for a sample of 5 Seyfert 1 disk winds \citep{B09}. It is easy to see that a smooth wind with a density profile of the form $n(r) = n_0 (r/r_0)^{-p}$ implies $p = (2 \alpha +1)/(\alpha +1)$,  corresponding to the range of $1 \lsim p \lsim 1.23$, which is in fact broadly consistent with other independent work on X-ray absorbers (e.g. F10; \citealt{Detmers11}; \citealt{F15}; F17; \citealt{F18}; \citealt{Trueba19}; \citealt{Laha16}).

For the purpose of investigating in detail the observed complex absorbers with subtle features, there  exists to date a set of sufficiently high flux objects with multi-epoch observations; e.g. \4u1630 \citep[e.g.][]{Miller15,Trueba19}, \gro1655 (e.g. \citealt{Miller06a,Miller08}; \citealt{Kallman09}; \citealt{NeilsenHoman12}; \citealt{Miller15}; \citealt{Shidatsu16}; F17), \h1743\ \citep[e.g.][]{Miller06b,Miller12,Miller15,Tomaru20} and \grs1915 \citep[e.g.][]{Ueda09,NeilsenLee09,Miller15,Miller16a,Neilsen20,Ratheesh21}, among other Galactic sources. To this end,  {\it Chandra}/HETGS has been instrumental in providing state-of-the-art dispersive spectroscopy of these sources; these are extremely useful in helping to unravel the underlying physical conditions of the observed ionized outflows. For example, it is generally understood that the disk winds detected in these transient sources are commonly characterized by the following canonical parameters; moderate column densities of $N_H \sim 10^{22}$ cm$^{-2}$, slow/moderate LoS outflow velocity of $v_{\rm out} \sim 100-1,000$ km/s and systematically high ionization parameter of $\log \xi \sim 4-6$ \citep[e.g.][]{Miller15}. Note, however, that exceptionally fast outflows have been claimed in some XRBs, for example, in IGR~J17091-3624 \citep[e.g.][]{King12} and GX~340+0 \citep[e.g.][]{Miller16b} (of significance much inferior to those of GRO~1655-40), which will have to be confirmed with higher S/N observations. Nonetheless, our current focus is placed on the ``canonical" disk-winds.

With \fexxv\ and \fexxvi\ being the most common and distinguishable absorption features in the XRB spectra, we outline below in more detail their properties.
The \fexxvi\ feature is a combination of a spin-orbit doublet between the K$\alpha_1$ transition at energy $E=6.973$ keV (with oscillator strength $f_{ij}=0.277$) and the K$\alpha_2$ one at $E=6.952$ keV (with $f_{ij}=0.139$); the \fexxv\ complex consists of a strong resonance line at $E=6.700$ keV, two intercombination lines at $E_{1,2}=6.682/6.668$ keV and a forbidden line at $E=6.637$ keV \citep[e.g.][]{Bianchi05,Tombesi11}. At energies below the Fe K complex, a wealth of atomic information can be further studied with gratings spectroscopies of a series of soft X-ray absorption signatures originating from various low-atomic number species (e.g. C, N, O, Ne, Ca, Mg, Si, S, Mn, Ar) of various charge states, which may be viewed as the Galactic-source analogs of AGN warm absorbers \citep[e.g.][]{ReynoldsFabian95,Reynolds97, McKernan07, Blustin05,Tombesi13,Laha14,Laha16}. 
%
%
The models of F17 have shown those lines to be consistent with observation in  support of the broader model properties.

In specific observations, \gro1655\ and \grs1915\ in particular have been given much attention given their well-structured winds of various ions (especially the strong \fexxv\ and \fexxvi\ lines) using phenomenological models. 
%
The exceptional high S/N data of multiple ions has further implied absorption at small distances (e.g. $r \lsim 10^5 R_g$) thus favoring 
%
%
magnetic-driving scenario in \gro1655 \citep[e.g.][but also see \citealt{Netzer06}]{Miller06a,Miller08} or perhaps in a hybrid form between thermal and magnetic driving \citep[e.g.][]{Everett05,NeilsenHoman12,Waters18}.
On the other hand, a detailed HD simulation of thermal winds predicts a global outflow density gradient that is   inconsistent with the observed AMD characteristics; i.e. the theoretical AMD predicted by the thermal wind simulations decreases with ionization parameter (i.e. $\alpha < 0$), whereas data clearly indicate $\alpha \sim 0 - 0.3 >0$ \citep[e.g.][]{B09}, a qualitatively fundamental discrepancy \citep[e.g.][]{Dyda17}. 

Motivated by the previous analyses, in this paper we investigate three exemplary BH XRBs, \4u1630\, \h1743\ and \gro1655\, in an attempt to model their absorbers across state transitions by utilizing the previously developed MHD wind model. Our study is thus focused specifically on  three critical points: (1) Modeling broadband multi-ion absorbers ranging from soft to Fe K band by incorporating the AMD method. (2) Constraining the global wind condition characterized primarily by a well-defined density structure during  state transition within a unique MHD-driving scenario. (3) Simulating prospective broadband absorption spectra to be obtained with the next generation of instruments such as {\it XRISM}/Resolve and {\it Athena}/X-IFU in anticipation of better identification of characteristic wind signatures across state transitions.
In \S 2, we overview a set of {\it Chandra}/HETGS data to be analyzed in this work. In \S 3, a brief introduction of the magnetically-launched disk-wind model is given. In \S 4, we present our spectral analysis and the bestfit results of \4u1630\, \h1743\ and \gro1655\ to constrain, for the first time, a large-scale wind condition; i.e. density gradient along a LoS and wind density normalization. Simulated spectra with {\it XRISM}/Resolve and {\it Athena}/X-IFU are also shown in \S 5. Finally in \S 6, we conclude with a summary and discussion

\begin{deluxetable}{l||ccccccccc}
\tabletypesize{\small} \tablecaption{Sources and Observational Log} \tablewidth{0pt}
\tablehead{Source (obsID)$^\ddagger$ & Wind & $M/\Msun$ & $\theta_{\rm obs}$ & Obs.   & Exposure$^\ast$     \\ & State$^\dagger$ &  & [deg] & Date$^\ast$  & [ks] }
\startdata
%
\4u1630 ({\tt obs13716})  & {\bf On} & 10 & 70 & 2012-1-26  & 29.3  \\
\4u1630 ({\tt obs14441})  & Off & - & - & 2012-6-3  & 19.0  \\ \hline
\gro1655 ({\tt obs5460})  & Off & 7 & 80 & 2005-3-12  & 14.9  \\
\gro1655 ({\tt obs5461})  & {\bf On} & - & - & 2005-4-1  & 44.5  \\ \hline
\h1743 ({\tt obs3803})  & {\bf On} & 10 & 70 & 2003-5-1  & 48.0  \\
\h1743 ({\tt obs3804})  & Off & - & - & 2003-5-28  & 40.5  \\
\h1743 ({\tt obs3805})  & {\bf On} & - & - & 2003-6-23  & 50.8  \\
\h1743 ({\tt obs3806})  & {\bf On} & - & - & 2003-7-30  & 50.2  \\
%
\enddata
\label{tab:tab1}
\vspace*{0.2cm}
$^\ast$ {\it Chandra}/HETGS data from {\tt TGCat} \citep[][]{Huenemoerder11}.
\\
$^\dagger$ Apparent appearance of winds (see \S 2 and \S 4).
\end{deluxetable}

\begin{deluxetable}{l||ccccccccccc}
\tabletypesize{\small} \tablecaption{Characteristic X-ray Continua for {\tt Wind-On} States. } \tablewidth{0pt}
\tablehead{Source (obsID) &   $N_H^{\rm Gal}$  & $\Gamma$ & $K_{{\tt po}}$ & $kT$ & $K_{{\tt diskbb}}$ & $L_{\rm ion}$ & Ref. \\ &  & & Norm & & Norm & & }
\startdata
%
%
\4u1630 ({\tt obs13716})  & 9.4 & - & - & 1.44  & 144.2 & 2.0 & [1]  \\ \hline
\gro1655 ({\tt obs5461})  & 0.74 & - & - & 1.42  & 485.9 & 5.0 & [2]  \\ \hline
%
%
\h1743 ({\tt obs3803})  & 2.3 & - & - & 1.32  & 615.3 & 2.8 & [3]  \\
\h1743 ({\tt obs3805})  & 2.3 & 2.6 & 0.07 & 1.15  & 771.8  & 2.9 & [3] \\
\h1743 ({\tt obs3806})  & 2.3 & - & - & 1.06  & 877.5 & 2.0 & [3]  \\
%
\enddata
\label{tab:tab2}
\vspace*{0.1cm}
\begin{flushleft}
{\bf Note}: The X-ray continua as given in the corresponding references for the wind-dominant state of each source along with their estimated ionizing luminosity $L_{\rm ion}$ (in $10^{38}$ erg/s). A set of characteristic spectral components includes a power-law ({\tt po}) of photon index $\Gamma$ and a multi-color disk blackbody ({\tt diskbb}) of temperature $kT$ (in keV) with the Galactic neutral column of $N_H^{\rm Gal}$ (in 10$^{22}$ cm$^{-2}$). [1] \cite{Miller15,Trueba19}.  [2] \cite{Miller06a,Miller08,Miller15}.  [3] \cite{Miller06b,Miller15}.
\end{flushleft}
\end{deluxetable}

\section{{\it The Chandra}/HETGS Data}

We utilize the archival {\it Chandra}/HETGS (first-order) data available from {\tt TGCat} (Transmission Grating Catalog and Archive) selectively choosing a set of representative epochs. These are known as {\it high/soft} states where clear wind signatures have been detected and {\it low/hard} states where very few and weak wind features are discernible as listed in {\bf Table~1}. However, some of the observations considered in this work do not necessarily meet the canonical spectral criteria of the states' properties (e.g. hardness ratio and disk flux) expected from the q-diagram.
In fact, the observed q-diagram often characterizes a locus of points with substantial inherent scatter and stochastic variation. 
%
Hence, we will refer in this paper to these observations as \windon\ state and  \windoff\ state rather than {\it high/soft} state and {\it low/hard} state, respectively, to avoid confusion. 
This naming convention of state distinction is based, for convenience, on the visual inspection of the presence of both soft ($\sim$ 1-4 keV) and Fe K absorbers, a fact that does not negate our findings of this paper\footnote[4]{Introducing a well-defined proxy such as, for example, ``line EW" would be very useful to more systematically quantify the variable strength of general absorption signatures for our future multi-epoch modeling.}. In Figures~\ref{fig:f4}-\ref{fig:f6} the distinction of the \windon/\windoff\ states is clear. The models (thin blue lines) do show some indication of absorption features that, however, is of much weaker amplitude than the local spectral noise).

As mentioned in \S 1, in our work we consider the prototype BH XRB absorbers detected in the well-analyzed sources, \4u1630, \gro1655 and \h1743, which are all likely viewed at high inclination. 

\4u1630\ ($M \sim 10\Msun$, $\theta \sim 70\deg$, $N_H^{\rm Gal} =9.4 \times 10^{22}$ cm$^{-2}$) has been
observed to exhibit recurring disk-winds in a number of {\it Chandra}/grating observations in its q-diagram \citep[e.g.][]{Trueba19}. Along the evolutionary track, we select a representative set of \windon\ state (obs13716) and \windoff\ state (obs14441) for comparison. In obs13716, a number of soft X-ray absorbers (e.g. Ca and Ar) at $\sim 3-5$ keV have been detected besides the usual Fe K absorbers. 

\gro1655\ ($M \sim 7\Msun$, $\theta \sim 60-80\deg$, $N_H^{\rm Gal} =7.4 \times 10^{21}$ cm$^{-2}$) has been intensively studied for the well-resolved broadband absorbers during its 2005 outburst (obs5461; e.g.\citealt{Miller06a,Miller08,Kallman09,NeilsenHoman12,Miller15}; F17). The source had been observed earlier, prior to obs5461 (less than a month earlier - obs5460), and exhibited only a weak absorption feature. The spectrum in the \windon\ state  (obs5461) contains a wealth of atomic transitions attributed to diverse species of ions of broad $\xi$ range at moderate velocities. The {\it Chandra} spectra in the $\sim 2-10$ keV band in these two epochs look very similar, perhaps because of their time proximity. However, the power-law continuum for $E>10$ keV is clearly harder during the \windoff\ state (obs5460) \citep[e.g.][]{NeilsenHoman12}. 

Lastly, \h1743\ ($M \sim 10\Msun$, $\theta \sim 70\deg$, $N_H^{\rm Gal} =2.3 \times 10^{22}$ cm$^{-2}$) has been observed multiple times including the 2003 outburst phase where the X-ray spectra clearly show a complex Fe K absorber variability during the \windon\ state (obs3803/3805/3806) as well as the \windoff\ state (obs3804) \citep[e.g.][]{Miller06b,Miller15,Tomaru20}. There exist a few more XRBs with appreciable soft X-ray absorption signatures \citep[e.g.][for GX~339-4]{Miller04}. However, our present study is focused on the above three sources to avoid other potentially complex spectral features such as (relativistic) disk reflection \citep[e.g.][]{Garcia15}.

One should be reminded that almost all these investigations and modeling in the past have been focused primarily on Fe K absorbers (i.e. \fexxv\ and \fexxvi) during {\it high/soft} state in the q-diagram.
Incorporating  physically-motivated theoretical disk-wind models in a  re-analysis of the data, we aim to establish, for the first time, the corresponding global wind properties in the \windon\ and \windoff\ states  transition by exploiting our  MHD disk-wind models  we have been developing.

\begin{figure}[t]
\begin{center}
\includegraphics[trim=0in 0in 0in
0in,keepaspectratio=false,width=4in,angle=-90,clip=false]{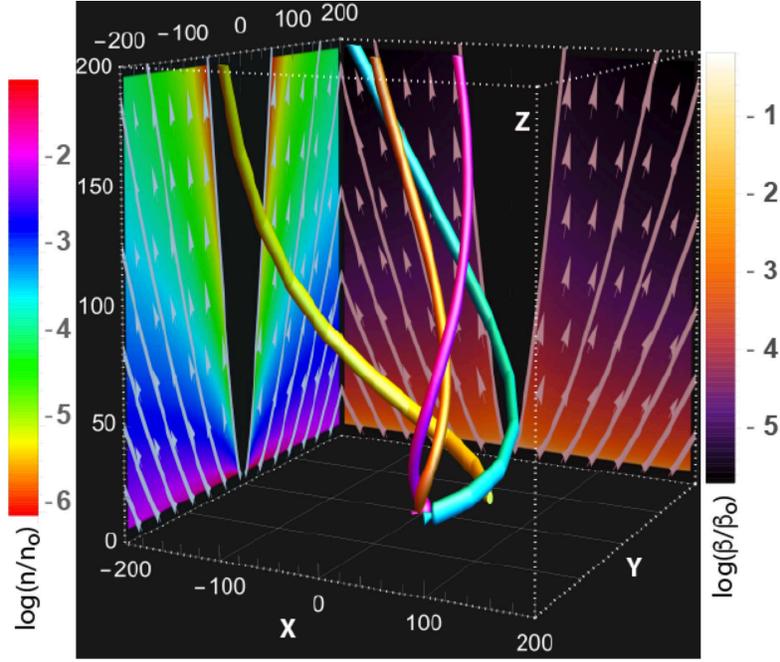}
\end{center}
\caption{Streamlines of a simulated MHD-wind for $p=1.3$ assuming a fiducial set of wind parameters  launched from three different locations on the disk surface. Poloidal distribution of normalized wind density $\log (n/n_{o})$ and plasma beta $\log (\beta/\beta_o)$ (where $\beta_o$ is a maximum value) are shown in color superimposed by wind velocity vector $\mathv(r,\theta)$ (white arrows) and poloidal projection of the magnetic field lines (thick white solid lines) in the innermost wind domain for $-200 R_g \le x \le 200 R_g$ and $-200 R_g \le y \le 200 R_g$. }
\label{fig:f1}
\end{figure}

\clearpage

\section{Model Description}

\subsection{Magnetized Winds}


The study of MHD-driven winds has a long history\footnote[5]{We refer the reader to the treatments of \citet{BP82} and \citet{CL94} and the more recent ones \citep[e.g.][]{Ferreira97,Jacquemin-Ide20}.}. We note that these winds are different from those of the standard radiation or thermally driven ones. In addition to their different launching mechanisms, they are launched across the entire disk domain that spans several decades in radius; i.e.  ranging from near the BH horizon to about half the distance to the companion star in XRB. 
As such, the solutions are naturally self-similar spanning the requisite number of decades in radius. For the purposes of our analysis, what really matters is the radial dependence of their density and radial (i.e. LoS)  velocity. These are given by
\begin{eqnarray}
n(r,\theta) \equiv n_{17} \left(\frac{r}{R_{\rm in}} \right)^{-p} f(\theta) ~~~{\rm and} ~~~ v_r(r,\theta) \sim v_{\rm K,in} \left(\frac{r}{R_{\rm in}}\right)^{-1/2} g_r(\theta), \label{eq:eqn2}
\end{eqnarray}
where $p$ denotes radial dependence of the wind, $n_{17}$ is the wind density normalization  (in the units of $10^{17}$ cm$^{-3}$) on the foot point of a streamline launched from the innermost radius at $r=R_{\rm in}$ in spherical coordinates $(r,\theta)$ and $v_{\rm K,in}$ is the Keplerian velocity at $r=R_{\rm in}$ being typically identified with the ISCO (see F10). We have also computed the angular dependences of the wind density and velocity components coupled to the field geometry, $f(\theta)$ and $g_i(\theta)$ (where $i=r,\theta,\phi$), respectively, by solving numerically the Grad-Shafranov equation (see, e.g., \citealt{CL94} for the original formulation and \citealt{F14} for more detailed wind morphology description). 
%
Note that $f(\theta)$ is maximum  at the equatorial disk plane (such that $f(90\deg) = 1$ by definition) and decreases rapidly with latitude (see F10) as approximately expressed by a simple fitting formula
\begin{eqnarray}
\log f(\theta) \simeq -12.2-3.3 \theta^{-1/2} + 3.2 \theta^{1/2}-0.28 \theta + 9.1 \theta^2 \times 10^{-4} \ , \label{eq:eqn3}
\end{eqnarray}
where $\theta$ is in degrees.   
%
Since $g_\theta(\theta)$ and $g_r(\theta)$ are both very small near the disk plane (i.e. $g_{r,\theta}(\pi/2) \ll 1$), $g_\phi(\theta)$ dominates the wind kinematics on the base of the wind. The poloidal velocity component rapidly increases indicating the escape of the wind particles as $g_{\phi}(\theta)$ decreases like $\sim (1/r) \sin \theta$ beyond the \Alfven radius to conserve wind angular momentum.
As noted in \cite{K12}, these equations are scale invariant and are applicable to Galactic as well as  AGN accretion disks, provided that the radial coordinate is scaled by the \sw radius and the luminosities and the mass flux rates to their Eddington values.

In {\bf Figure~\ref{fig:f1}} we present a three-dimensional (3D) visualization of a fiducial MHD disk-wind morphology focused on the innermost region of $200 R_g \times 200 R_g$ domain. Shown are three different streamlines that are launched from different locations on the disk surface. Also shown are the poloidal projection of the wind density $\log [n(r,\theta)/n_{o}]$ and the plasma $\beta (\equiv P_{\rm gas}/P_{\rm mag})$, $\log [\beta(r,\theta)/\beta_o]$, normalized to their maximum values, $n_o$ and $\beta_o$ at $r=R_{\rm in}$, respectively. The poloidal  geometry of magnetic field lines (solid white lines) along with wind velocity $\mathv(r,\theta)$ (white arrows) are also shown. It is clear that the wind density is highest near the disk surface (i.e. the equator) that provides a reservoir of the outflowing plasma.
The calculated plasma $\beta \equiv P_{\rm gas}/P_{\rm mag}$ is seen to decrease by orders of magnitude from the disk surface towards the funnel region as the plasma gas pressure drops faster than that of the magnetic field, thus effectively stabilizing and collimating ionized outflows.
Disk-winds are efficiently accelerated along a field line mostly before the outflow reaches the \Alfven point (see F10). Because of the unique 2D wind morphology (unlike a quasi-spherical 1D radiation-driven one), a given LoS will ``cut through" regions of widely varying values of $N_H, \xi$ and $v_{\rm out}$; the the outer part of the wind 
can be 5-6 orders of magnitude larger than $R_{\rm in}$. This property is essential in producing an AMD (see eqn.~(\ref{eq:eqn1})) in agreement with observations.  While modeling the Fe K absorbers alone does not provide much information about the density gradient $p$, multi-ion broadband absorbers do provide useful insight into characterization of the global wind condition.

Assuming a fiducial set of  parameters considered in our series of disk-wind modeling of AGNs \citep[see][]{F18} and BH XRBs (see F17), our previous studies have indicated that the value $p \sim 1.2$ is statistically favored in general (F10; F17; \citealt{F18}), and this obtained value of $p \sim 1.2$ is indeed consistent with the observed value from the AMD analyses using multiple ions of various charge state in Seyfert 1 AGNs \cite[e.g.][]{B09,Detmers11}.
%


\subsection{Spectral Modeling of Broadband Absorbers}

We employ the wind model described above to simulate the underlying MHD wind structure mainly governed by the radial density slope ($p$) and the density normalization ($n_{17}$). Based on the previous investigations, we consider a wide range of parameter space spanned by $0.9 \le p \le 1.5$  and $0.01 \le n_{17} \le {128}$. Note that $p=0.9$ wind provides an increasing column per decade in distance, while $p=1.5$ wind corresponds to the \cite{BP82} type of MHD-wind density profile with column density that decreases like $N_H \propto r^{-1/2}$ or alternatively $N_H \propto \xi$.
Our current approach follows directly the methodology adopted in F10. For photoionization calculations with {\tt XSTAR} \citep[][]{KallmanBausas01}, we use the reported X-ray continua in the literature as   ionizing spectra consisting primarily of a power-law of photon index $\Gamma$ and MCD radiation of the innermost temperature $kT$ as listed in {\bf Table~\ref{tab:tab2}}.

\begin{figure}[t]
\begin{center}
\includegraphics[trim=0in 0in 0in
0in,keepaspectratio=false,width=3.3in,angle=-0,clip=false]{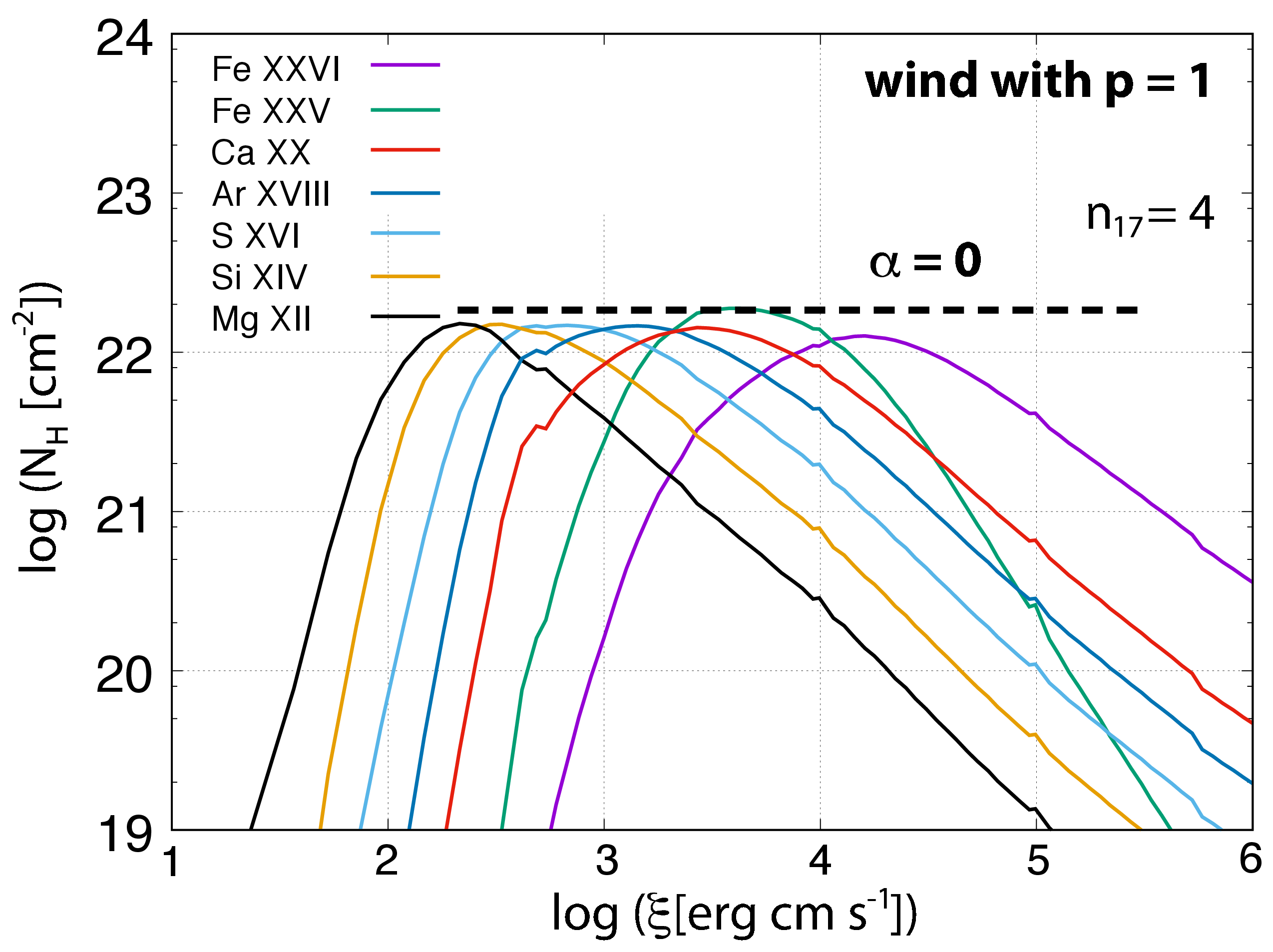}\includegraphics[trim=0in 0in 0in
0in,keepaspectratio=false,width=3.3in,angle=-0,clip=false]{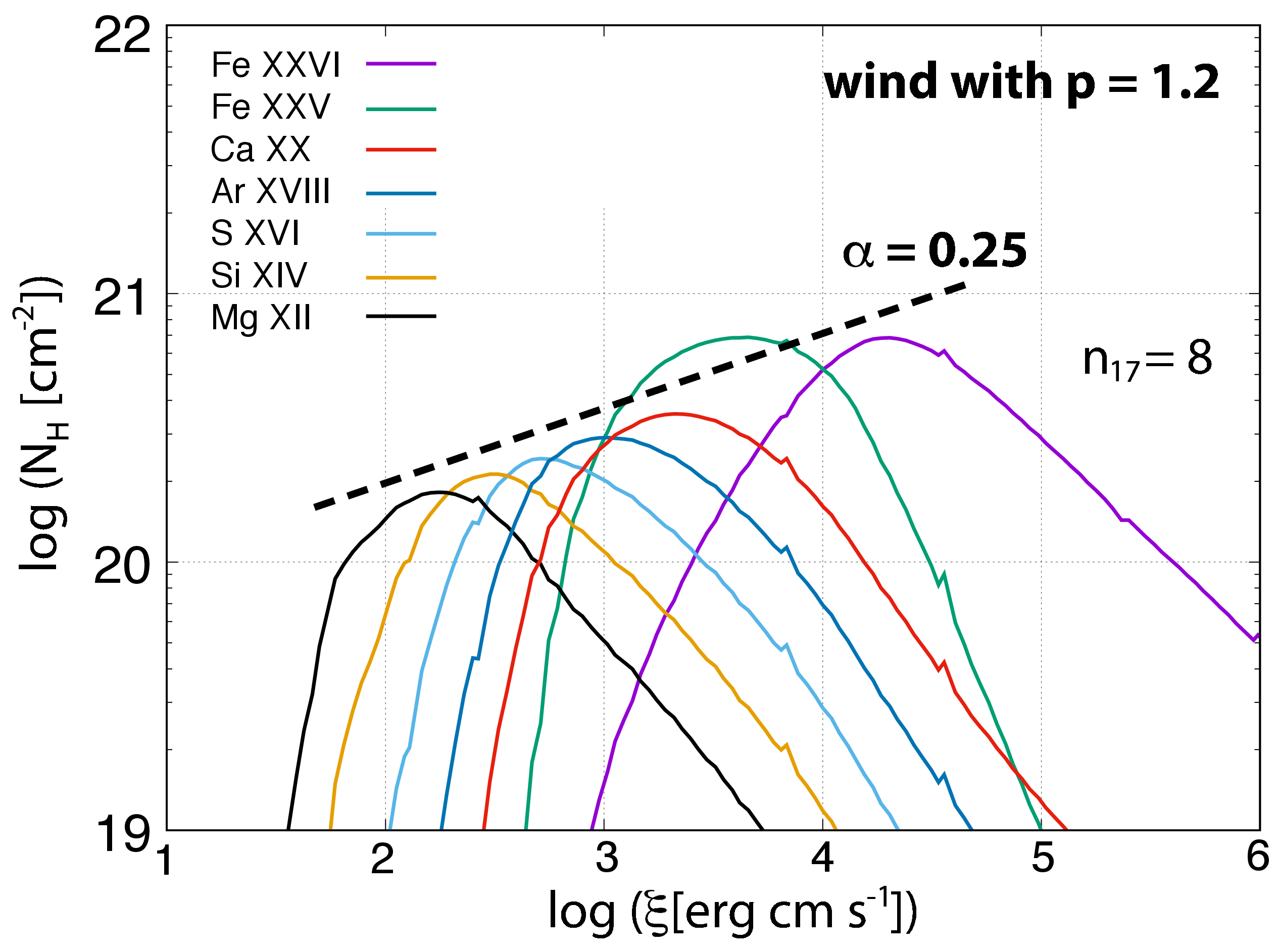}
\end{center}
\caption{Examples of calculated absorption measure distribution (AMD) for a series of major ions from fiducial winds with $p=1$ (i.e. $\alpha=0$) and $n_{17}=4$ (left) while  $p=1.2$ (i.e. $\alpha=0.25$) and $n_{17}=8$ (right) where $p=(2\alpha+1)/(\alpha+1)$. See \S 1. }
\label{fig:f2}
\end{figure}

The local ionic column $N_{\rm ion}$ is computed with {\tt XSTAR}  under thermal equilibrium, whereas the photo-absorption cross section $\sigma_{\rm abs}$ is separately calculated using
the usual Voigt profile as a function of photon frequency $\nu$ and  the line broadening factor  $\Delta \nu_\ell \approx  (\Delta v_{r}/c) \nu_c$ relative to the centroid (rest-frame) frequency $\nu_c$. 
We show in {\bf Figure~\ref{fig:f2}} examples of calculated AMDs from fiducial wind solutions with $p=1$ (i.e. $\alpha=0$) and $n_{17}=4$ (left) while  $p=1.2$ (i.e. $\alpha=0.25$) and $n_{17}=8$ (right). It is clearly demonstrated that a series of major ions (Fe, Ca, Ar, S, Si and Mg among others) that we consider are progressively formed in response to irradiating X-rays according to the radial density slope, $p$. As expected, the (hydrogen-equivalent) column $N_H$ is almost uniformly distributed over many decades in $\xi$ for $p=1$ winds, whereas the column gradually decreases with decreasing $\xi$ for $p=1.2$ winds according to the expected relation of $p=(2\alpha+1)/(\alpha+1)$ as previously observed \citep[e.g.][]{HBK07,B09,Detmers11,Trueba19}. Our spectral modeling thus comports with  this AMD property as well.

As another key feature of the model, the LoS (radial) shear velocity of the wind naturally provides the line broadening rather than an arbitrarily assumed (thermal) turbulent motion. As often suggested \citep[e.g.][]{Kallman09,Miller15,Garcia15}, super-solar abundances for heavy ions (e.g. Fe, S and Si) are also considered as another set of free parameters in the range of $1 \le A_{\rm Fe, S, Si} \le 3$. One can then calculate the line optical depth  $\tau_\ell = \sigma_{\rm abs} N_{\rm ion}$ to simulate the spectrum. Intrinsic source parameters such as the continuum spectral shape, the disk inclination angle $\theta$ and the X-ray luminosity  $L_{\rm ion}$, while highly uncertain,  are also used for creating a template library of synthetic spectra with photoionization calculations.
It should be reminded that our goal is to test the hypothesis that  disk-wind is persistently present throughout state transition in BH XRBs by utilizing the MHD-driven wind model for spectral fits over multi-epoch data. To this end, we are not focused much on a rigorous determination of the underlying continua and it is beyond the scope of our current work to further investigate those in detail.

\begin{deluxetable}{l|lllllllll}
\tabletypesize{\small} \tablecaption{Bestfit Model Parameters of the MHD-Wind} \tablewidth{0pt}
\tablehead{  & Slope & Density & Iron & Sulphur & Silicon & Goodness \\
Source (obsID)$^\dagger$  & $p$ & $n_{17}$  & $A_{\rm Fe}$ & $A_{\rm S}$ & $A_{\rm Si}$  & $\chi^2$/dof \\ &  & [$10^{17}$ cm$^{-3}$] &  &  & }
\startdata
%
\4u1630 ({\bf obs13716}) & $1.41_{-0.01}^{+0.07}$ & $3.87_{-0.8}^{+7.2}$ & $3.0_{-1.1}^{p}$ & $1.0_{p}^{+0.18}$ & $1.0_{p}^{+0.62}$ & 1691/2455  \\
\4u1630 ({\tt obs14441}) & $1.46_{-0.19}^{p}$ & $1.0_{p}^{0.42}$ & $1.37_{p}^{+0.58}$ & $1.88_{p}^{p}$ & $2.69_{-1.84}^{p}$ & 2156/2458  \\ \hline
\gro1655 ({\tt obs5460}) & $1.39_{-0.05}^{+0.04}$ & $0.089_{-0.012}^{+0.075}$ & $1.44_{-0.31}^{+0.32}$ & $1.16_{p}^{+0.92}$ & $1.55_{-0.43}^{p}$ & 5749/4335   \\
\gro1655 ({\bf obs5461}) & $1.34$ & $3.0$ & $3.0$ & $1.0$ & $1.0$ & 27271/4335   \\ \hline
\h1743 ({\bf obs3803}) & $1.19_{-0.02}^{+0.05}$ & $0.013_{-0.001}^{+0.001}$ & $1.95_{-0.13}^{+1.01}$  & $1.00_{p}^{+0.14}$   & $3.00_{-0.83}^{p}$ & 8577/4341  \\
\h1743 ({\tt obs3804}) & $1.31$ & 0.019   &  1.19 & 1.05   & 2.99 & 5899/2921   \\
\h1743 ({\bf obs3805}) & $1.19_{-0.003}^{+0.04}$ & $0.0117_{-0.001}^{+0.001}$ & $1.47_{-0.21}^{+0.22}$  & $1.0_{p}^{+0.53}$ & $3.0_{-1.0}^{p}$ & 4605/2681  \\
\h1743 ({\bf obs3806}) &  $1.19_{-0.005}^{+0.004}$ & $0.011_{-0.001}^{+0.001}$ & $2.78_{-0.72}^{p}$  & $1.0_{p}^{+0.94}$   & $3.0_{-1.0}^{p}$ & 4338/2683  \\
\enddata
\label{tab:tab3}
\begin{flushleft}
Unless otherwise stated, we assume the solar abundances for all elements.
\\
$^\dagger$ We note \windon\ state in {\bf bold} font.
``$p$" denotes that values are pegged at hard limit.
\end{flushleft}
\end{deluxetable}

\begin{figure}[t]
\begin{center}$
\begin{array}{cc}
\includegraphics[trim=0in 0in 0in
0in,keepaspectratio=false,width=3.3in,angle=-0,clip=false]{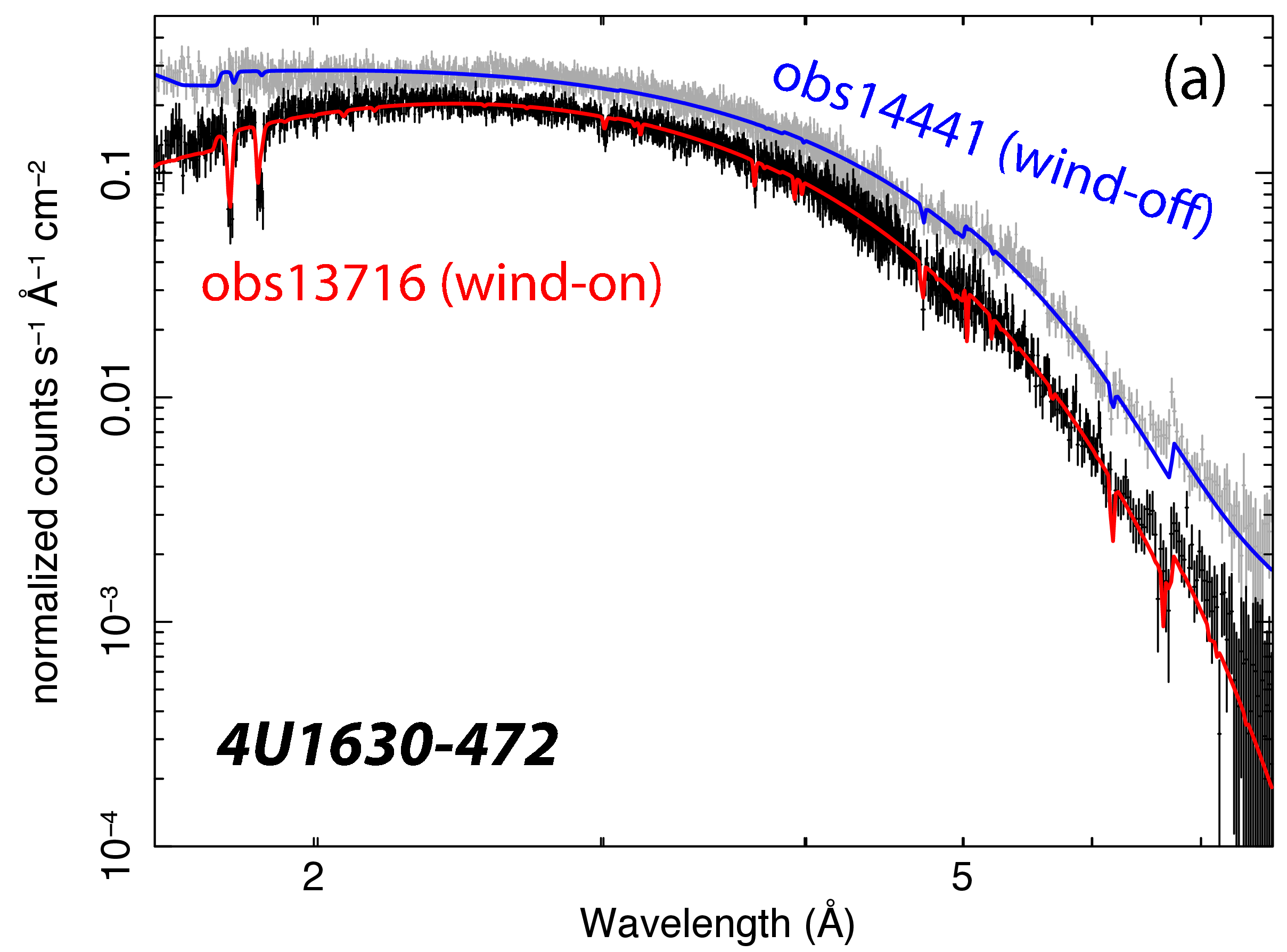}
\includegraphics[trim=0in 0in 0in
0in,keepaspectratio=false,width=3.3in,angle=-0,clip=false]{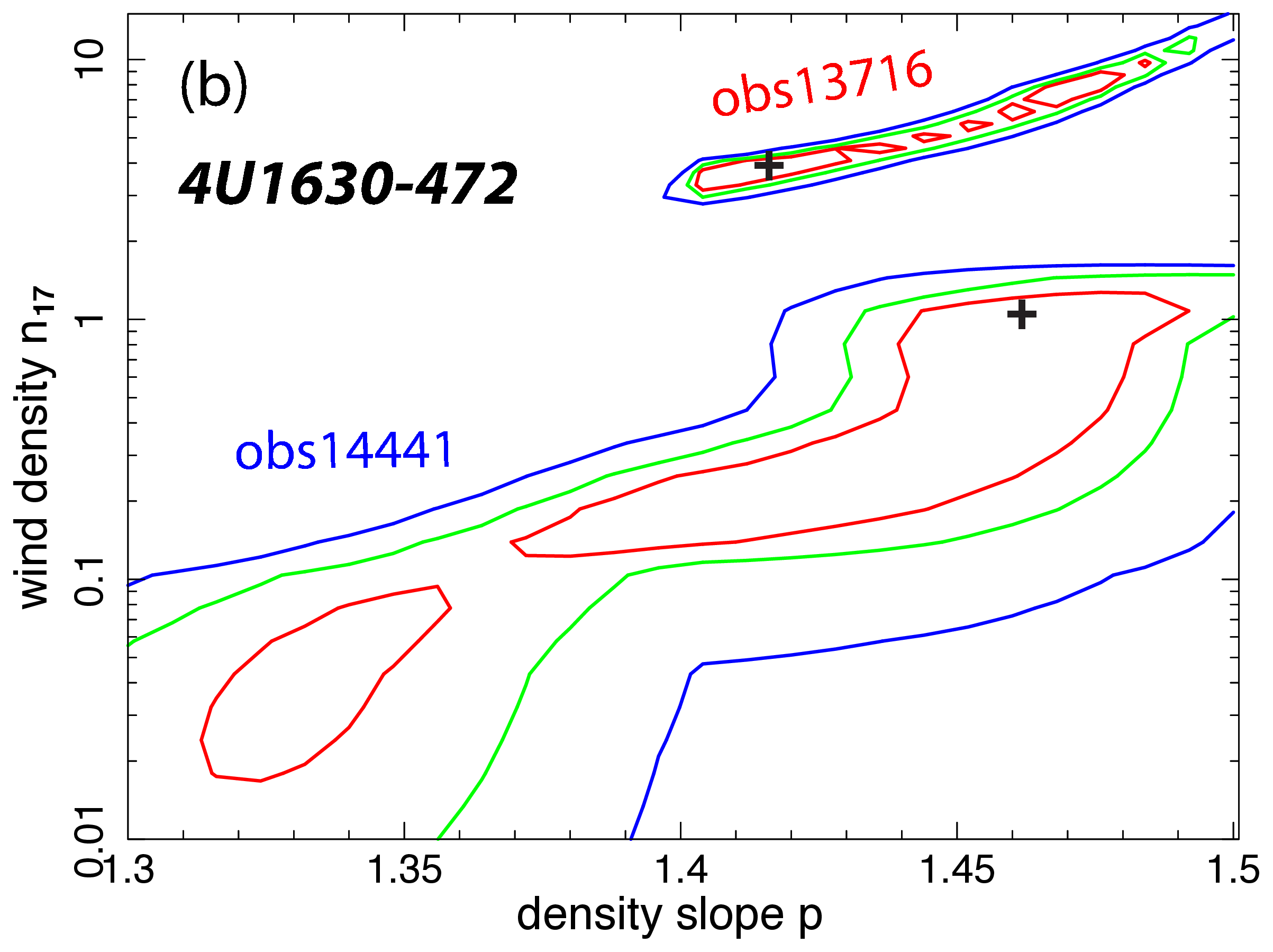}
\end{array}$
\end{center}
\caption{Bestfit MHD-wind models for (a) the observed {\it Chandra}/HETGS spectra of \4u1630\ and  (b) its confidence contour map (68\% in red; 90\% in green; 99\% in blue) in {\tt obs13716} ({\tt wind-on}) and {\tt obs14441} ({\tt wind-off}). See {\bf Table~\ref{tab:tab3}} for the besfit model parameters. }
\label{fig:f3}
\end{figure}

\section{Results}

\subsection{Broadband Fit with MHD-Wind Models}

Our spectral analyses are performed using {\tt XSPEC} v12.10.0c \citep{Arnaud96} based on $\chi^2$-minimization method.
As the dispersive instrument is configured with a binning of the same size in wavelength, we show the {\it Chandra}/HETGS spectra in {\it wavelength}. All the fits are made to the first-order grating spectra.

\begin{figure}[t]
\begin{center}$
\begin{array}{cc}
\includegraphics[trim=0in 0in 0in
0in,keepaspectratio=false,width=3.3in,angle=-0,clip=false]{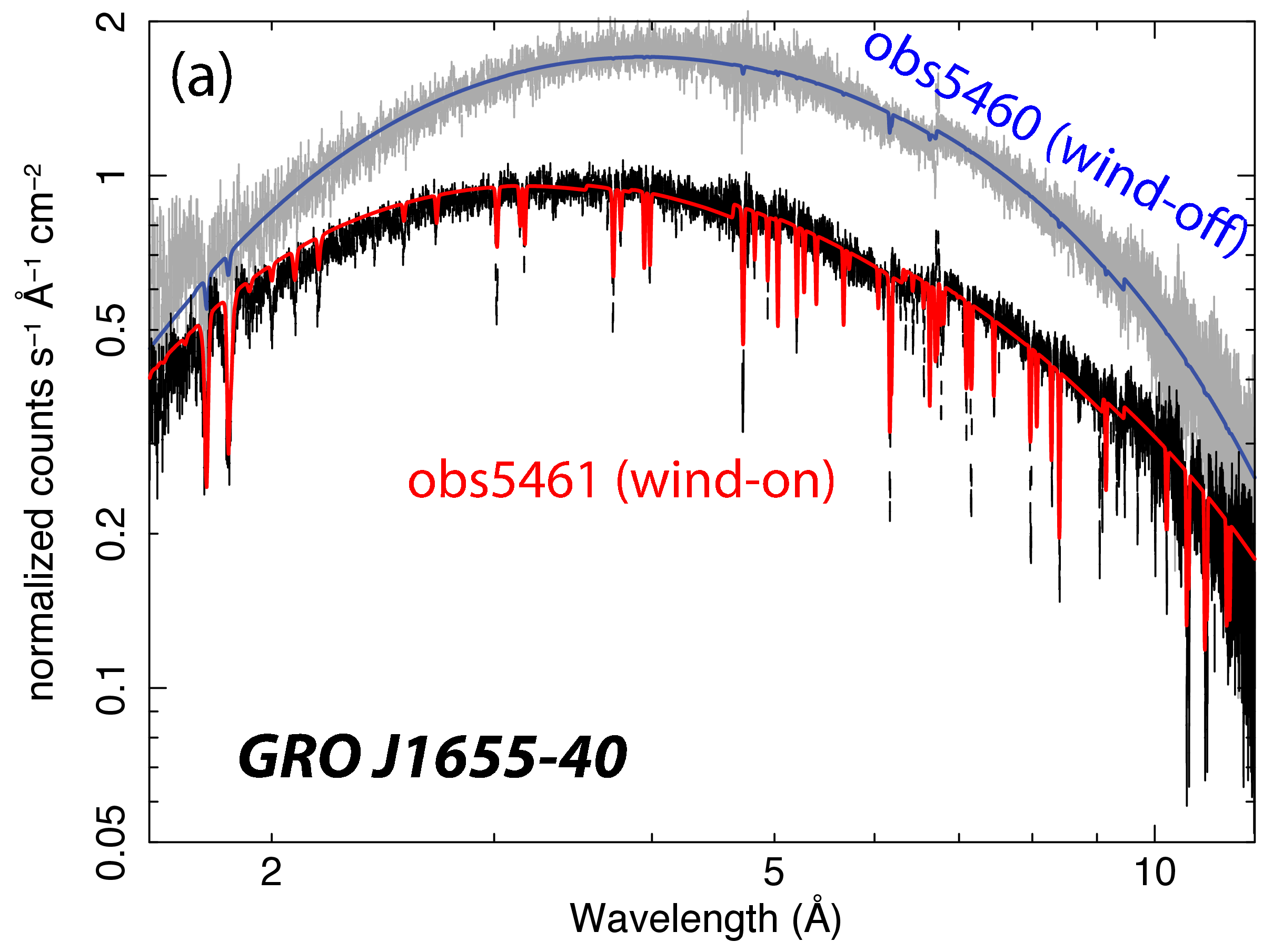}
\includegraphics[trim=0in 0in 0in
0in,keepaspectratio=false,width=3.3in,angle=-0,clip=false]{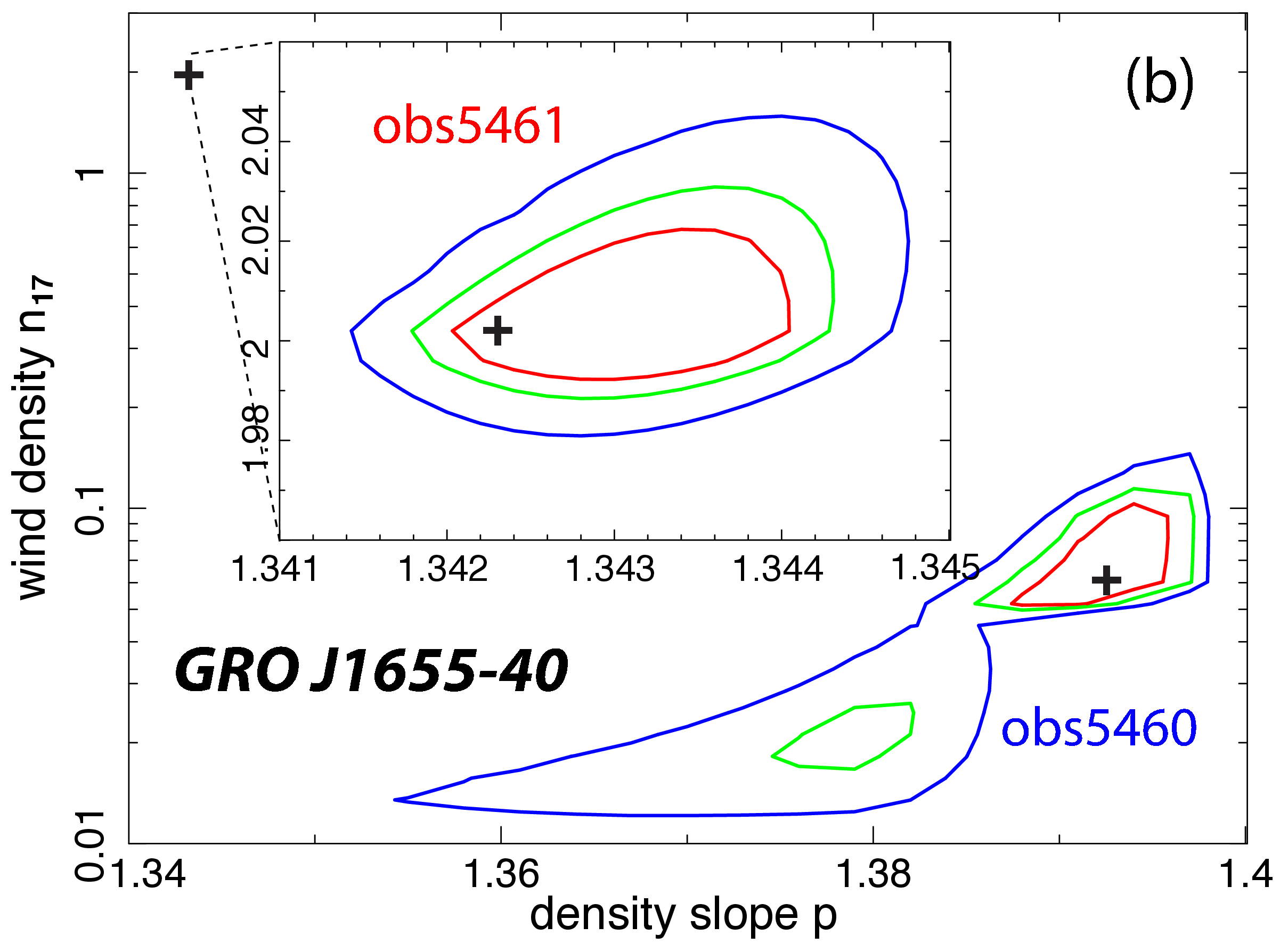}
\end{array}$
\end{center}
\caption{Bestfit MHD-wind models for (a) the observed {\it Chandra}/HETGS spectra of \gro1655\ and (b) its confidence contour map (68\% in red; 90\% in green; 99\% in blue) in {\tt obs5460} ({\tt wind-off}) and {\tt obs5461} ({\tt wind-on}). The former spectrum is vertically offset for presentation purpose. See {\bf Table~\ref{tab:tab3}} for the besfit model parameters. }
\label{fig:f4}
\end{figure}

\begin{figure}[t]
\begin{center}$
\begin{array}{cc}
\includegraphics[trim=0in 0in 0in
0in,keepaspectratio=false,width=3.3in,angle=-0,clip=false]{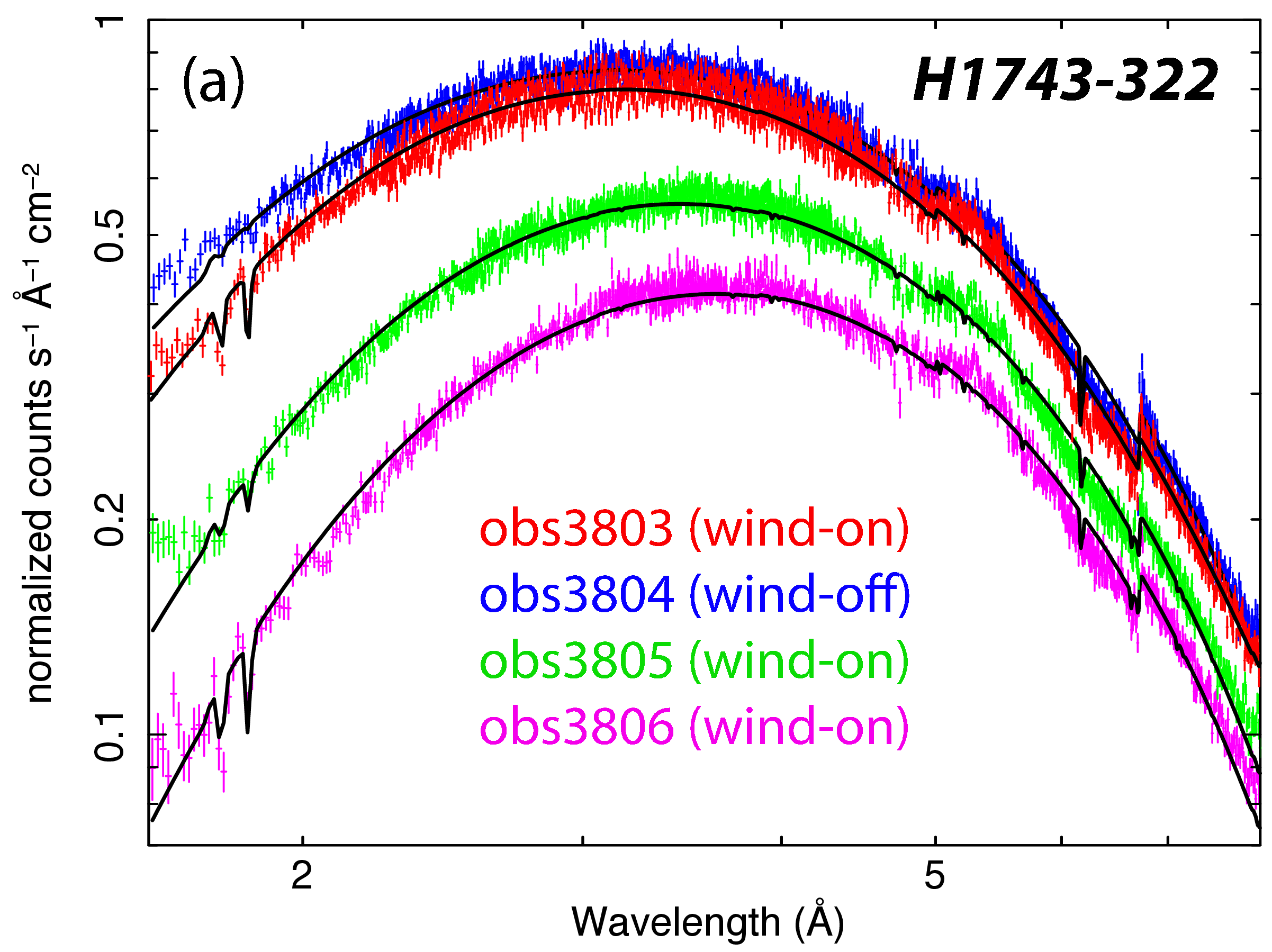}
\includegraphics[trim=0in 0in 0in
0in,keepaspectratio=false,width=3.3in,angle=-0,clip=false]{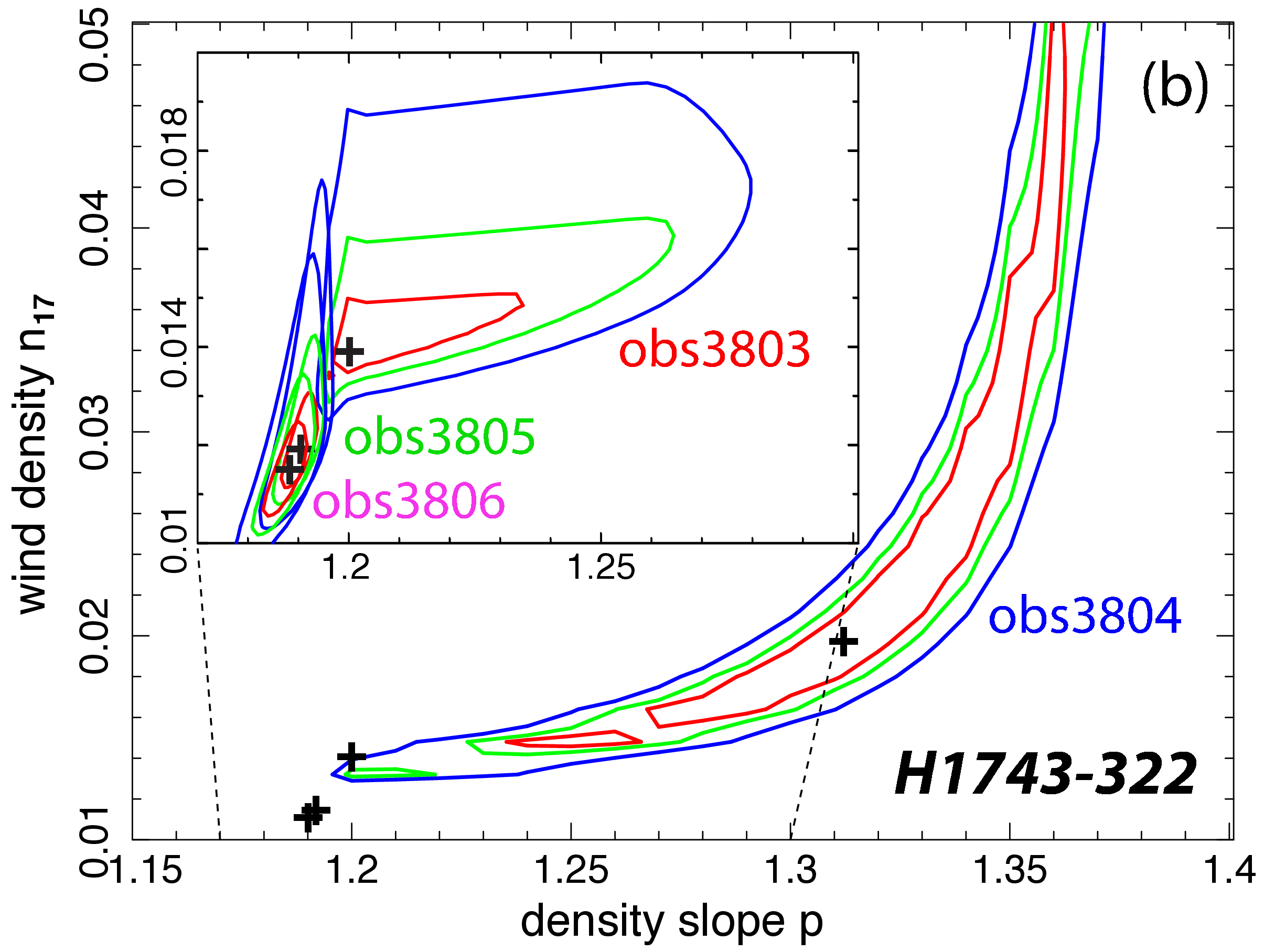}
\end{array}$
\end{center}
\caption{Bestfit MHD-wind models for (a) the observed {\it Chandra}/HETGS spectra of \h1743\ and (b) its confidence contour map (68\% in red; 90\% in green; 99\% in blue) in {\tt obs3803} (\windon), {\tt obs3804} (\windoff), {\tt obs3805} (\windon) and {\tt obs3806} (\windon). See {\bf Table~\ref{tab:tab3}} for the besfit model parameters. }
\label{fig:f5}
\end{figure}

Following the literature for individual sources (see {\bf Table~\ref{tab:tab2}} for details), we first fix the continuum components  in  $\sim 1-10$ keV band, i.e. the power-law and MCD components.
Subsequently, we apply the computed wind component as a multiplicative table model for a broadband spectral fit.
The obtained bestfit broadband spectra for each state (i.e. \windon\ and \windoff\ state) for each source are presented in {\bf Figures~\ref{fig:f3}-\ref{fig:f5}} along with the contour plots of ($p,n_{17}$). The bestfit parameters are listed in {\bf Table~\ref{tab:tab3}}. The wind parameters, in passing from the \windon\ to the \windoff\
states, exhibit the following {\it three} broad trends: (1) the wind density slope $p$ increases, (2) the density  normalization $n_{17}$ (significantly) decreases  and (3) super-solar abundances for Fe ($A_{\rm Fe}>1$) are favored. 
%
%
We point out that even a small change in the slope $p$ can result in a significant difference between the  expected ionic columns of soft X-ray absorbers (i.e. ions with lower $\xi$ at larger $r$) and those from Fe K lines (i.e. ions with higher $\xi$ at smaller $r$), considering that they are separated both in $\xi$ and $r$ as described in equation~(\ref{eq:eqn1}).  
%
More specifically, the wind density normalization $n_{17}$ varies  quite a bit by factors of a few to over an order of magnitude indicating a significant mass flux change between these states. The data systematically favor super-solar metal abundances, especially Fe during the \windon\ state in every source, consistent with the past analysis \citep[e.g.][]{Kallman09}.
Note also that both bestfit values of $n_{17}$ and $A_{\rm Fe}$  obtained from the absorption features in this work roughly agree with  the values obtained independently from relativistic disk reflection spectroscopy in BH XRB accretion \citep[e.g.][]{Garcia16}.

From the derived contour maps, it is inferred that the bestfit solution in each case is well constrained, particularly in the density normalization parameter $n_{17}$.
It is also clearly demonstrated that the density slope $p$ spans a rather restricted range of values during \windon\ state (i.e. $1.2 \lsim p \lsim 1.4$) in  good agreement with the independent studies with AGN warm absorbers using AMD analyses \citep[e.g.][]{B09,Detmers11,Laha14,Laha16,F18}.
Another  spectral change associated with the wind transition is the apparent suppression of the observed soft X-ray absorbers,  while the Fe K absorbers remain  present even in the \windoff\ state (most notably seen in \gro1655\ and \4u1630; see also
\citealt[][]{NeilsenHoman12}). This trend is also hinted at the wind transition of GRS~1915+105 \citep[e.g.][]{Ratheesh21}. The above variable nature of the derived wind parameters ($p,n_{17}$) in {\bf Table~\ref{tab:tab3}} is thus a manifestation of the potential bi-modality of the large-scale wind  properties. It is apparent that it would be hard to draw similar conclusions by focusing on the properties of a single ion (i.e. Fe K absorbers only) in a conventional approach of a Gaussian line fit. To follow the variable nature of the observed broadband absorbers, we further discuss our modeling as a function of the wind model parameters in \S 4.2.
 
In our analysis, we allow ion abundances to vary from one state to the other within a given source as shown in Table~3. However, it is also physically reasonable that the intrinsic quantities like abundances are less likely to drastically change. Assuming that the ``wind-on" state would better reflect more realistic ion abundances (due to stronger absorbers), one could instead freeze abundances at the same values obtained from the corresponding ``wind-on" states. Under this assumption, 
%
%
we set $A_{\rm Fe} = 3.0, A_{\rm S} = A_{\rm Si} = 1.0$ in \gro1655, $A_{\rm Fe} = 3.0, A_{\rm S} = A_{\rm Si} = 1.0$ in \4u1630\ and $A_{\rm Fe} = 2.78, A_{\rm S} = 1.0, A_{\rm Si} = 3$ in \h1743\ such that the ion abundances are  identical between the ``wind-on" and ``wind-off" states for each source. As a result, we find $(p, n_{17}) = (1.48_{-0.02}^{+0.01}, 1.31_{-0.29}^{+0.75})$ or $(1.36_{-0.02}^{+0.01}, 0.0187_{-0.005}^{+0.008})$ (where these are almost degenerate) in \gro1655\ (obs5460), $(1.47_{-0.26}^{p}, 0.10_{p}^{+0.17})$ in \4u1630\ (obs14441) and $(1.36, 0.049)$ in \h1743\ (obs3804) for the new bestfit solutions; i.e. either the density slope steepens or the density normalization significantly decreases in the \windoff\ states for all 3 sources as previously found (in Table 3) regardless of whether abundances are fixed or not. Hence, our conclusion still holds that the wind conditions must change across the wind state transition in a way described above.

To complement the bestfit results presented above, we list in {\bf Table~\ref{tab:tab4}} some characteristic parameters  of the Fe K absorbers of \4u1630\ \windon\ state (obs13716) and \h1743\  \windon\ state (obs3803) that are estimated from a theoretical template spectrum in the model. A quantity with the subscript ``p" denotes a value at a LoS distance where the peak (maximum) column is locally produced. The velocity range $\Delta (v_{1/2})$ is computed over the radial extent over which 50\% of the peak column is being produced. We calculate the innermost (minimum) distance $r_{\rm min}$ along a LoS that produces 50\% of the peak column, together with wind temperature $T_p$, column $N_H$ and EW. Because of the instrumental degradation of the {\it Chandra} gratings, however, the derived values from observations may not exactly match, but these are broadly consistent with the general characteristics of the observed BH XRB winds \citep[e.g.][]{King14,Trueba19} as well as predictions from other wind models \citep[e.g.][]{Tomaru20}.

\begin{deluxetable}{l||ccccccc}
\tabletypesize{\small} \tablecaption{Derived Characteristics of Fe K Absorbers in a Template Model} \tablewidth{0pt}
\tablehead{Quantities & $\log \xi_p$ $^\dagger$  & $\Delta (v_{1/2})$ $^\ddagger$ & $\log (r_{\rm min}/R_g)$ $^\diamond$ & $\log T_p$ $^\dagger$ & $N_H$ & EW \\
& [erg~cm~s$^{-1}$] & [km~s$^{-1}$] & & [K] & [$10^{22}$ cm$^{-2}$] & [eV] }
\startdata
\4u1630\ (obs13716) &      &           &     &     &      &   \\
\fexxv\  & 3.67 & 304 - 358 & 6.8 & 6.2 & 0.93 & 21.1 \\
\fexxvi\ & 4.26 & 323 - 627 & 5.2 & 6.5 & 1.6 & 22.4 \\ \hline
\h1743\ (obs3803)   &      &           &     &     &      &   \\
\fexxv\  & 3.62 & 306 - 339 & 6.9 & 6.0 & 2.0 & 21.6 \\
\fexxvi\ & 4.35 & 321 - 419 & 5.9 & 6.1 & 1.5 & 11.8 \\
\enddata
\label{tab:tab4}
\vspace*{0.0cm}
$^\dagger$ Value at a LoS distance where the column peaks.  \\
$^\ddagger$ Range of value over which a 50\% of the peak column is being produced along a LoS. \\
$^\diamond$ LoS distance where a 50\% of the peak column is produced.
\end{deluxetable}

\begin{figure}[t]
\begin{center}$
\begin{array}{cc}
\includegraphics[trim=0in 0in 0in
0in,keepaspectratio=false,width=3.3in,angle=-0,clip=false]{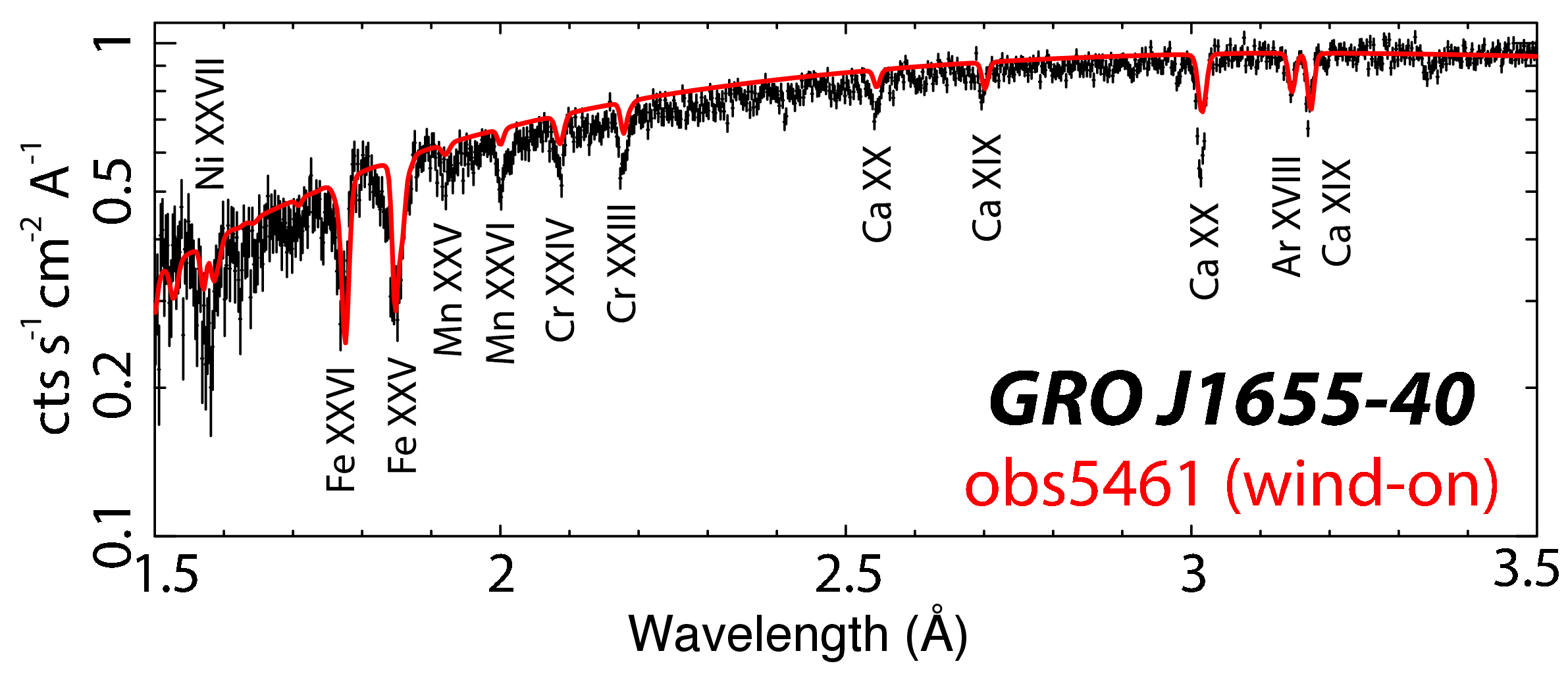}\includegraphics[trim=0in 0in 0in
0in,keepaspectratio=false,width=3.3in,angle=-0,clip=false]{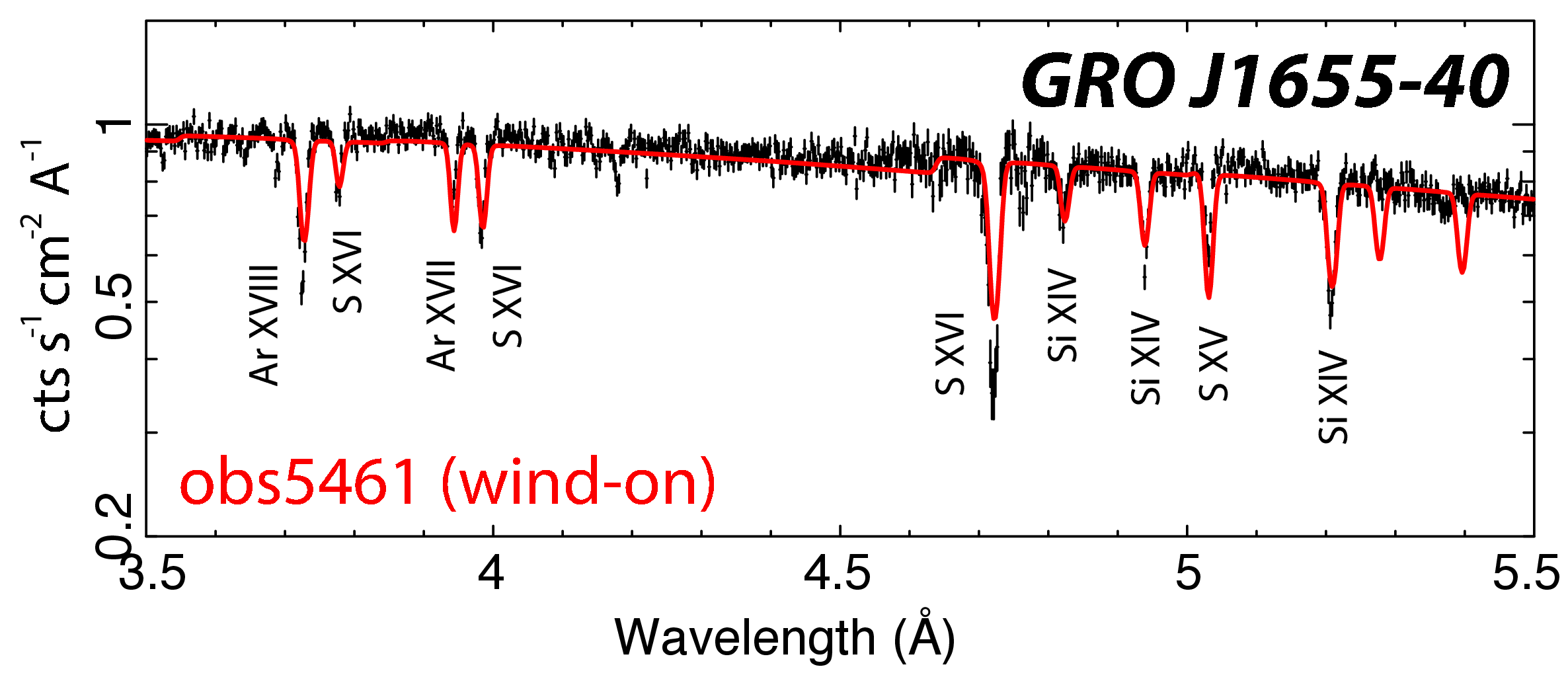}
\\
\includegraphics[trim=0in 0in 0in
0in,keepaspectratio=false,width=3.3in,angle=-0,clip=false]{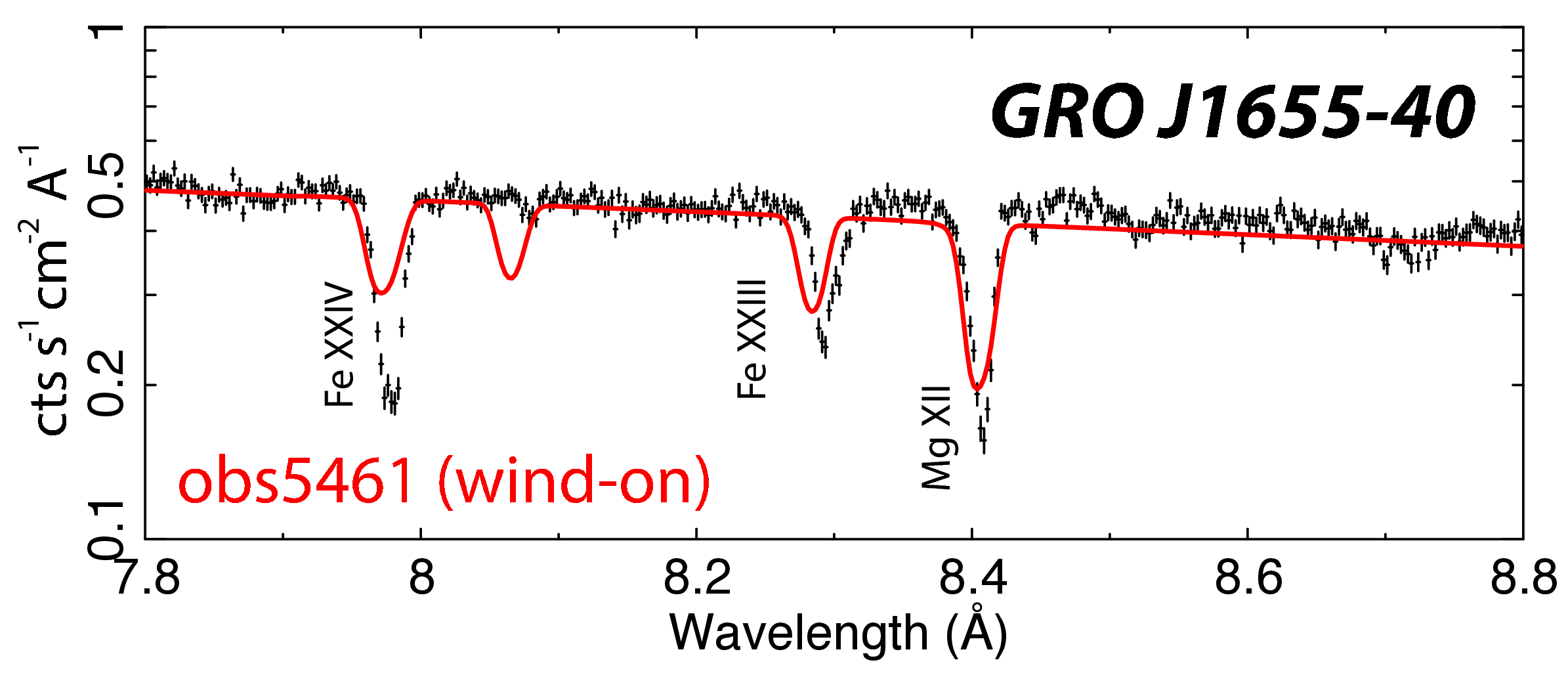}\includegraphics[trim=0in 0in 0in
0in,keepaspectratio=false,width=3.3in,angle=-0,clip=false]{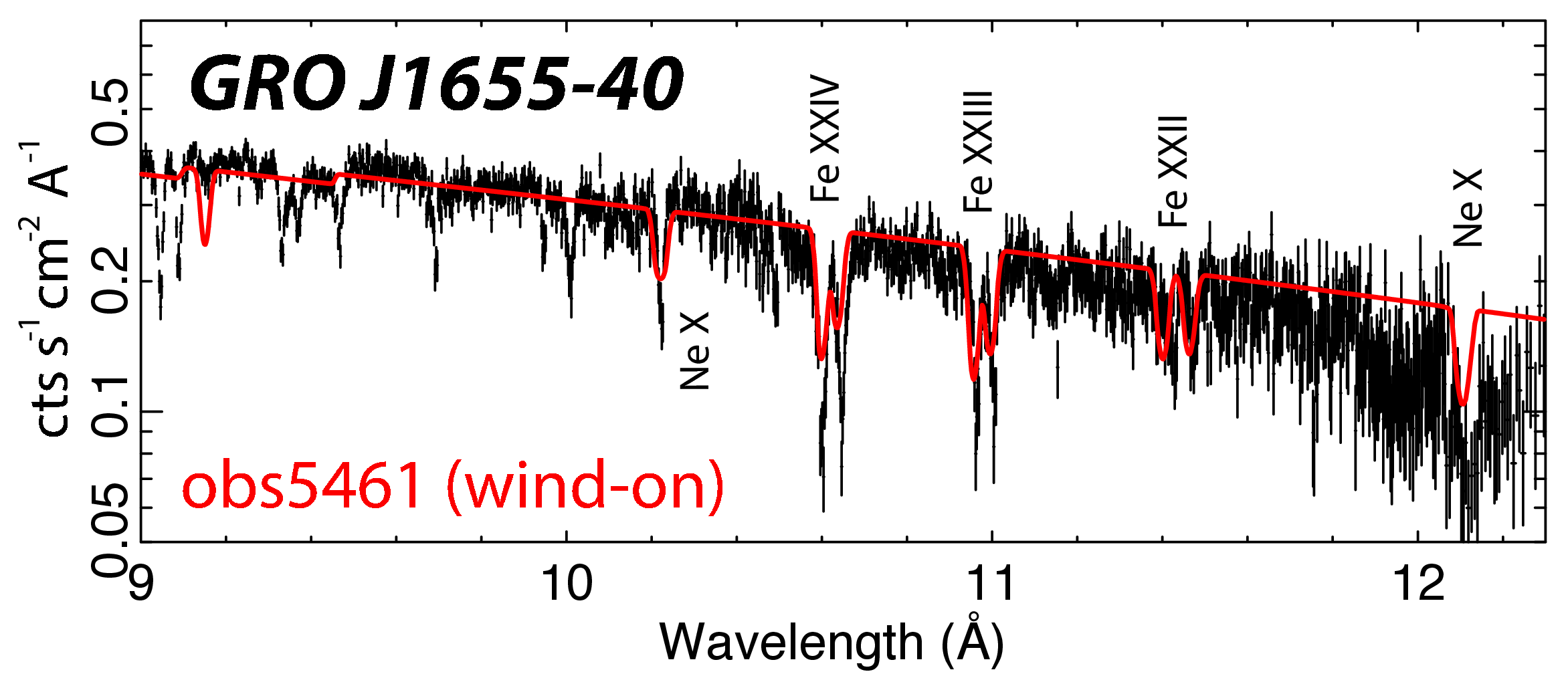}
\end{array}$
\end{center}
\caption{Zoom-in view of the 45ks {\it Chandra}/HETGS spectra of \gro1655\ for {\tt obs5461} (\windon) with the bestfit MHD-wind model in red. See {\bf Table~\ref{tab:tab3}} for the besfit model parameters. }
\label{fig:f6}
\end{figure}

\subsection{Implications from the Multi-Ion Modeling with AMD}

\subsubsection{Broadband Spectral Modeling during \windon\ State}

As demonstrated by the bestfit broadband spectra in {\bf Figures~\ref{fig:f3}-\ref{fig:f5}}, our spectral model can self-consistently describe the observed  multi-ion wind including soft X-ray features.
We now take a closer look at the individual absorption lines with focus on the \windon\ state in each source.

\begin{figure}[t]
\begin{center}$
\begin{array}{cc}
\includegraphics[trim=0in 0in 0in
0in,keepaspectratio=false,width=3.3in,angle=-0,clip=false]{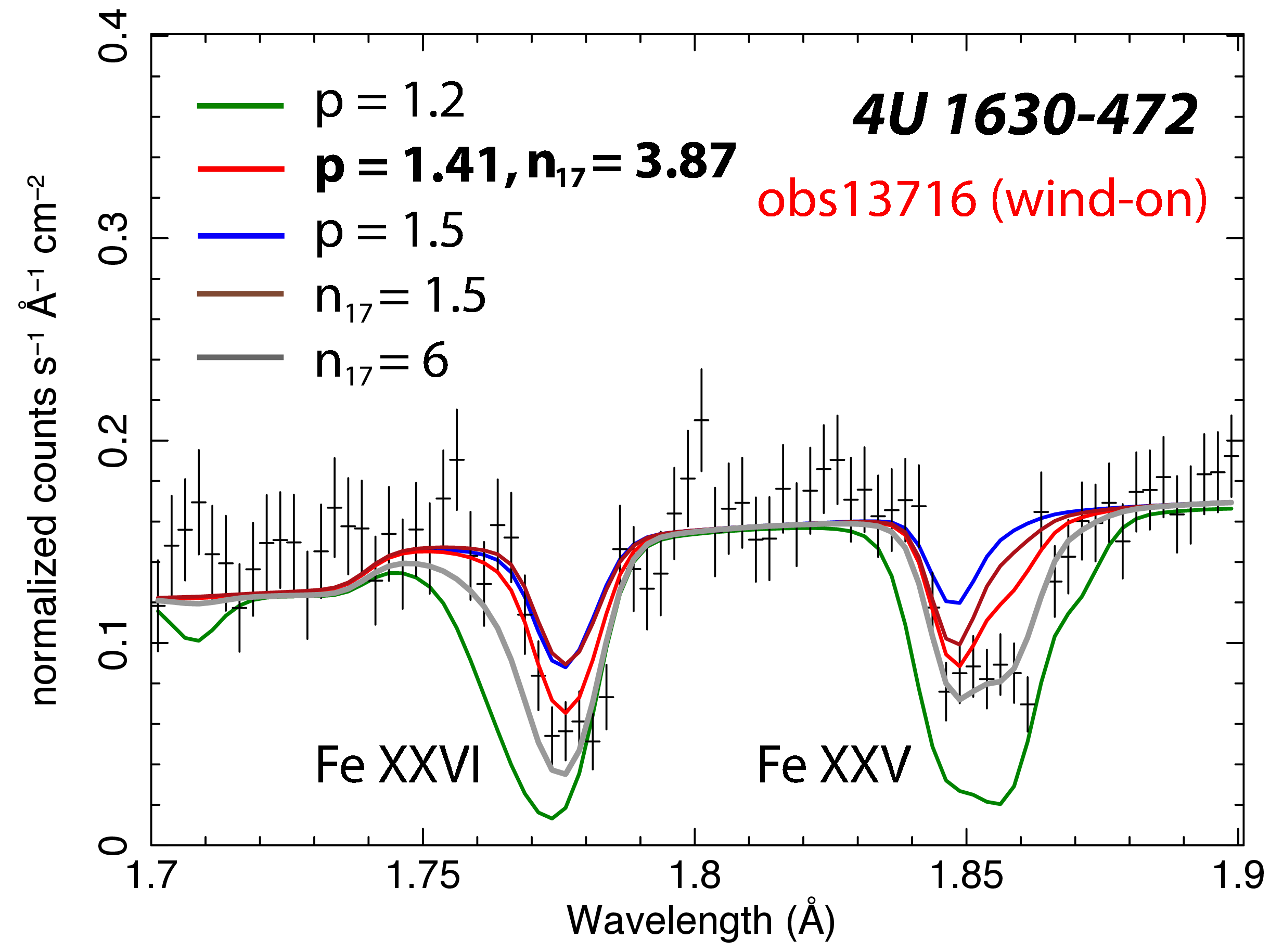}\includegraphics[trim=0in 0in 0in
0in,keepaspectratio=false,width=3.3in,angle=-0,clip=false]{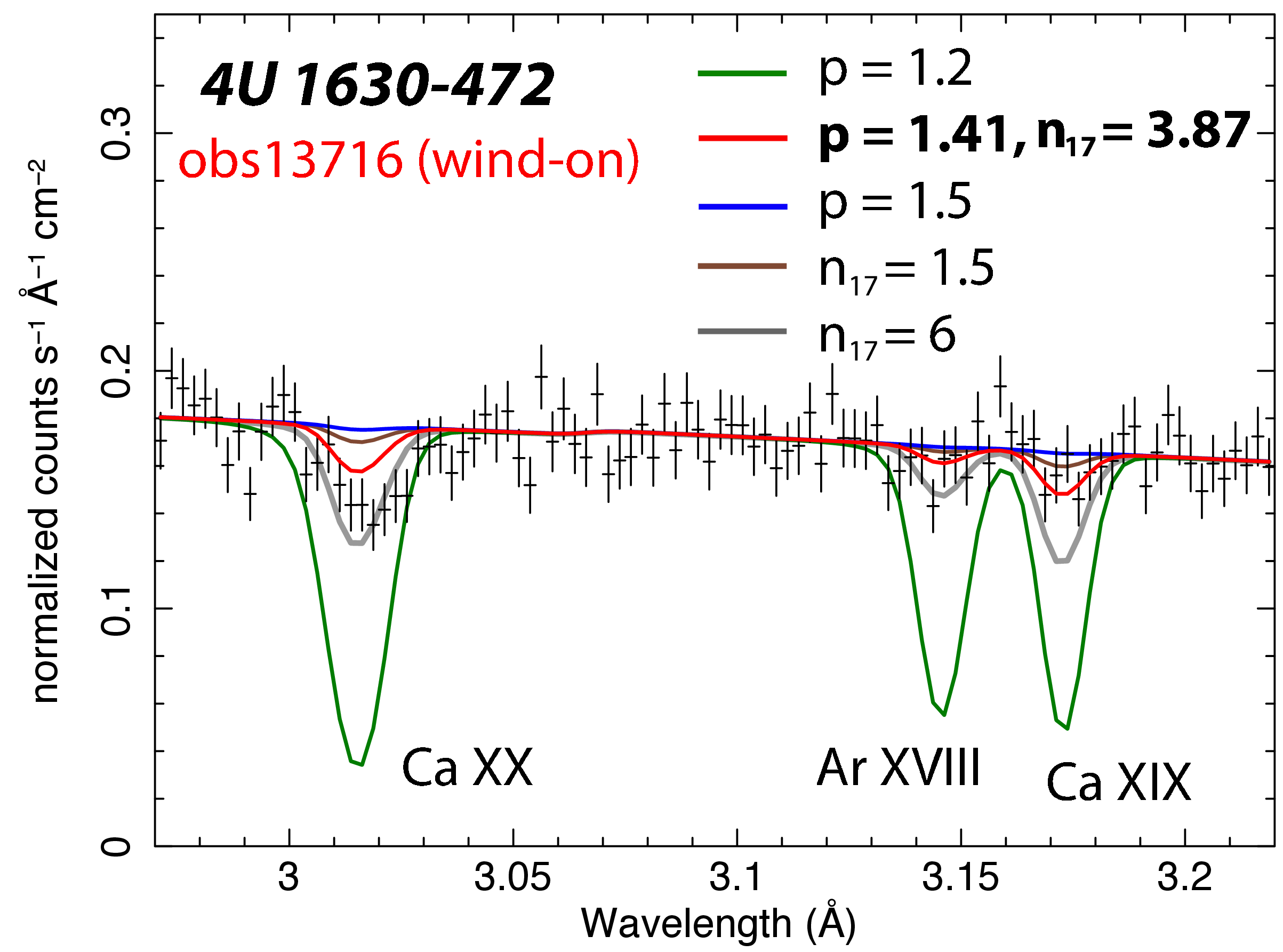}
\\
\includegraphics[trim=0in 0in 0in
0in,keepaspectratio=false,width=3.3in,angle=-0,clip=false]{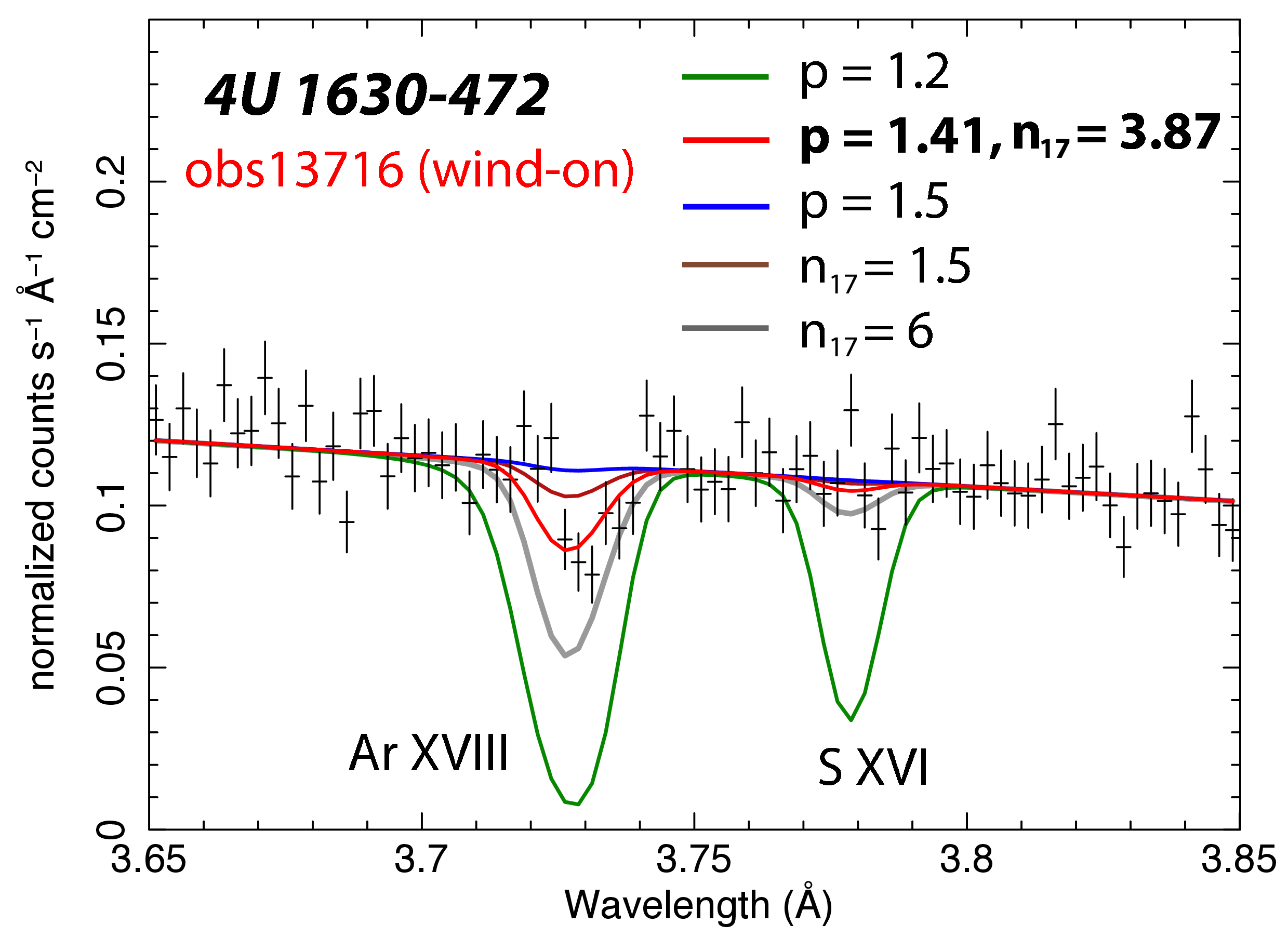}\includegraphics[trim=0in 0in 0in
0in,keepaspectratio=false,width=3.3in,angle=-0,clip=false]{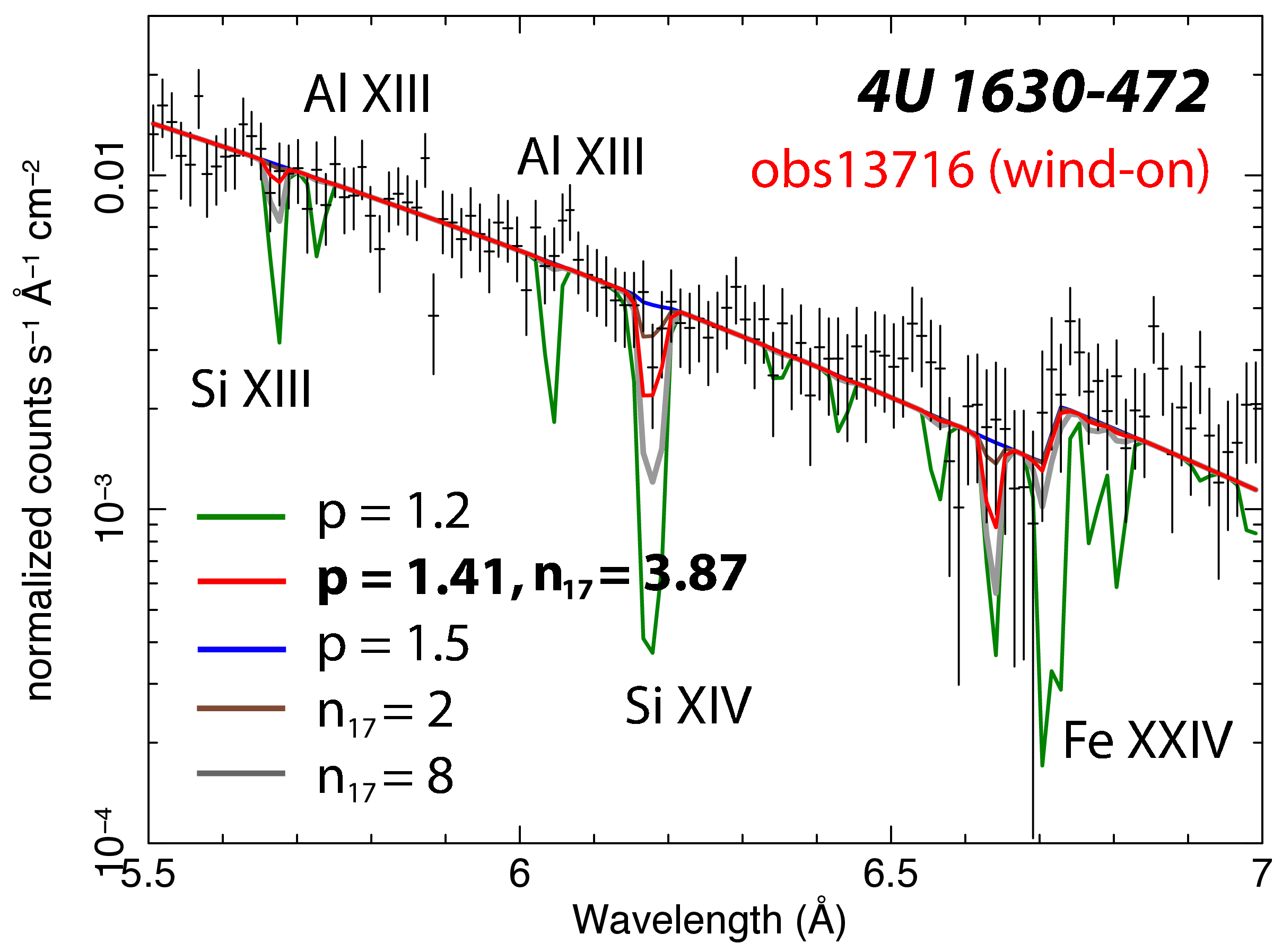}
\end{array}$
\end{center}
\caption{Zoom-in view of the {\it Chandra}/HETGS spectra of \4u1630\ for {\tt obs13716} (\windon) for different  wind model parameters ($p, n_0$) with the bestfit model in red. See {\bf Table~\ref{tab:tab3}} for the besfit model parameters. }
\label{fig:f7}
\end{figure}
\begin{figure}[t]
\begin{center}$
\begin{array}{cc}
\includegraphics[trim=0in 0in 0in
0in,keepaspectratio=false,width=3.3in,angle=-0,clip=false]{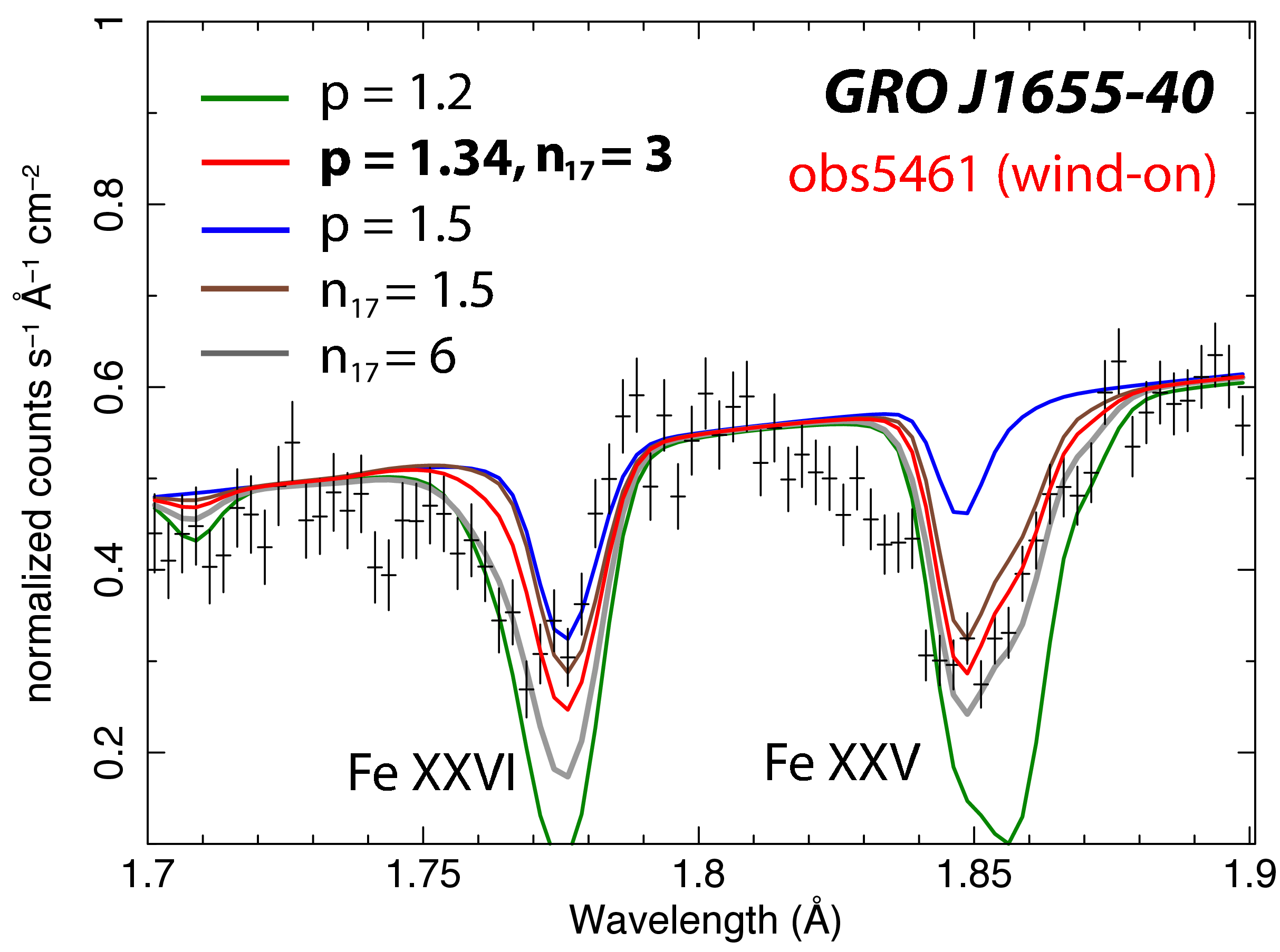}\includegraphics[trim=0in 0in 0in
0in,keepaspectratio=false,width=3.3in,angle=-0,clip=false]{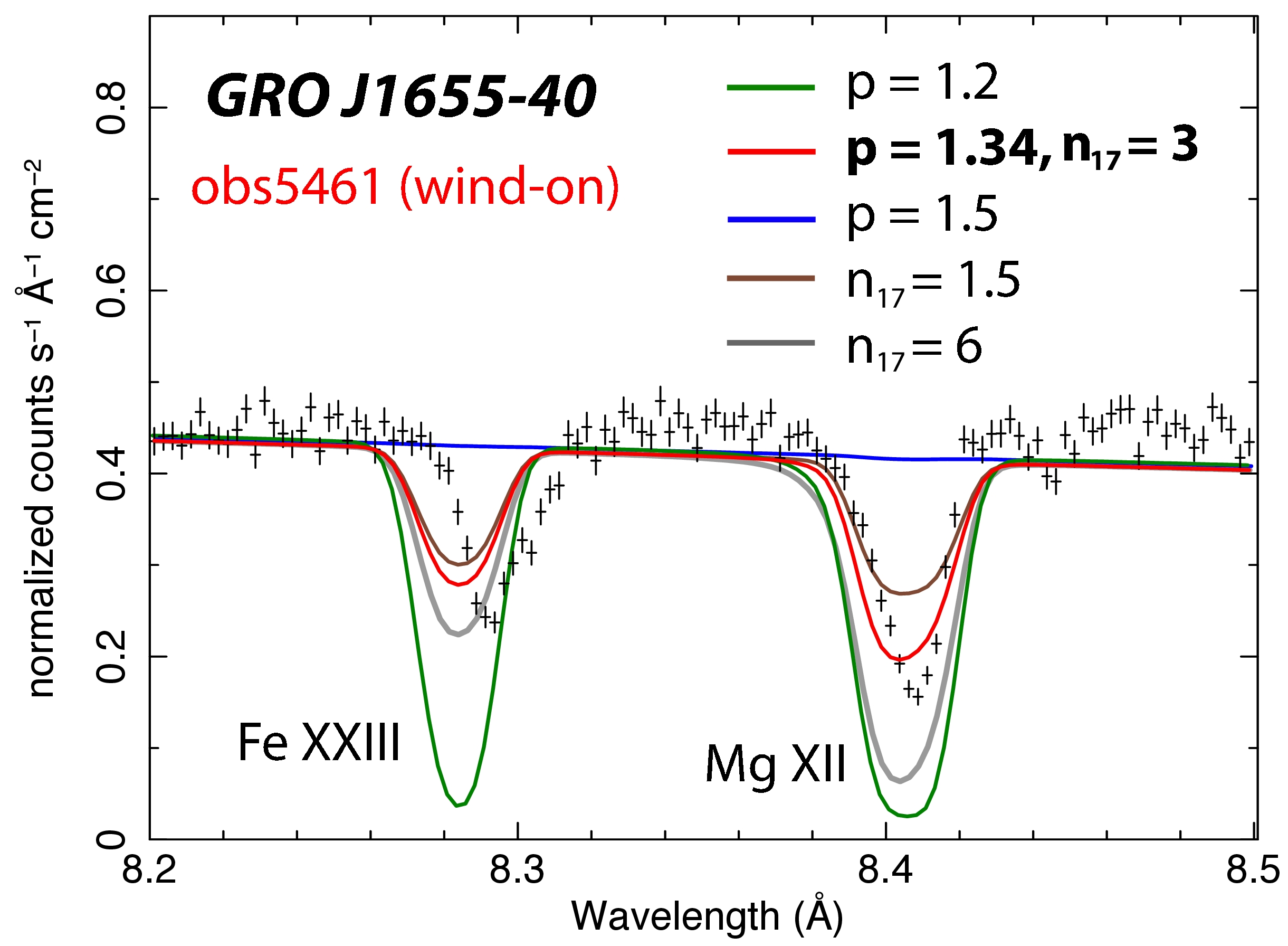}
\end{array}$
\end{center}
\caption{Zoom-in view of the {\it Chandra}/HETGS spectra of \gro1655\ for {\tt obs5461} (\windon) for different  wind model parameters ($p, n_0$) with the bestfit model in red. See {\bf Table~\ref{tab:tab3}} for the besfit model parameters. }
\label{fig:f8}
\end{figure}
\begin{figure}[t]
\begin{center}$
\begin{array}{cc}
\includegraphics[trim=0in 0in 0in
0in,keepaspectratio=false,width=3.3in,angle=-0,clip=false]{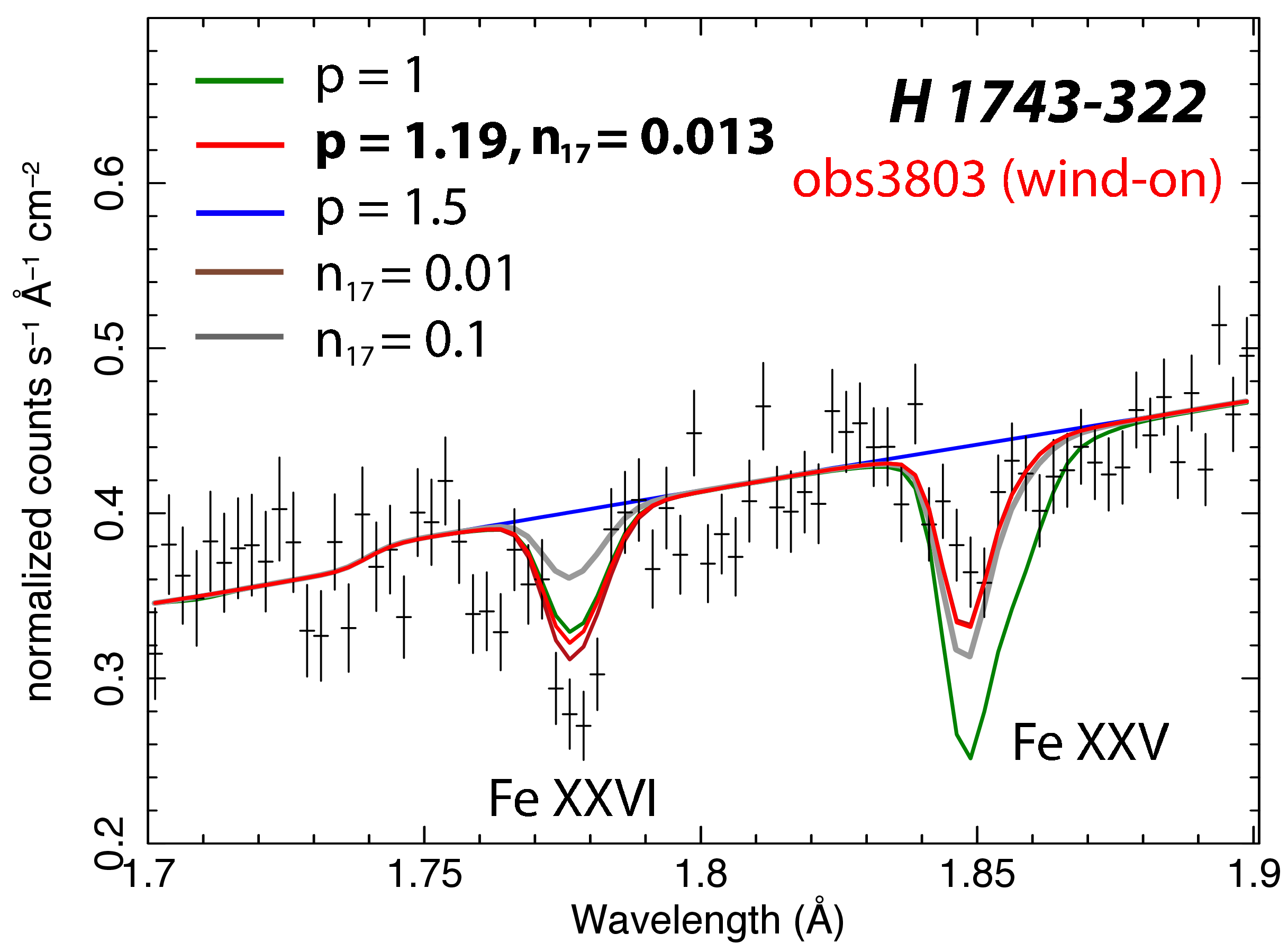}\includegraphics[trim=0in 0in 0in
0in,keepaspectratio=false,width=3.3in,angle=-0,clip=false]{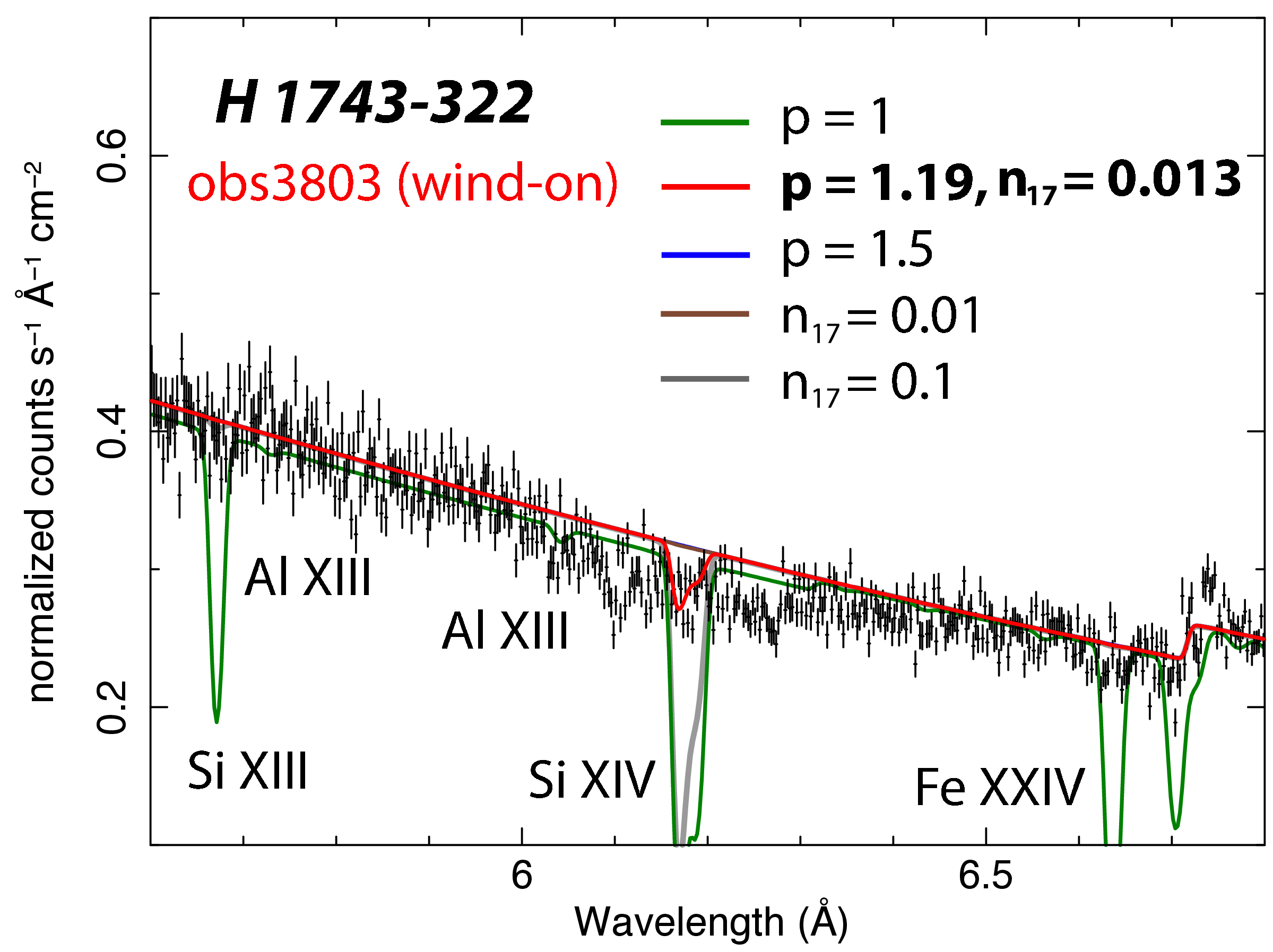}
\end{array}$
\end{center}
\caption{Zoom-in view of the {\it Chandra}/HETGS spectra of \h1743\ for {\tt obs3803} (\windon) for different  wind model parameters ($p, n_0$) with the bestfit model in red. See {\bf Table~\ref{tab:tab3}} for the besfit model parameters. }
\label{fig:f9}
\end{figure}

In {\bf Figures~\ref{fig:f6}-\ref{fig:f9}} we show  relatively stronger absorption signatures in each source during the \windon\ state. Among the three sources, \gro1655\ exhibits the largest number  of well-identified multi-ion absorbers almost in the entire X-ray band. These are shown in {\bf Figure~\ref{fig:f6}} where we also present  the bestfit model of $p=1.34$ and $n_{17}=3.0$ (in red). The centroid line wavelengths and widths for a series of the modeled lines are in excellent agreement with this high S/N data set\footnote[6]{ Note that some weak lines are missed, as expected, since the model is focused mainly on H-like/He-like strong features thus lacking those weak line components.} (see also F17 for the original spectral modeling of \gro1655\ with the MHD wind scenario as discussed later in \S 6).
%
%
The bestfit model for \gro1655\ in  \windon\ state (obs5461) clearly favors super-solar abundances for Fe ($A_{\rm Fe}=3$) as listed in {\bf Table~\ref{tab:tab3}}, thus reducing the need of higher wind density and shallower density slope. On the other hand, we see that the absorbers at longer wavelengths are less successfully modeled due to the obtained steeper slope, which tends to under-produce the ionic columns at larger distances where the soft X-ray absorbers are preferentially formed.

\clearpage

It is worth noting again that all these multi-wavelength ions are mutually coupled together within the same continuous wind calculated with the model, and hence their ionic properties (e.g. column, density, ionization, velocity) are  uniquely determined by a global wind condition combined with photoionization equilibrium. Our approach of spectral modeling is therefore very stringent about tightly constraining those ionic parameters from data, leaving little freedom for arbitrary fine-tuning\footnote[7]{This assumes that the wind density profile is well represented by a {\it single} power-law; it is quite conceivable that local deviations from this function may also be present.}.

\subsubsection{Model Dependencies of Multi-Ion MHD Disk-Winds}

We further compare the dependencies of the model for both Fe K and soft X-ray band in comparison with data. In {\bf Figure~\ref{fig:f7}} we present Fe K and soft X-ray absorbers in \4u1630\ \windon\ state for various model parameters ($p,n_{17}$). 
The bestfit model given by $p=1.41_{-0.01}^{+0.07}$ and $n_{17}=3.87_{-0.8}^{+7.2}$ does a good job at fitting these lines simultaneously. Neither a steeper slope $p$ nor lower density $n_{17}$ would fit better because that would  underproduce the required columns. This is especially true for, for instance, \arxviii, because they are predominantly produced at larger distances due to the lower ionizing threshold (compared to that for Fe K) and thus the inferred ionic column is more sensitive to the values of $(p,n_{17})$. In contrast, either a shallower slope or higher density would inevitably overproduce the ionic column being completely inconsistent with observations. Note also that the centroid line velocity is weakly sensitive to these parameters.

{\bf Figure~\ref{fig:f8}} shows similar model dependencies of the predicted spectrum between Fe K absorbers and \mgxii/\fexxiii\ lines in the \gro1655\ \windon\ state. It is clear that different models with even a small difference in $p$ can be successfully discriminated in the high S/N data of this source. Here again, our spectral modeling unambiguously demonstrates the power of our AMD-based multi-ion wind model, i.e. the subtle balance between the Fe K absorbers and their soft X-ray counterparts (see {\bf Figure~\ref{fig:f2}} in \S 3.2). Consequently, a \citet{BP82} type of MHD wind structure as well as spherical winds are safely ruled out, while strongly favoring a much shallower density profile to be viable with the observed broadband absorbers.

Lastly, we assess a similar parameter dependence of the modeled spectrum of  \h1743\ \windon\ state in {\bf Figure~\ref{fig:f9}}. As implied from the broadband spectrum in {\bf Figure~\ref{fig:f5}}, \h1743\ does not appear to possess strong soft X-ray absorbers even during the \windon\ state. For comparison, we show \alxiii/\sixiii/\sixiv/\fexxiv\ band as  marginal wind proxies. In order to be simultaneously consistent with these absorbers, we find a relatively shallow slope $p=1.19_{-0.02}^{+0.05}$ with low density $n_{17} = 0.013_{-0.001}^{+0.001}$ also favoring super-solar Fe/Si abundances.

In the current {\it Chandra}/HETGS first-order data in general, however, the expected finer spectral features such as the spin-orbit doublet in \fexxvi\ line, for example, is not quite resolved due to the limited resolving power and photon statistics (see, however, \citealt{Miller15} for the third-order data analysis). We shall discuss this point later in \S 5 with microcalorimeter simulations.

\begin{figure}[t]
\begin{center}$
\begin{array}{cc}
\includegraphics[trim=0in 0in 0in
0in,keepaspectratio=false,width=3.3in,angle=-0,clip=false]{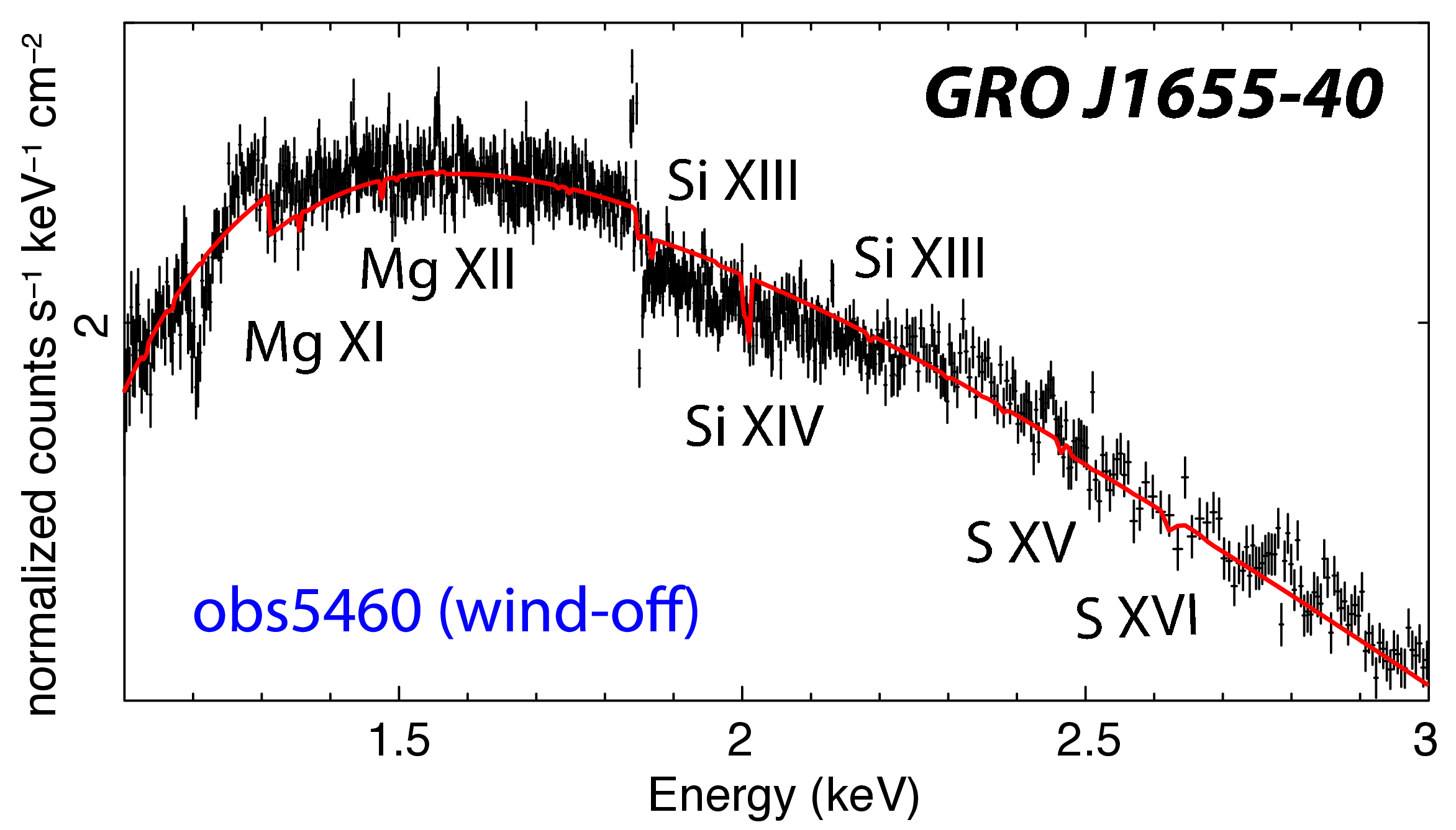}\includegraphics[trim=0in 0in 0in
0in,keepaspectratio=false,width=3.3in,angle=-0,clip=false]{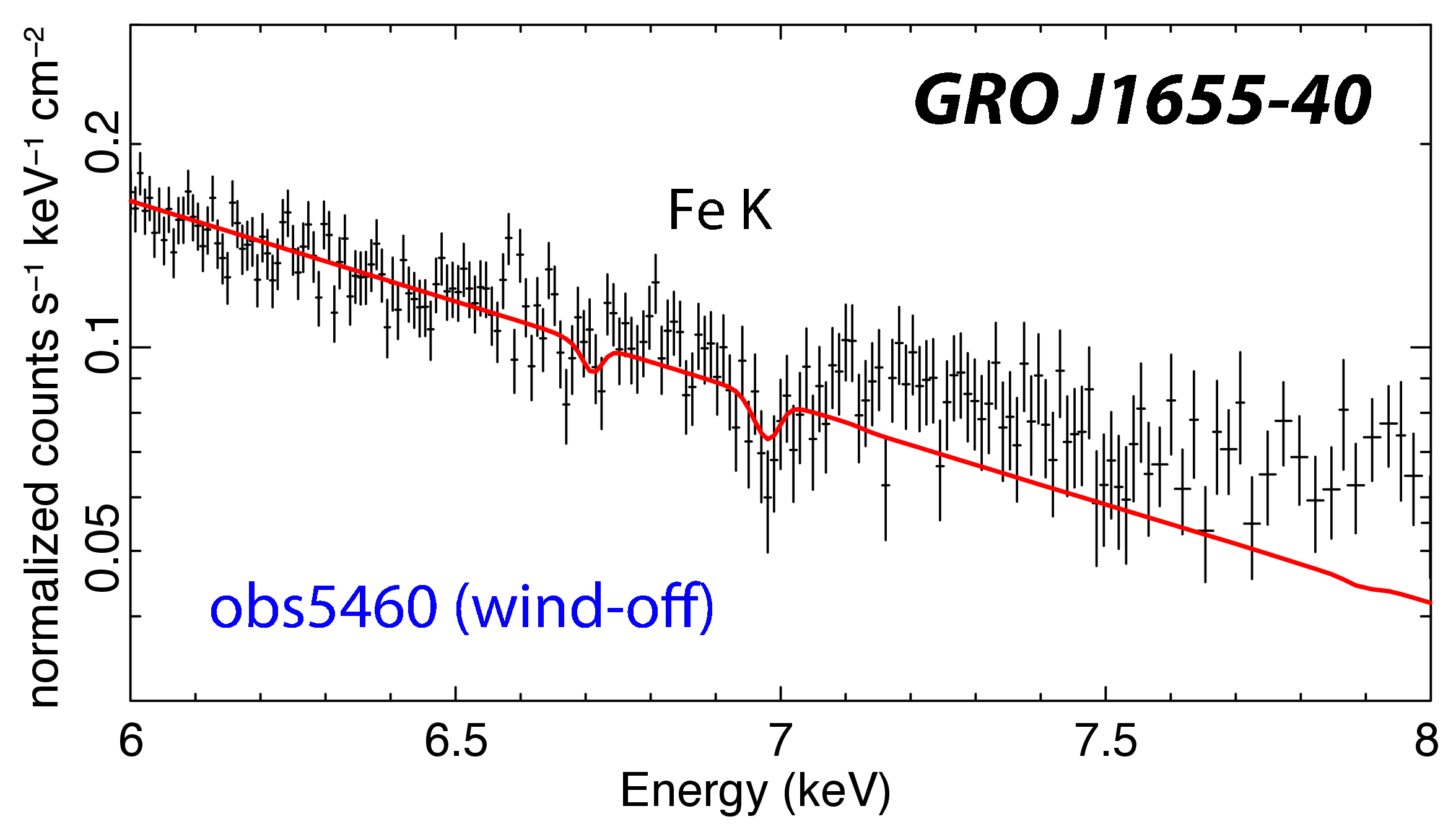}
\end{array}$
\end{center}
\caption{Zoom-in view of the {\it Chandra}/HETGS spectra of \gro1655\ for {\tt obs5460} (\windoff) with the bestfit MHD-wind model.}
\label{fig:f10}
\end{figure}


\subsubsection{Multi-Ion Absorbers during the \windoff\ State of \gro1655}

In the canonical evolution of transient BH XRBs in general, these sources normally transition into a {\it low/hard} state
\citep[e.g.][]{DoneGierlinskiKubota07}
in which outflow absorbers are often found to become much weaker (if not physically absent) in the spectrum. However,
the presence of (at least) Fe K absorbers are in fact hinted in the observed spectra even during the \windoff\ state in these BH XRBs as seen in {\bf Figures~\ref{fig:f3}-\ref{fig:f5}}.


We have modeled obs5460 data from \gro1655\ during its \windoff\ state (see {\bf Fig.~\ref{fig:f4}}). The ionized wind is found to have a steeper density slope and orders of magnitude lower wind density as listed in {\bf Table~\ref{tab:tab3}}. In order to better unpack the elusive nature of the underlying disk-wind in \gro1655\ \windoff\ state, we show the soft band ($\sim 1-3$ keV) and Fe K band spectra in {\bf Figure~\ref{fig:f10}}. While not strong, the Fe K features are marginally discerned as also shown previously \citep[e.g.][]{NeilsenHoman12}. On the other hand, the soft X-ray absorbers (e.g. \mgxi, \mgxii, \sixiii, \sixiv, \sxv\ and \sxvi) appear to be too weak for a robust identification even with grating instruments, even though the MHD wind model implies their persistent presence in the source broadband spectrum. Hence, it is  conceivable that insufficient statistics could easily account for the absence of absorber features in the soft energy band. In our model, these soft X-ray absorbers are formed at larger distances where the wind column density becomes sufficiently low ($n \propto r^{-1.39}$) to obscure their presence in the spectrum.  This point shall be illustrated again later with simulated spectra in \S 5.

\subsection{Other Potential Effects of X-Ray Absorbers}

While our models provide specific spectral fits to the particular sources discussed above, they also allow us to explore the effects on the spectra by varying a number of model parameters, thereby examining their potential effects on states or objects not included in our spectral analysis. To this end, we investigate below the dependences of the absorber's spectra on (1) truncation of the disk size where winds are launched, (2) inclination/viewing angle and (3) wind kinematics.


\subsubsection{Disk Truncation: $R_{\rm in}$ and $R_{\rm out}$}

We assume the innermost wind launching radius to coincide with that of the ISCO for a \sw BH\footnote[8]{While higher BH spin in general allows for smaller ISCO, this would also imply higher radiative efficiency. It is not obvious whether these effects would be observable in the absorption spectra. 
} 
; i.e. $R_{\rm in}/R_g=6$ as a fiducial non-truncated case corresponding to a {\it high/soft} state where the Comptonized power-law component is essentially negligible energetically. 
As mentioned above, it has been suggested (and observationally supported) that the truncation radius recedes away from the ISCO with decreasing luminosity while the system transitions into its {\it low/hard} state \citep[e.g.][]{Esin97,Ponti12}. It has been pointed out in \cite{K15,K19} that such a behavior is allowed dynamically for disk-wind configurations with mass flux decreasing  toward the BH along with decreasing luminosity. 

To mimic this behavior of the {\it inner} disk truncation radius, we terminate the inner extent of the disk at $R_{\rm in}/R_g = 10$, 30, 50 and 100 (maximum truncation) with photoionization calculations performed using the \windoff\ state X-ray spectrum and correspondingly smaller accretion rate, luminosity and $n_{17}$. We find that the calculated EWs of the absorption features effectively decrease with the decrease of $n_{17}$ due to the overall decrease of the wind column density, virtually disappearing at the lower luminosity during harder states similar to those in {\bf Figures~\ref{fig:f3}-\ref{fig:f5} and \ref{fig:f10}}.
We have also conducted another exercise by mechanically removing the inner part of the wind  within those radii at $R_{\rm in}$ considered above, while not changing anything else in the model. In this case, we see that the observed absorption features are essentially unaffected because the absorbers, especially those in the soft X-ray band, are produced predominantly at larger distances away from the BH.    
 

\begin{figure}[t]
\begin{center}$
\begin{array}{cc}
\includegraphics[trim=0in 0in 0in
0in,keepaspectratio=false,width=3.0in,angle=-0,clip=false]{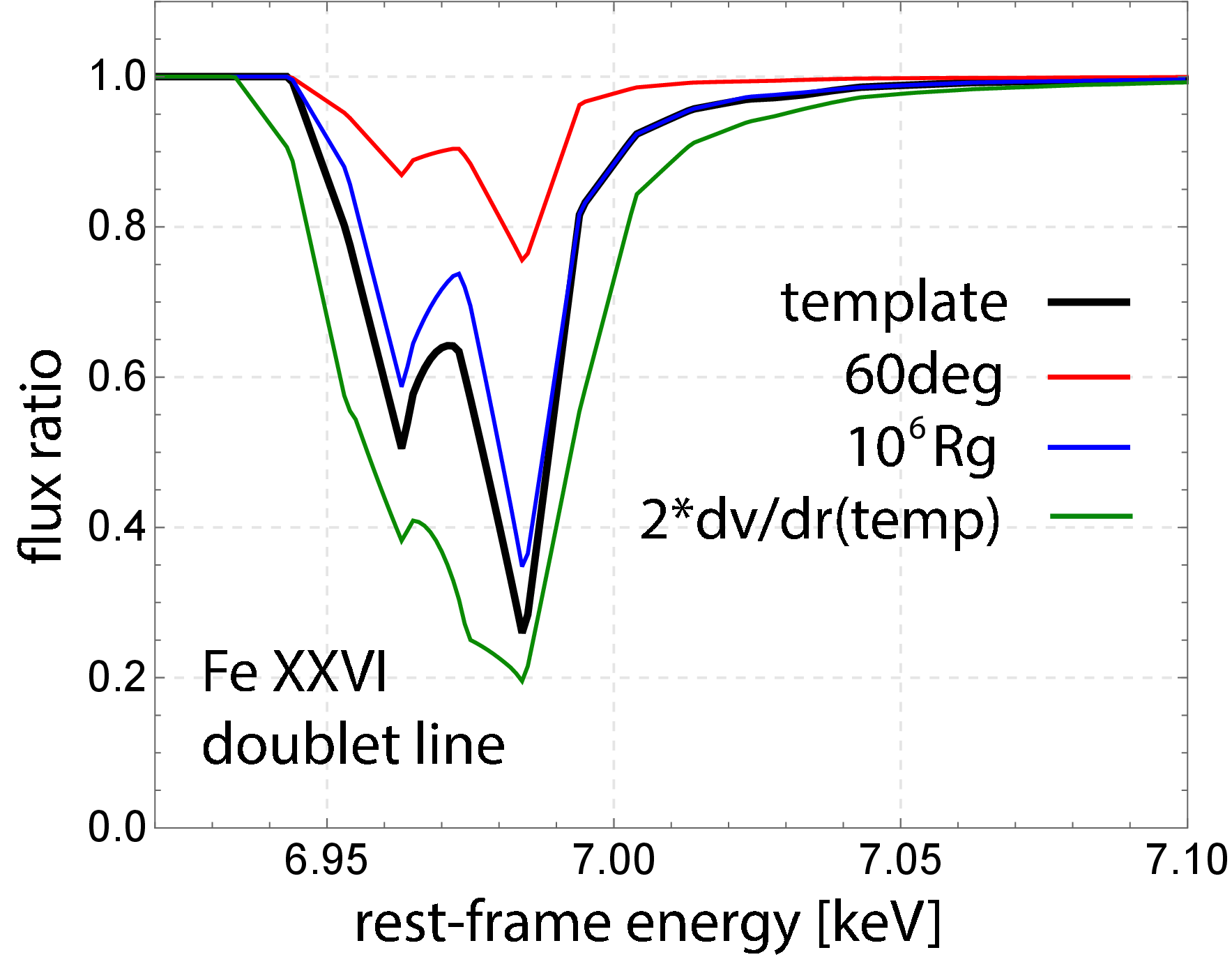}\includegraphics[trim=0in 0in 0in
0in,keepaspectratio=false,width=3.0in,angle=-0,clip=false]{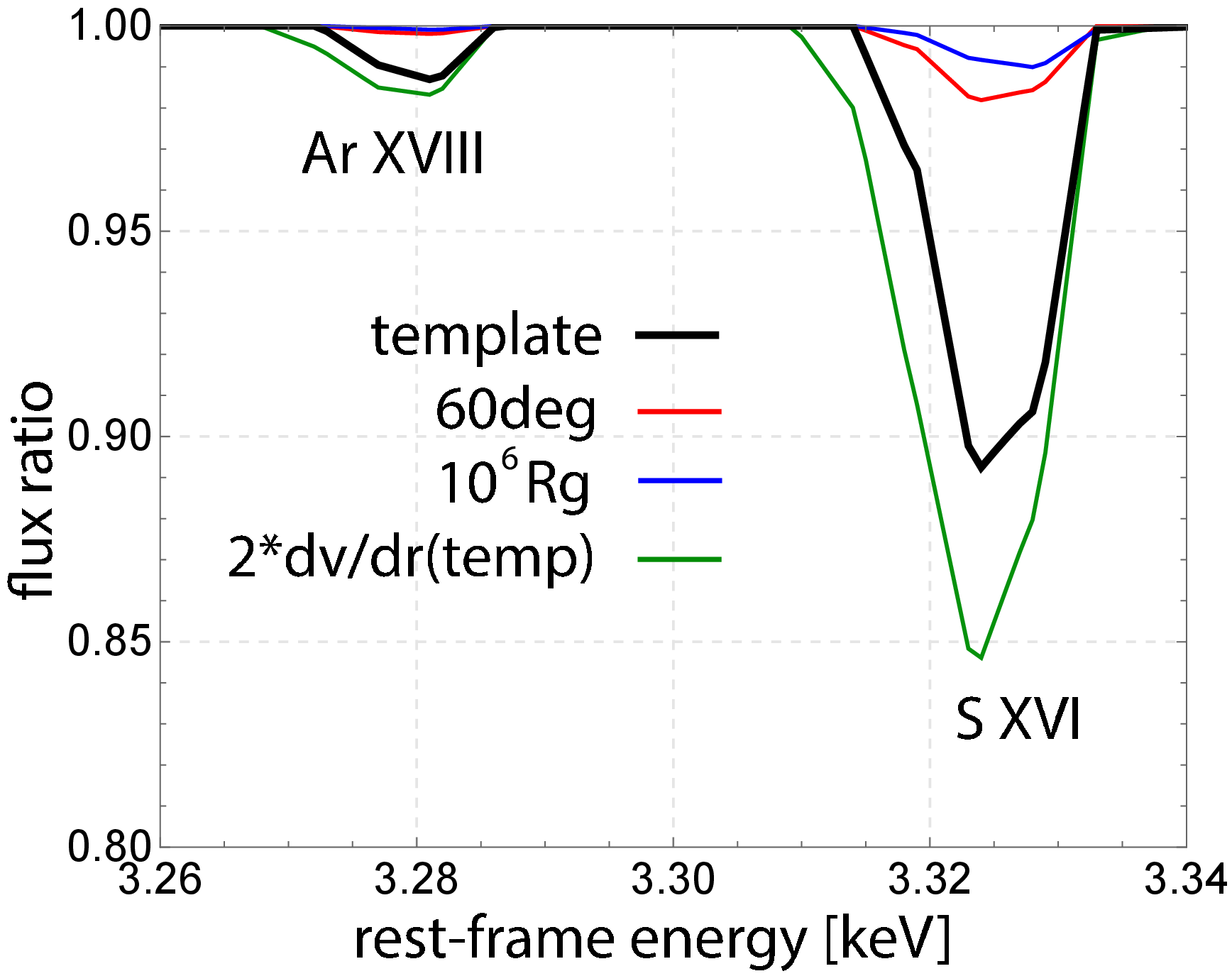}
\end{array}$
\end{center}
\caption{Dependences on wind properties of the theoretical template spectra (with $\sim 1$ eV resolution) from the model: template (black), $\theta=60\deg$ (red), $R_{\rm out}/R_g=10^6$ (blue), and $2(dv/dr)_{\rm temp}$ (green) for Fe K features and soft x-ray ions for comparison. }
\label{fig:f11}
\end{figure}

On the other hand, the outer disk radius at $r=R_{\rm out}$ is somewhat independent of the state of the system and depends on the size of the binary orbit, ranging from object to object\footnote[9]{The binary separation is estimated to be on the order of $\sim 10^{11-13}$ cm in these BH XRBs \citep[e.g.][]{McClintockRemillard06}.}. 
In {\bf Figure~\ref{fig:f11}} we present a set of theoretical model profiles of $\sim 1$ eV theoretical resolution, for both (a) Fe K and (b) soft X-ray features for comparison. One of the template calculations used for analysis (in black) corresponds to $\theta=70\deg$ and $R_{\rm out}/R_g=10^7$ as fiducial values for  \windon\ state spectrum of \4u1630. A reduction of $R_{\rm out}$ to $10^6 R_g$ (in blue) is found to affect mainly the lower ionization ions, for example, \arxviii\ and \sxvi\, which are formed at the outer, lower $\xi$ wind region, while the higher ionization species like \fexxv/\fexxvi\ remain largely unaffected. 
%
%
%
At this resolution, we see the doublet structure of \fexxvi\ line which remains almost unchanged with variation in $R_{\rm out}$. This is not the case for the lower ionization ions, \arxviii\ and \sxvi, formed farther out; i.e. their EW decreases significantly with a decrease in $R_{\rm out}$. It will be of interest to search for such a dependence among sources of different binary orbital sizes.


\subsubsection{Inclination}

In {\bf Figure~\ref{fig:f11}}, we also show the effects of decreasing inclination angle to $60\deg$ (in red). The effects of this parameter have a dual role: (i) The main effect is the decrease in column density with decreasing $\theta$ given by the function $f(\theta)$ (see eqn.~(\ref{eq:eqn3})) leading to a shallower absorption feature and (ii) the outflow velocity and its projection along the observer's LoS changing the EW. It is apparent that the decrease in column provides a more dramatic effect especially for the lower $\xi$ ions with  \arxviii\ being almost absent in the spectrum.
The doublet line signature of \fexxvi\ remains preserved. Similarly to the effect of $R_{\rm out}$, the detection of the soft X-ray absorbers could be challenging in this case.

\subsubsection{Wind Kinematics}

Besides the dependencies discussed above, the model absorption features depend also on the wind kinematics. We probe this effect by including a larger velocity gradient $dv/dr$ in our calculation of the Voigt profile. As noted in our earlier works (see F10), this gradient naturally replaces the parameter {\tt vturb}, commonly employed in similar modeling with {\tt xstar} in the literature \citep[e.g.][]{Miller08,NeilsenHoman12,DiazTrigo14,Trueba19,Tomaru20}. To examine the effect of this gradient $dv/dr$, we double its value in our spectral calculations (in green) as shown in {\bf Figure~\ref{fig:f11}}. As expected, the increase in the gradient broadens and deepens the absorption line profiles of all ions. The case of \fexxvi\ is quite interesting because the doublet troughs (separation of $\sim 20$ eV) are almost blended so that it would be hard to discern even in these theoretical calculations with $\sim 1$ eV resolution. 
Such an increase in the velocity gradient $dv/dr$ might actually be possible if, for a given wind density, the ionizing luminosity drops sufficiently to bring the ionization front for \fexxvi\ closer to the BH, hence resulting in higher velocity ($v \propto r^{-1/2}$) and higher velocity gradient ($dv/dr \propto r^{-3/2}$). Such a behavior may be possible to observe with the upcoming {\it XRISM} mission. 
Once again, even through the currently available third-order {\it Chandra} gratings data, \fexxvi\ doublet line and weak soft X-ray absorbers are very hard to securely resolve even if not entirely impossible \citep[e.g.][]{Miller15}. Future microcalorimater observations made available with {\it XRISM}/Resolve and {\it Athena}/X-IFU will shed light on these signatures as  demonstrated in \S 5.

\begin{figure}[t]
\begin{center}$
\begin{array}{cc}
\includegraphics[trim=0in 0in 0in
0in,keepaspectratio=false,width=3.3in,angle=-0,clip=false]{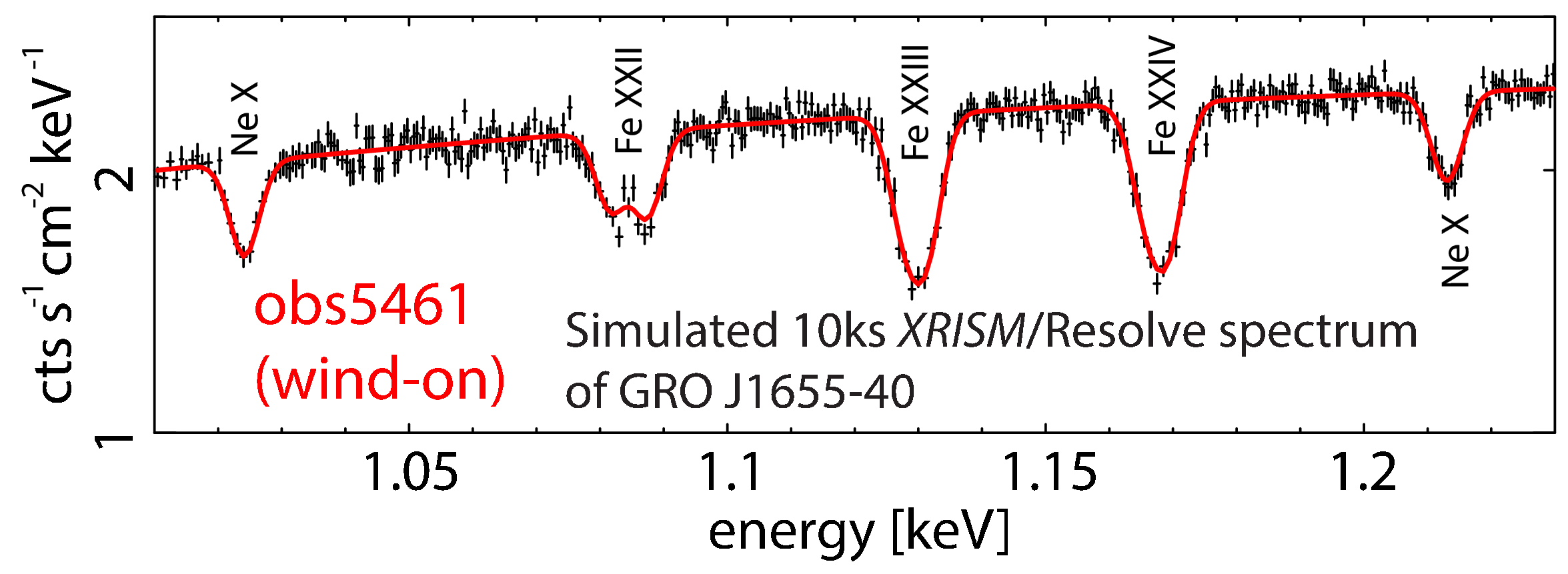}\includegraphics[trim=0in 0in 0in
0in,keepaspectratio=false,width=3.3in,angle=-0,clip=false]{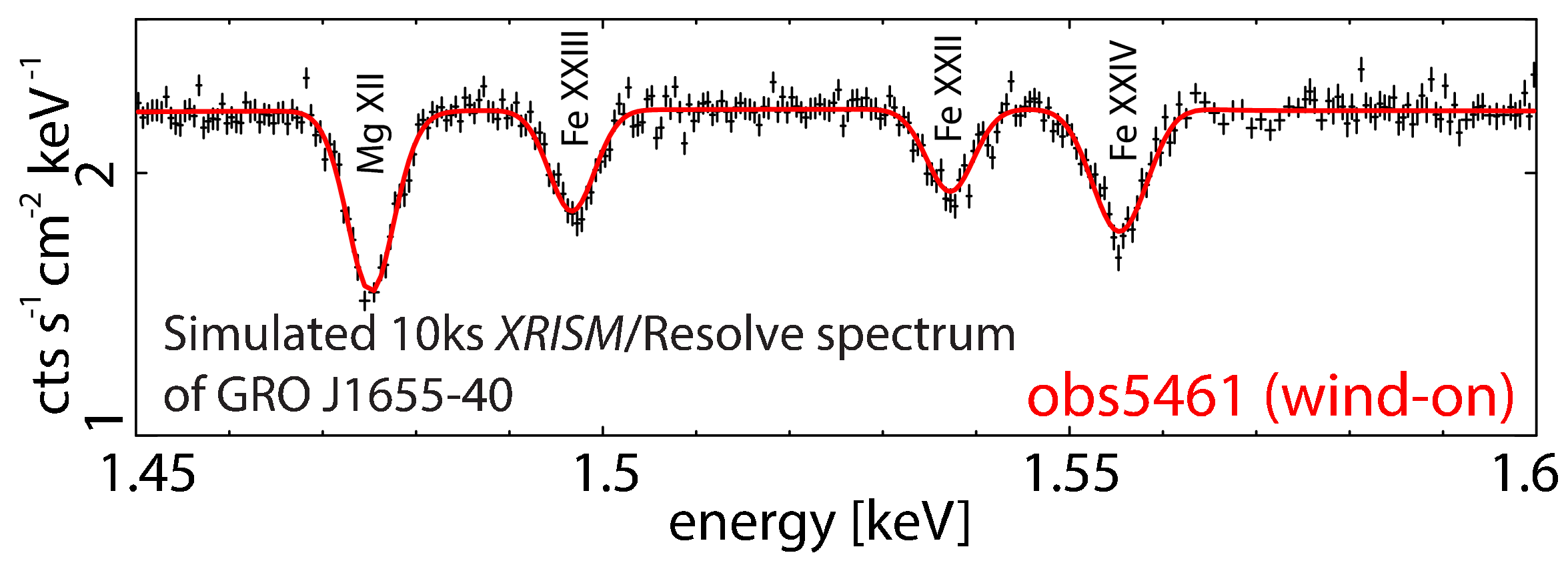}
\\
\includegraphics[trim=0in 0in 0in
0in,keepaspectratio=false,width=3.3in,angle=-0,clip=false]{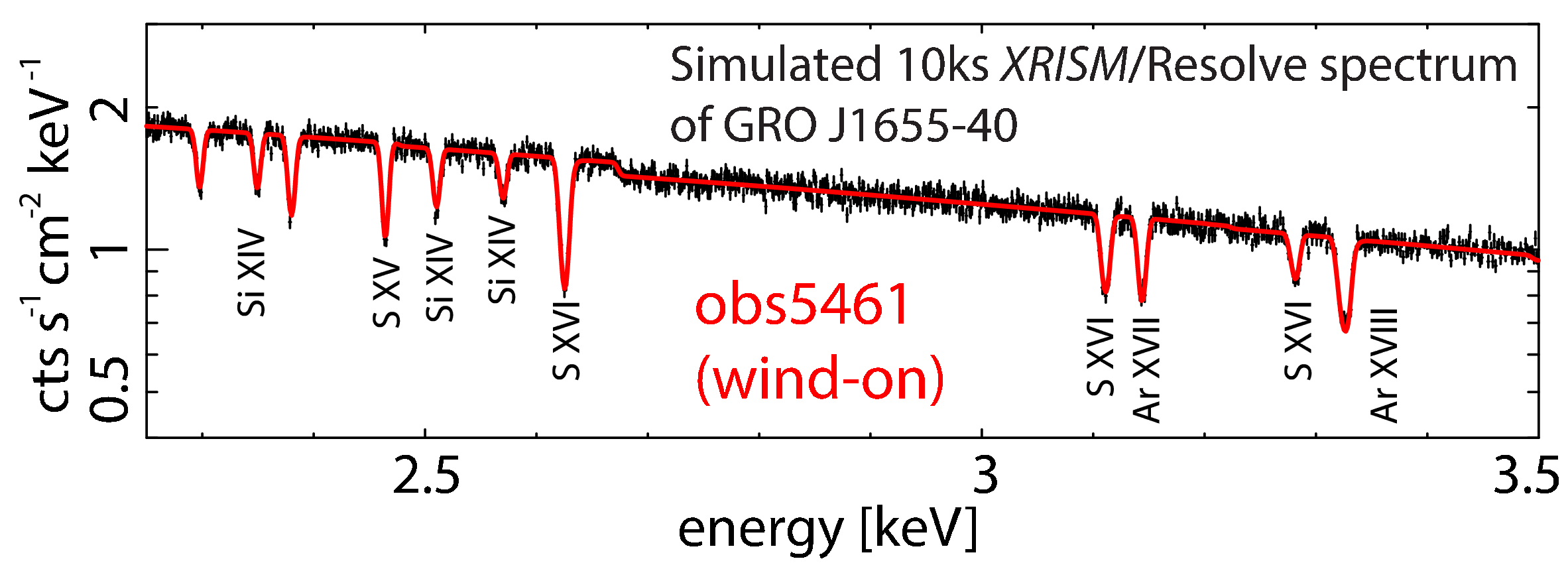}\includegraphics[trim=0in 0in 0in
0in,keepaspectratio=false,width=3.3in,angle=-0,clip=false]{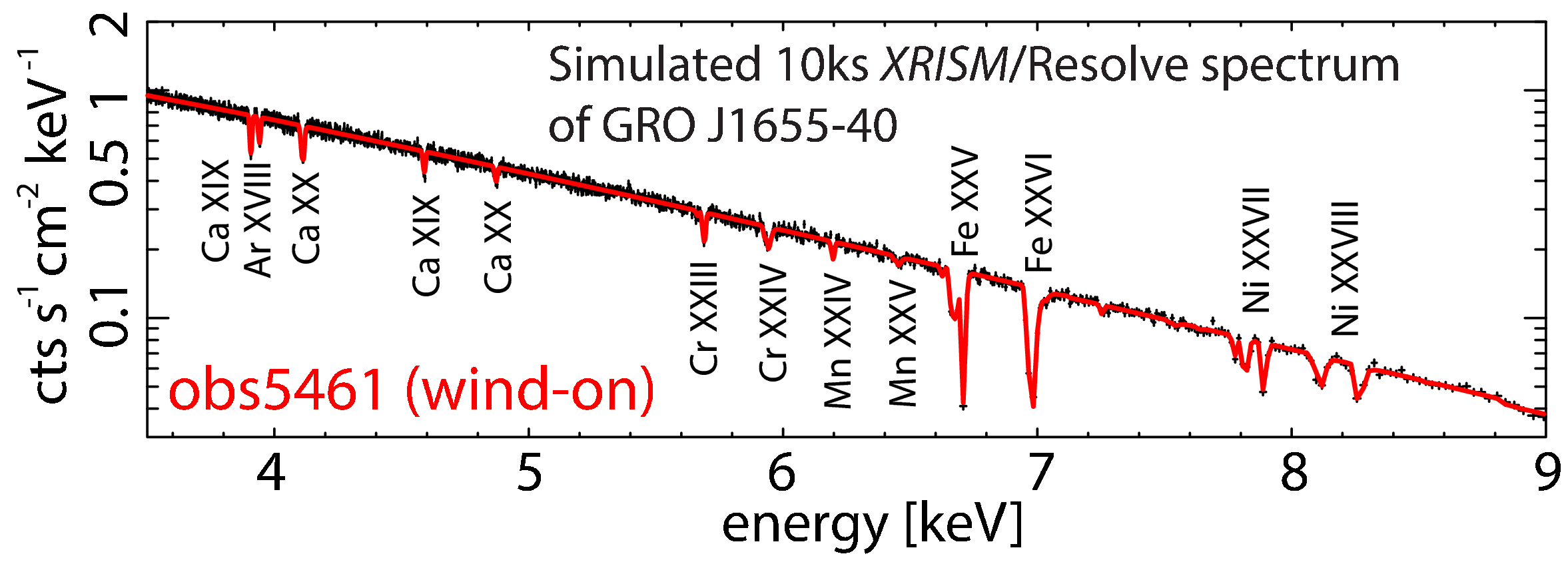}
\end{array}$
\end{center}
\caption{Simulated 10ks {\it XRISM}/Resolve spectra of \gro1655\ for {\tt obs5461} with the bestfit MHD-wind model.}
\label{fig:f12}
\end{figure}

\section{{\it XRISM}/Resolve and {\it Athena}/X-IFU Simulations with MHD Disk-Winds}

The upcoming X-ray missions, {\it XRISM} and {\it Athena}, with their microcalorimeter instruments with higher  sensitivity are expected to revolutionize X-ray spectroscopy and, as such, also our understanding of the physics of X-ray absorbers.
In anticipation of these missions, we provide in this section a number of spectral simulations of our bestfit model spectra of both the \windon\ and \windoff\ states discussed above with the convolution of {\it XRISM}/Resolve and {\it Athena}/X-IFU responses, {\tt xarm\_res\_h5ev\_20170818.rmf} and {\tt XIFU\_CC\_BASELINECONF\_2018\_10\_10.rmf}, respectively.

The constant energy resolution $\Delta E$ of microcalorimeters, compared to the constant resolving power $\lambda /\Delta \lambda$ of {\it Chandra} gratings, provides a resolution advantage of these detectors at the higher energy transitions (with a break-even point at $E \sim 3$ keV). Since every bin has the same size in units of energy for microcalorimeters, we show simulated spectra in {\it energy} in this section.


\begin{figure}[t]
\begin{center}$
\begin{array}{cc}
\includegraphics[trim=0in 0in 0in
0in,keepaspectratio=false,width=3.3in,angle=-0,clip=false]{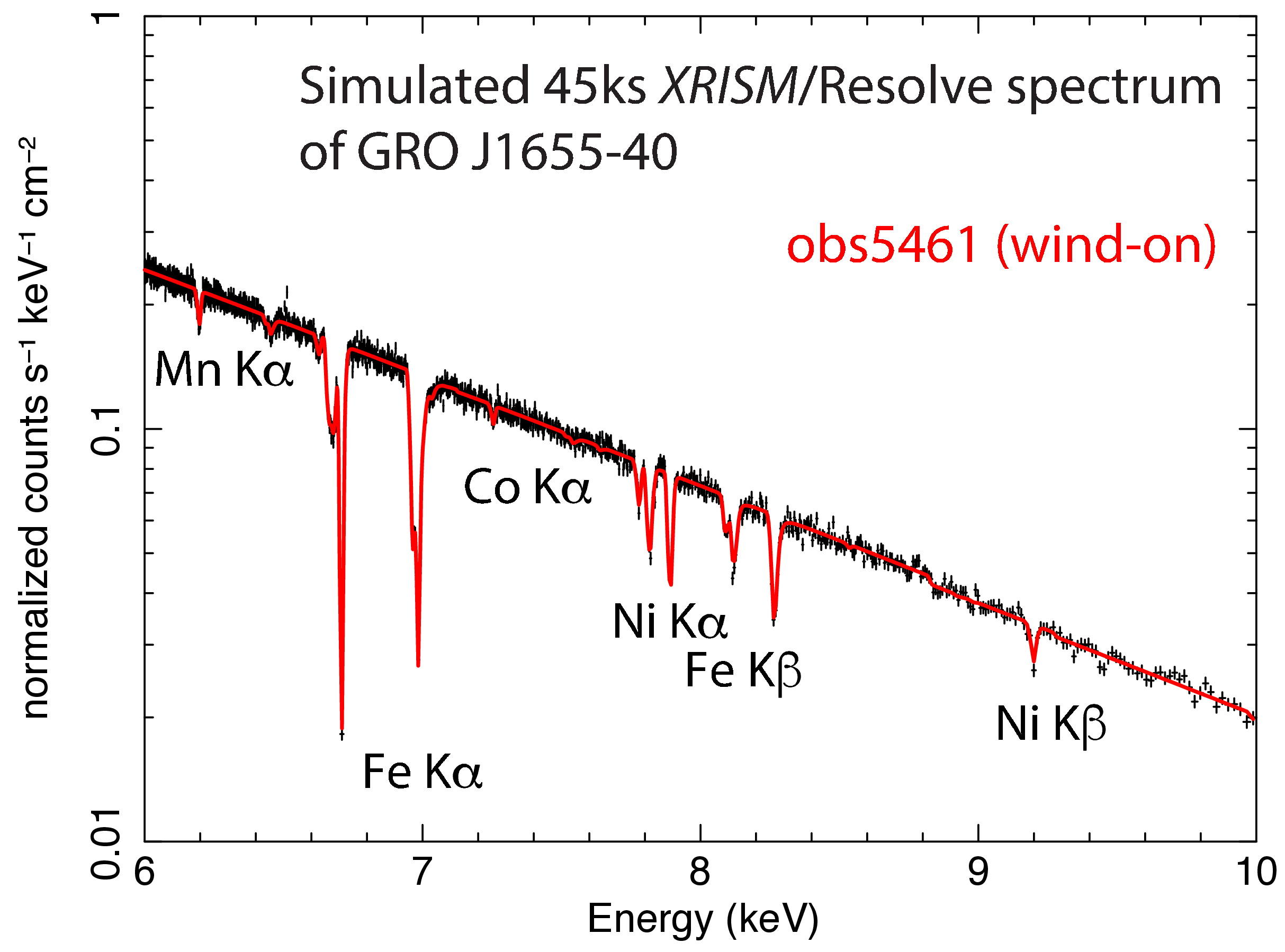}
\end{array}$
\end{center}
\caption{Zoom-in view (Fe K band and beyond) of simulated 45ks {\it XRISM}/Resolve spectra of \gro1655\ for {\tt obs5461} with the same bestfit MHD-wind model in Figure~\ref{fig:f11}.}
\label{fig:f13}
\end{figure}

\subsection{\gro1655\  during \windon\ State}

Considering first the spectrum of \gro1655\ in the  \windon\ state ({\tt obs5461}), currently the one with the highest multitude and significance of detected features, we present in {\bf Figure~\ref{fig:f12}} a simulated 10ks {\it XRISM}/Resolve broadband spectrum ($\sim 5$ eV resolution) for its bestfit model corresponding to {\bf Figures~\ref{fig:f6}} and {\bf \ref{fig:f8}}. We see that the the expected transitions of Cr K, Mn K, Fe K, and Ni K are far more clearly discerned due to the superior sensitivity of the instrument at these energies only with 10ks exposure compared to the 45ks {\it Chandra}/HETGS observation.

\begin{figure}[t]
\begin{center}$
\begin{array}{cc}
\includegraphics[trim=0in 0in 0in
0in,keepaspectratio=false,width=3.3in,angle=-0,clip=false]{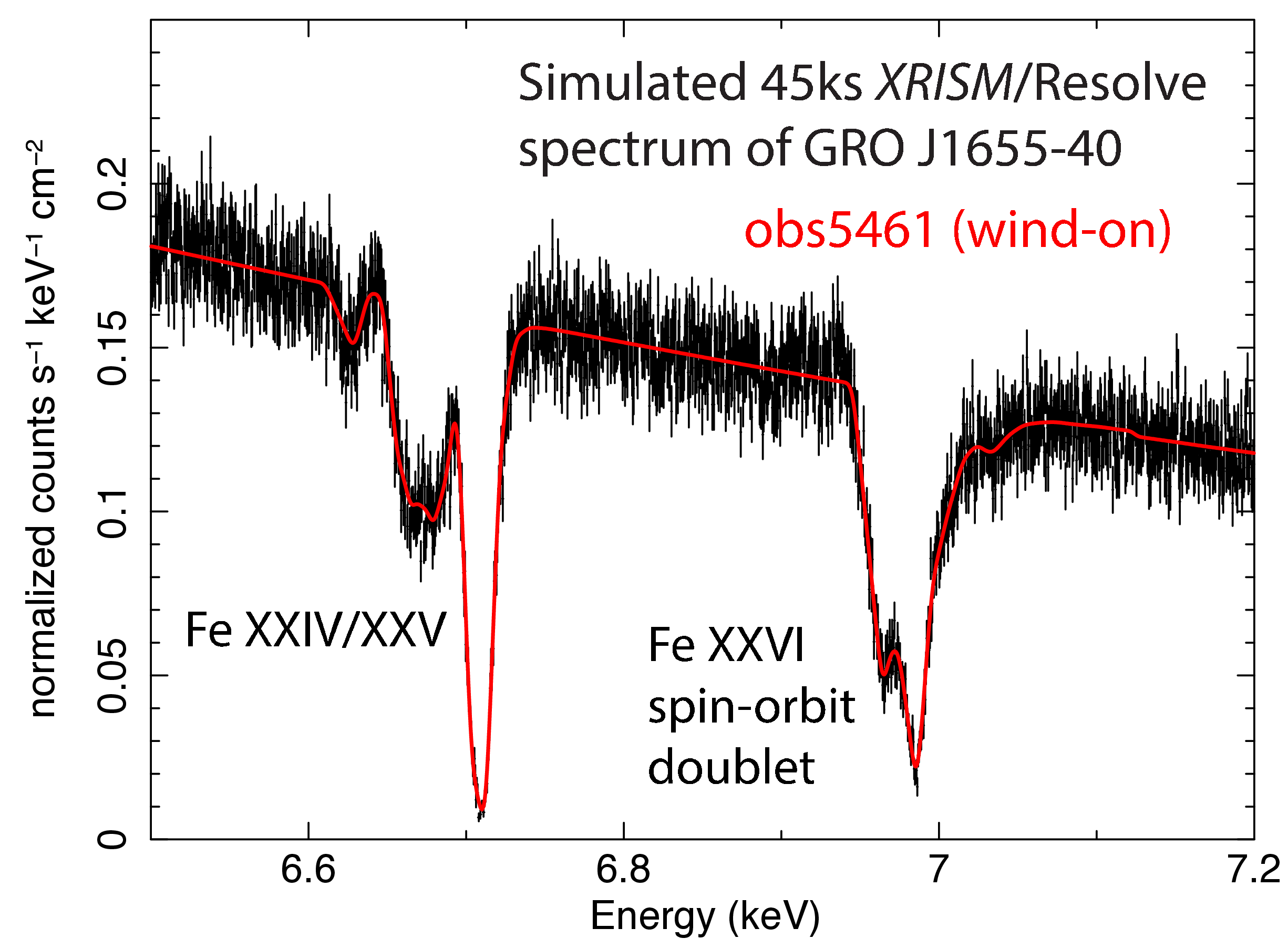}
\includegraphics[trim=0in 0in 0in
0in,keepaspectratio=false,width=3.3in,angle=-0,clip=false]{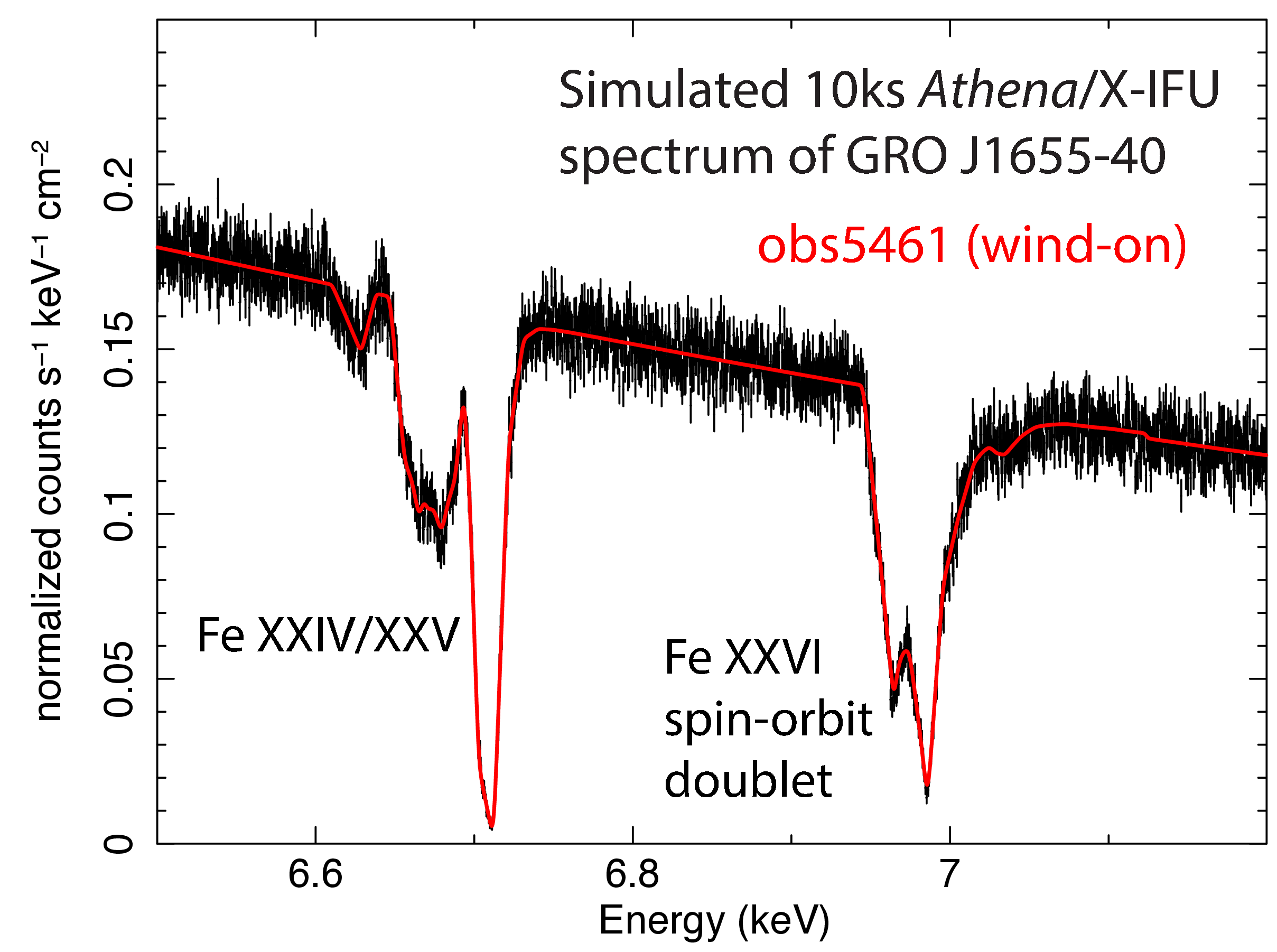}
\end{array}$
\end{center}
\caption{Simulated Fe K band spectra of \gro1655\ with 45ks {\it XRISM}/Resolve observations and
10ks {\it Athena}/X-IFU observations for {\tt obs5461} \windon\ state.}
\label{fig:f14}
\end{figure}

A longer {\it XRISM}/Resolve exposure of 45ks  would afford additional benefits as shown in {\bf Figure~\ref{fig:f13}}: Besides the strong Fe K${\alpha}$ lines, K$\alpha$/K$\beta$ features from high-Z metals (e.g. Co, Ni) would be also resolved. Focused on Fe K band in {\bf Figure~\ref{fig:f14}}, a spin-orbit doublet structure of \fexxvi\ line is clearly visible with 45ks {\it XRISM}/Resolve exposure along with a blended feature from \fexxiv/\fexxv\ at lower energy. We also show a comparison of the same simulated Fe K spectra of \gro1655\ with a simulated 10ks {\it Athena}/X-IFU spectrum (3168-pixel array and 2.5 eV resolution). Both simulations unambiguously show 
the \fexxvi\ spin-orbit doublet, naturally, with the {\it Athena} observation being better than that of {\it XRISM} owing  to {\it Athena}'s large effective area. As noted earlier in \S4.3 (see {\bf Fig.~\ref{fig:f11}}), resolving this feature depends on the wind velocity gradient and thus the \fexxvi\ doublet feature can provide additional information on the location of its formation.

\begin{figure}[t]
\begin{center}$
\begin{array}{cc}
\includegraphics[trim=0in 0in 0in
0in,keepaspectratio=false,width=3.3in,angle=-0,clip=false]{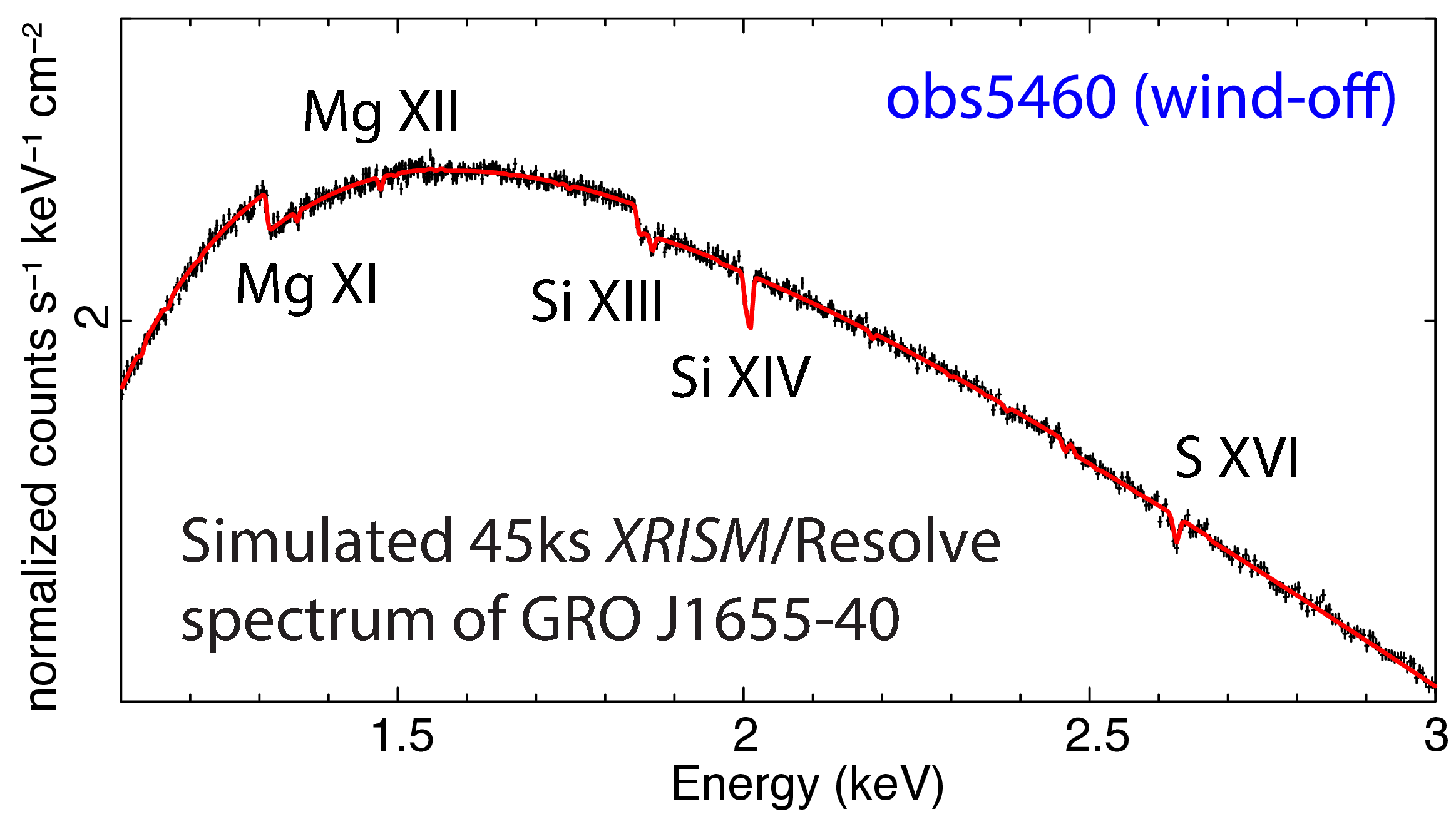}\includegraphics[trim=0in 0in 0in
0in,keepaspectratio=false,width=3.3in,angle=-0,clip=false]{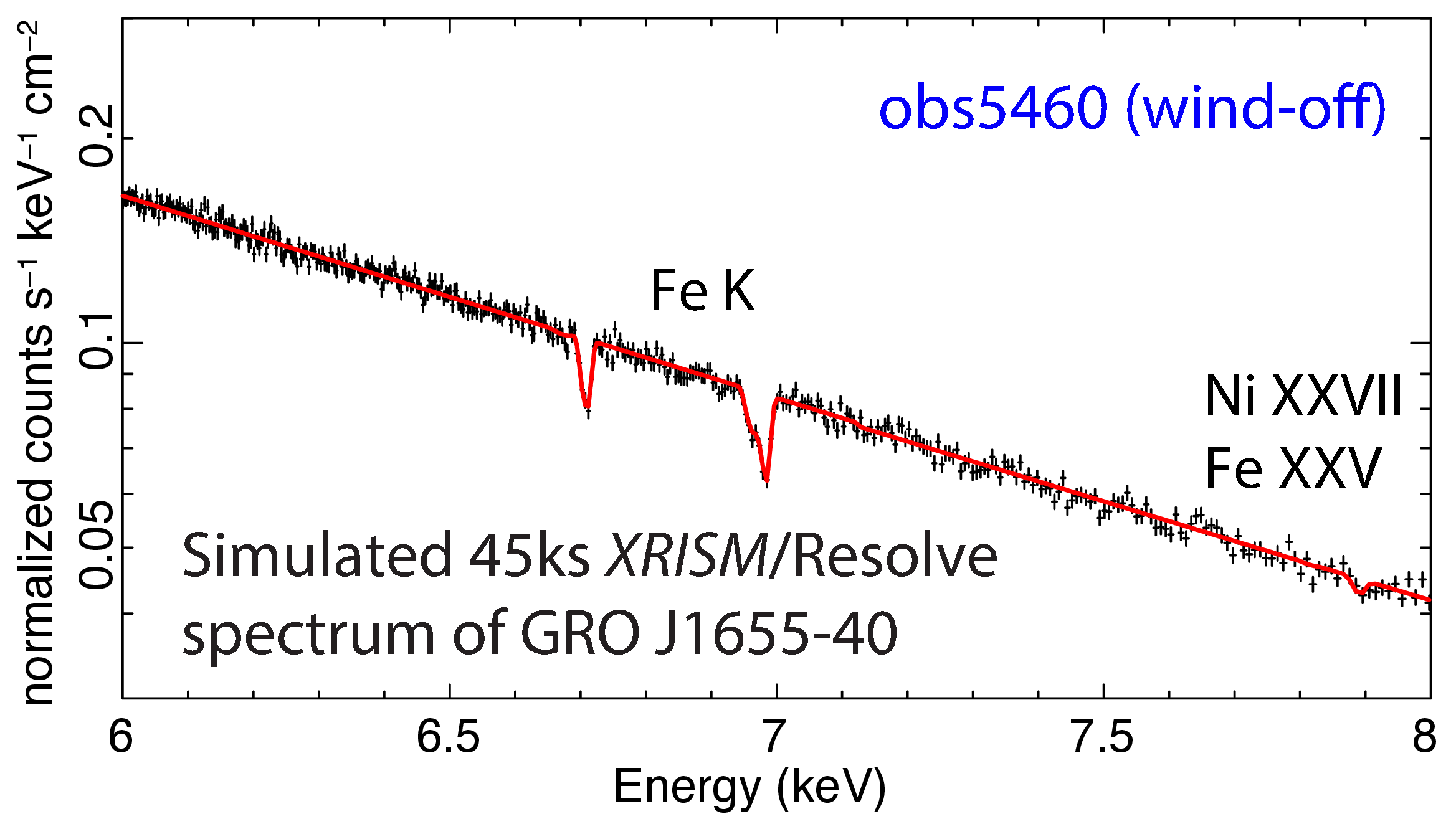}
\end{array}$
\end{center}
\caption{Simulated 45ks {\it XRISM}/Resolve spectrum of \gro1655\ \windoff\ state (obs5460) with the bestfit MHD-wind model.}
\label{fig:f15}
\end{figure}

\subsection{\gro1655\ during \windoff\ State}

The \gro1655\ \windoff\ state gives  some hints of Fe K absorptions {\bf (see Figure~\ref{fig:f10})} as found similarly in the past analyses (e.g. \citealt{NeilsenHoman12}) while its physical parameters not being very well constrained. Based on the bestfit model  of \gro1655\ during \windoff\ state in {\bf Figures~\ref{fig:f4} and \ref{fig:f10}}, we have also performed a 45ks {\it XRISM}/Resolve simulation in {\bf Figure~\ref{fig:f15}}. 
The soft X-ray lines would be better captured with 45ks exposure including \sixiv\ and \sxvi\ lines in addition to Fe K lines. Again, simultaneous detections of the broadband multi-ion absorbers would be the key for a deeper understanding of X-ray winds.

\begin{figure}[t]
\begin{center}$
\begin{array}{cc}
\includegraphics[trim=0in 0in 0in
0in,keepaspectratio=false,width=3.3in,angle=-0,clip=false]{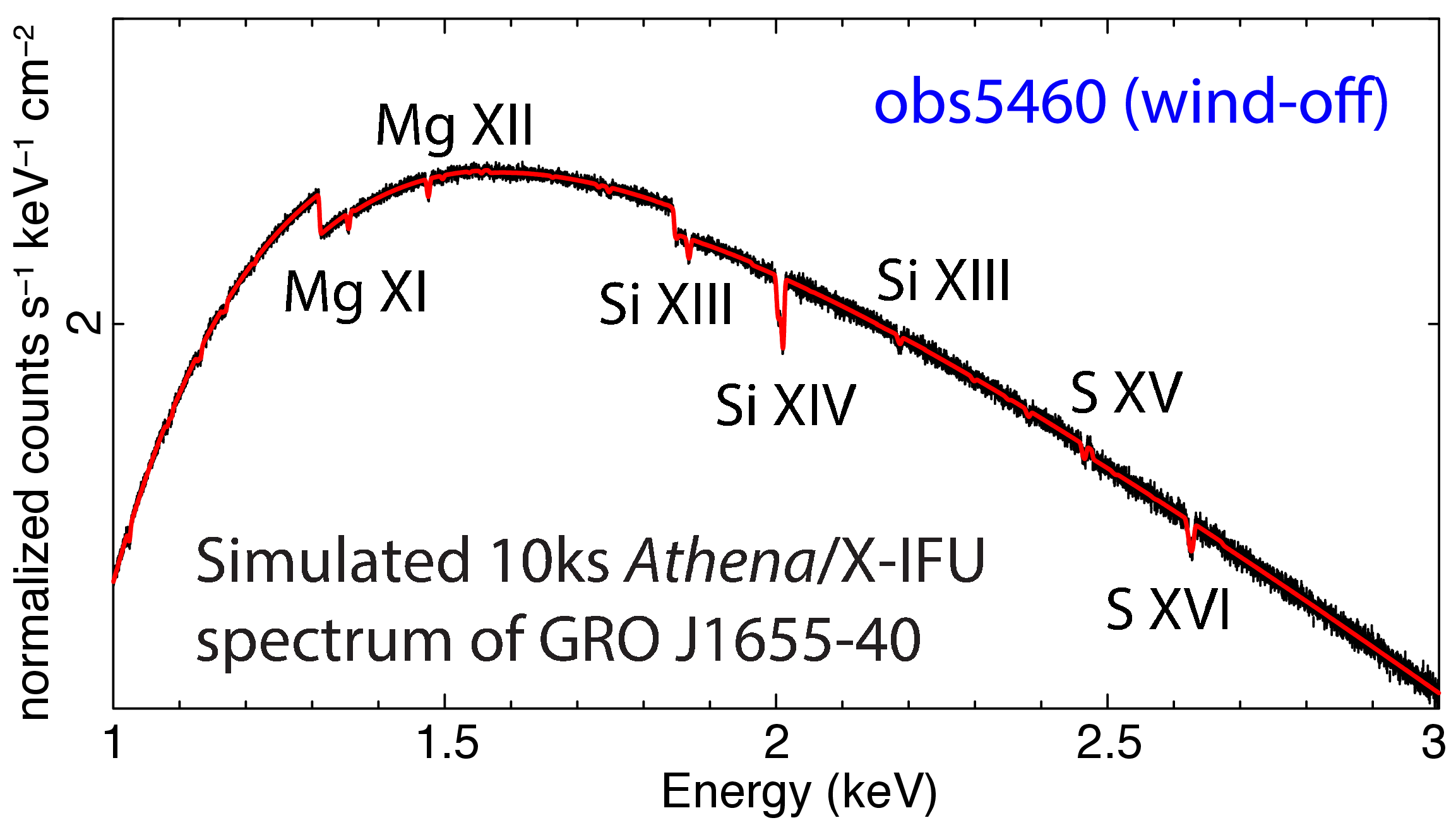}\includegraphics[trim=0in 0in 0in
0in,keepaspectratio=false,width=3.3in,angle=-0,clip=false]{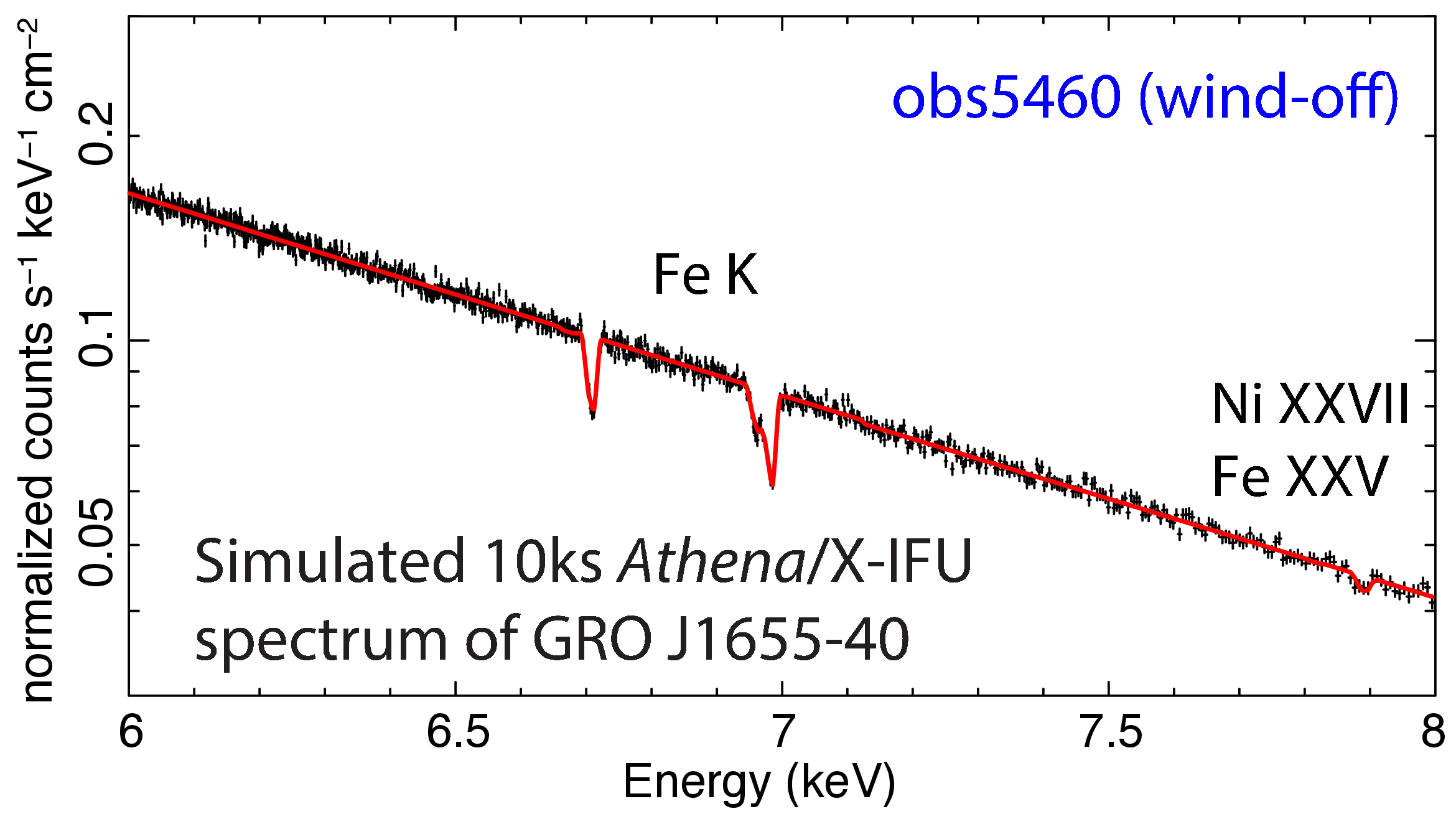}
\end{array}$
\end{center}
\caption{Zoom-in view of 10 ks simulated {\it Athena}/X-IFU spectra of \gro1655\ for {\tt obs5460} with the bestfit MHD-wind model.}
\label{fig:f16}
\end{figure}

\begin{figure}[t]
\begin{center}$
\begin{array}{cc}
\includegraphics[trim=0in 0in 0in
0in,keepaspectratio=false,width=3.3in,angle=-0,clip=false]{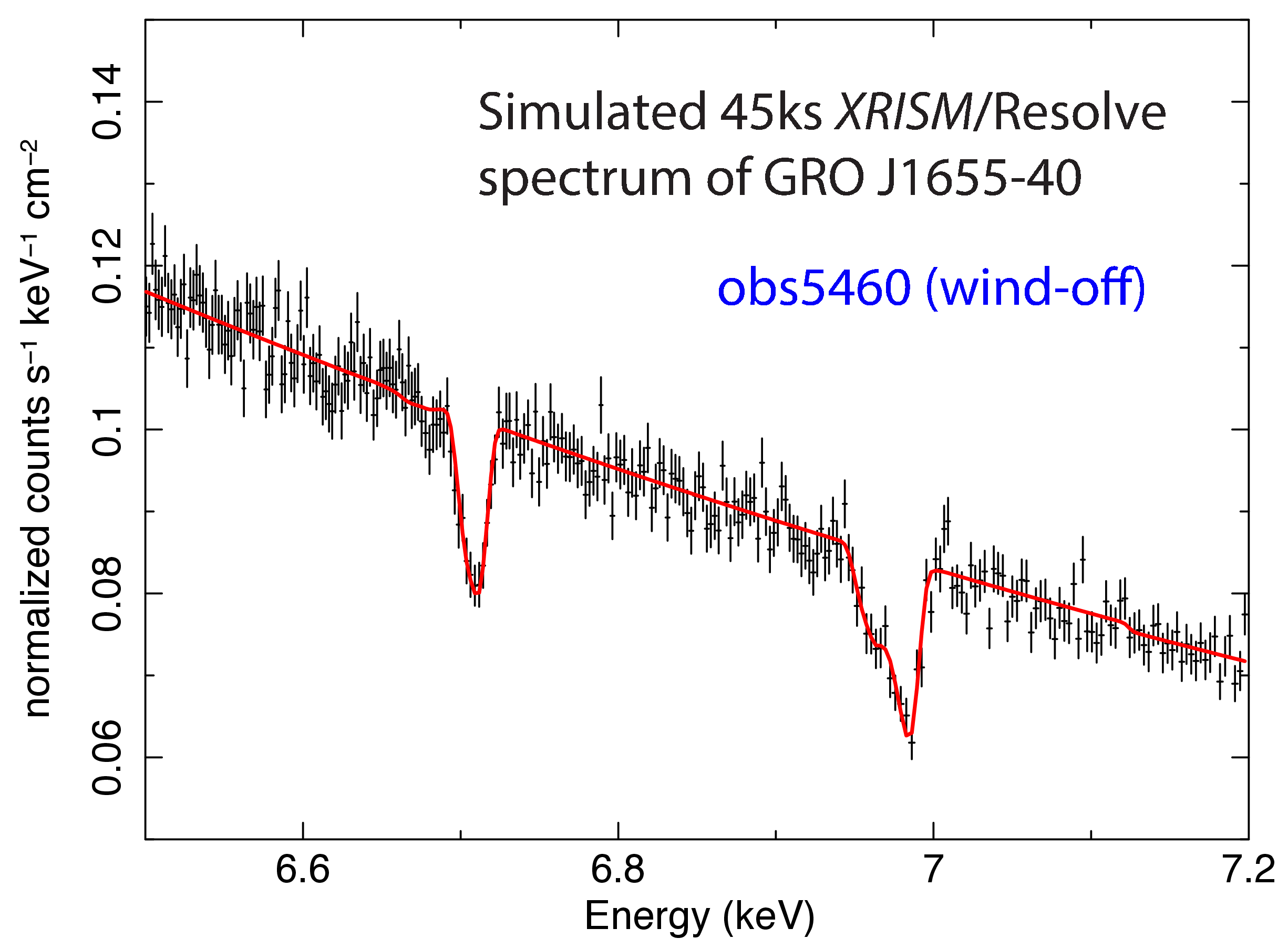}
\includegraphics[trim=0in 0in 0in
0in,keepaspectratio=false,width=3.3in,angle=-0,clip=false]{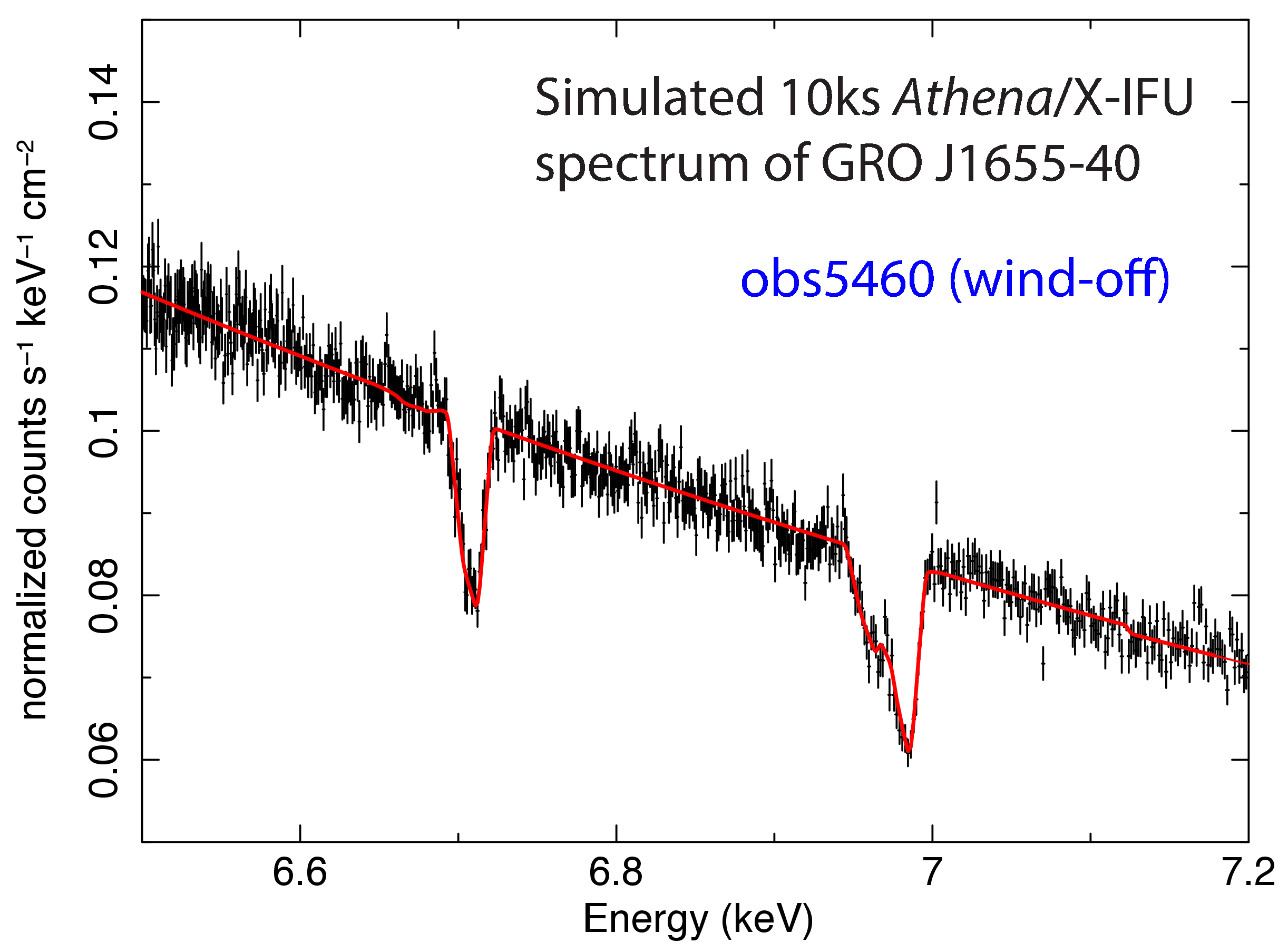}
\end{array}$
\end{center}
\caption{Simulated Fe K band spectra of \gro1655\ with 45ks {\it XRISM}/Resolve observations and
10ks {\it Athena}/X-IFU observations for {\tt obs5460} \windoff\ state.}
\label{fig:f17}
\end{figure}

Fully exploiting a larger collecting area and finer energy resolution with {\it Athena}/X-IFU, we further present a 10ks
{\it Athena}/X-IFU spectrum of \gro1655\ during  \windoff\ state in {\bf Figure~\ref{fig:f16}}. In comparison with 45ks {\it XRISM}/Resolve simulations, the soft X-ray absorbers would be robustly identified with such a short exposure. With a zoom in the Fe K band of both simulations,  {\bf Figure~\ref{fig:f17}}  shows that the Fe K features can be  clearly detectable, although the expected doublet feature of \fexxvi\ line would be ambiguous. The detection of the multiple ions is thus paramount for the determination of the wind's radial density structure, which would otherwise remain unconstrained with the detection of Fe K features alone.

\subsection{\4u1630\ during \windon\ State}

\begin{figure}[t]
\begin{center}$
\begin{array}{cc}
\includegraphics[trim=0in 0in 0in
0in,keepaspectratio=false,width=3.3in,angle=-0,clip=false]{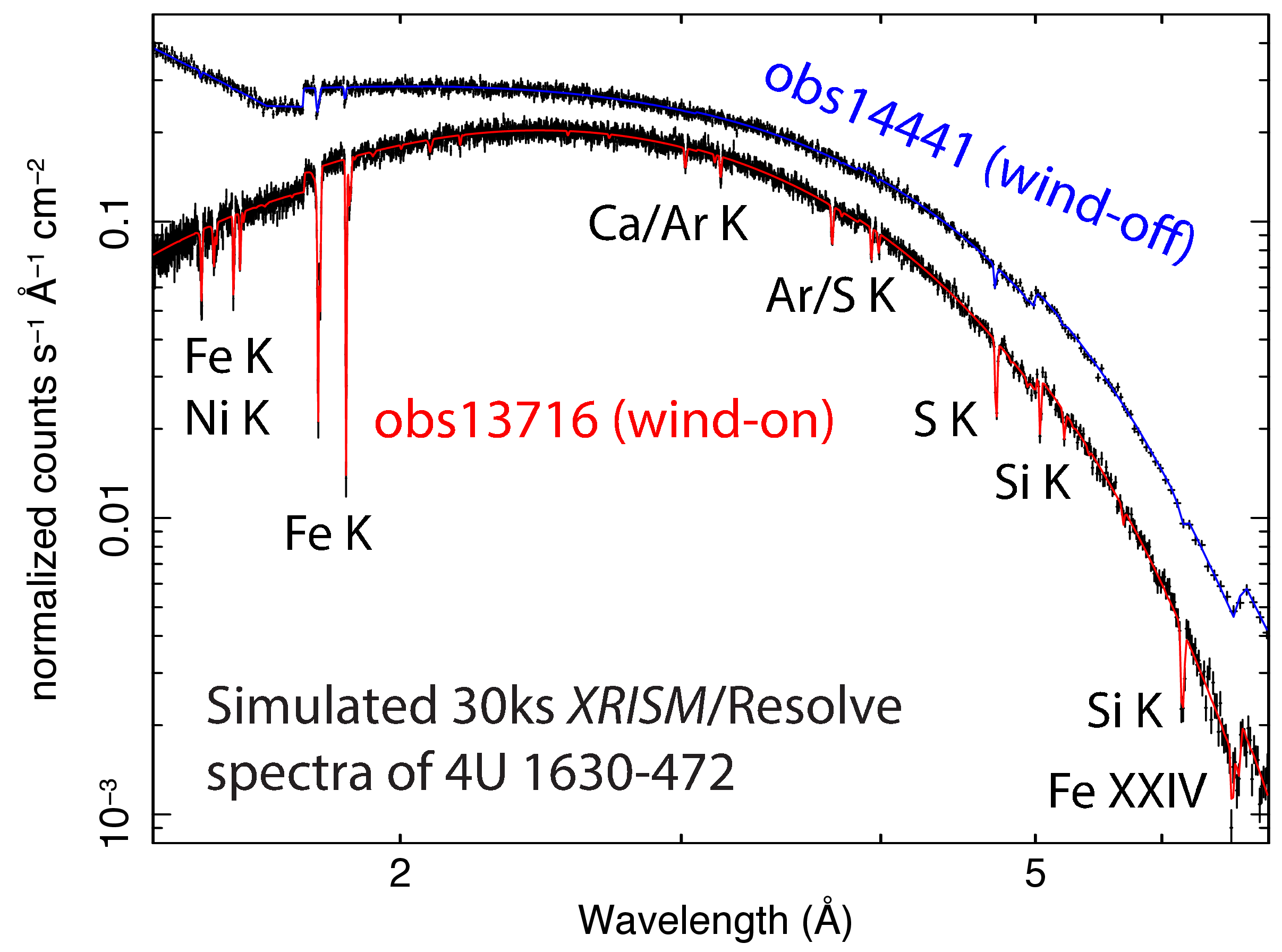}\includegraphics[trim=0in 0in 0in
0in,keepaspectratio=false,width=3.3in,angle=-0,clip=false]{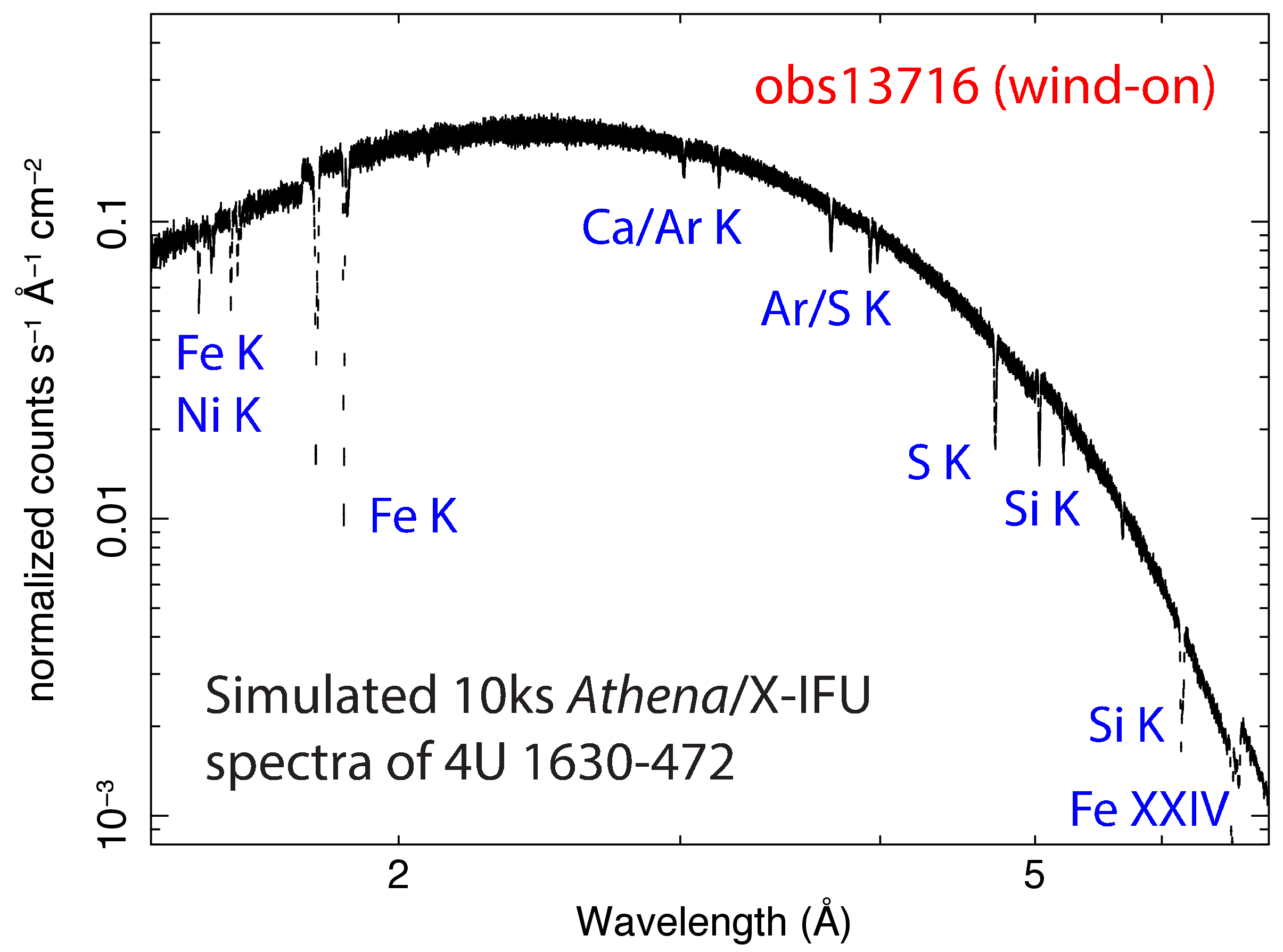}
\end{array}$
\end{center}
\caption{Simulated spectra of \4u1630\ with 30ks {\it XRISM}/Resolve and 10ks {\it Athena}/X-IFU (showing only simulated data for clarity) based on the obtained bestfit MHD-wind model in comparison with {\bf Figure~\ref{fig:f3}}.}
\label{fig:f18}
\end{figure}

Finally, in {\bf Figure~\ref{fig:f18}} we show simulated broadband spectra of \4u1630\ with 30ks {\it XRISM}/Resolve exposure (same exposure for  \windon\ state (obs13716) with {\it Chandra}/HETGS) and 10ks {\it Athena}/X-IFU exposure. An excellent improvement in photon statistics is clearly demonstrated in both cases with more confident line identification being made possible especially in the soft X-ray band. In particular, the detection of Fe K absorbers and S K line, for example, even during  \windoff\ state (obs14441) would be made much clearer in comparison with {\it Chandra} grating data. These simulations are thus making a strong case for a promising capability of microcalorimeters to be onboard these missions.

\section{Summary \& Discussion}

In this paper, we conducted spectral analyses of the X-ray outflow features  
using the first-order {\it Chandra}/HETGS archival data with a focus on three exemplary Galactic BH XRBs, namely \gro1655, \4u1630\ and \h1743, all of which exhibit the \windon\ and \windoff\ states. Our goal is to determine the wind properties associated with these states within the framework of the photoionized MHD disk-wind models of multi-ion absorbers employed in our earlier works (F10; F17; \citealt{F18}). 

\begin{figure}[t]
\begin{center}$
\begin{array}{cc}
\includegraphics[trim=0in 0in 0in
0in,keepaspectratio=false,width=3.0in,angle=-0,clip=false]{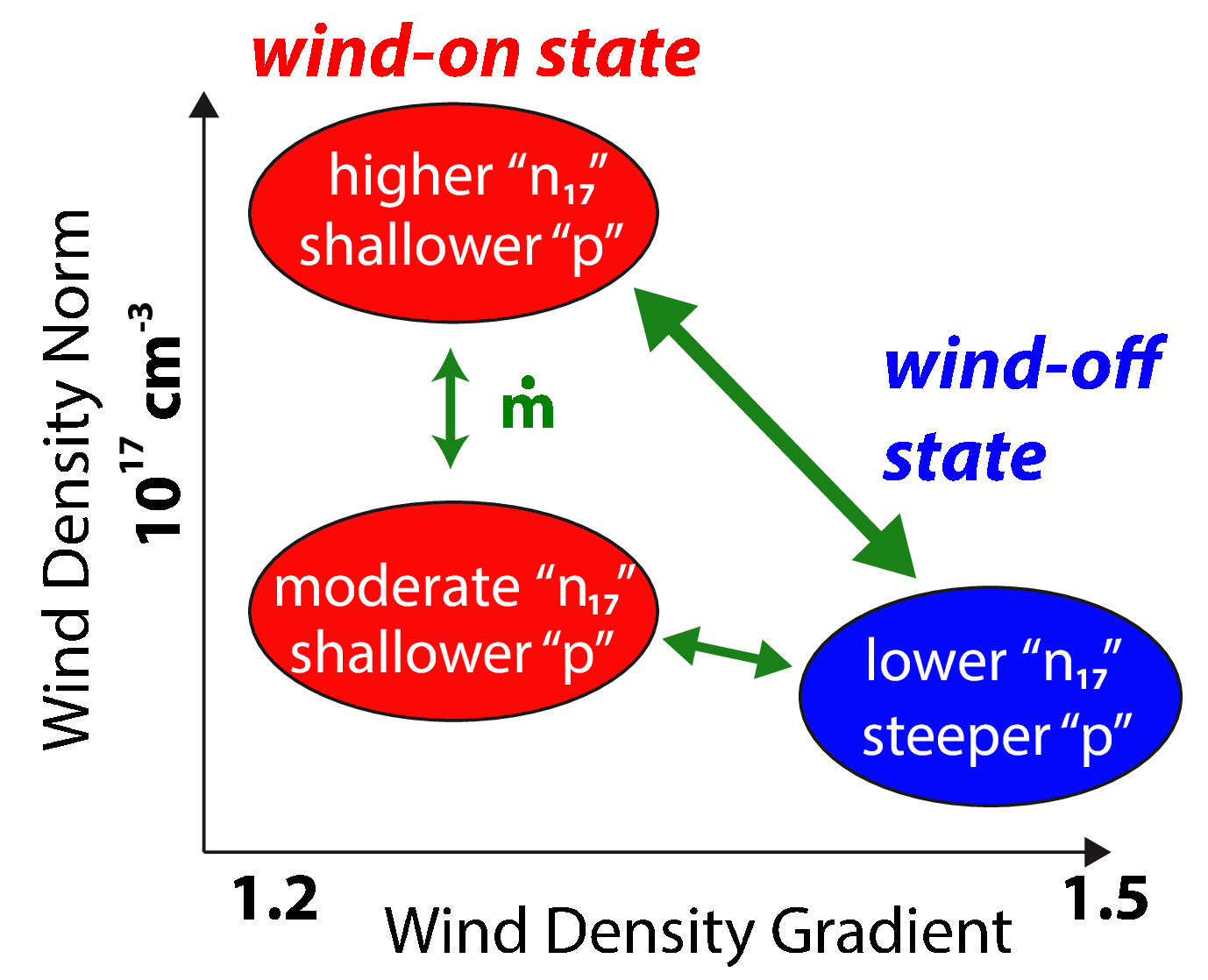}
\end{array}$
\end{center}
\caption{A schematic diagram showing a suggested bi-modal evolution of physical parameters ($p, n_{17}$) as X-ray disk-winds undergo state transition. }
\label{fig:f19}
\end{figure}

Assuming the underlying winds to have the self-similar structure discussed in our earlier works, namely, the power-law density profiles of equation~(2), we fit both the \windon\ and \windoff\ state spectra of each source to determine the underlying physical conditions associated with the wind transition. 
%
With our MHD wind model, primarily characterized by two defining physical quantities, (1) the wind density profile index, $p$, and (2) its density normalization, $n_{17}$, supplemented by relevant elemental abundances,  
we find that the slope $p$ lies in the  range of $1.2 \lesssim p \lesssim 1.5$, while the normalization in the range of $0.01 \lesssim n_{17} \lesssim 1$ across the wind transition (see {\bf Table~3}). Although the noted range in $p$ is spread among all states of the three sources, for each source the value of $p$ is consistently larger in \windoff\ than in its \windon\ state; i.e. the large-scale wind density gradient becomes steepened while its density normalization is reduced during the \windoff\ state almost consistently for all the sources.
Super-solar abundances for heavier elements such as Fe, S and Si are also favored in our modeling. This modality of the wind transition suggested in this work between \windon\ and \windoff\ states is schematically shown in {\bf Figure~\ref{fig:f19}}, indicating qualitatively how the disk-wind properties inherently change across the state transitions. 
%
These results strongly suggest that the apparent change in absorber properties between the \windon\ and \windoff\ state is caused not only by the photoionization balance (i.e. the change in ionizing spectrum), but also by a restructuring of the internal properties of the global wind, as previously conjectured \citep[e.g.][]{NeilsenHoman12,Ponti12}, while the disk wind itself is persistently present regardless of the spectral state in our model \citep[see also][]{Higginbottom20}.
%
%
%
Our model thus does not imply the speculation of ``wind dichotomy" across the observed wind transition. 


In principle, the density normalization $n_{17}$ should be closely related to mass accretion rate in the disk, whereas the radial density gradient $p$ reflects a large-scale structure of the wind that is perhaps guided by magnetic field configuration. As our wind calculations treat the disk as a boundary condition (to provide the materials and the origin of the global magnetic field) in this work, however, a theoretical identification of the physical cause of this trend is beyond the scope of this work. Fully global MHD simulations incorporating both accretion physics and outflowing plasma should be able to provide more insight into the expected trend across the wind transition from the first principle and test whether this modality suggested in the current analysis is indeed consistent or not.

We note that our present results for \gro1655\  \windon\ state (obs5461; $p=1.34$ and $n_{17}=3.0$) is slightly different from our previous findings in F17 ($p=1.2$ and $n_{17}=9.3$). This difference can be simply understood by the fact that the current wind model allows also for a choice of the heavier ion abundances, namely $1 \le A_{\rm Fe,S,Si} \le 3$, whereas the model of F17 was fixed at the solar abundances ($A_{\rm Fe,S,Si} = 1$). The different fitting procedures then suggest the presence of degeneracies among the model parameters which would likely require additional observations of higher sensitivity to resolve. The advantage of higher abundances is that they allow for a lower density normalization and slightly steeper density slope for the wind given in F17, resulting in a smaller wind mass flux. This steeper density profile, however, results in slightly under-estimated columns at longer wavelength as indeed seen in {\bf Figures~\ref{fig:f6} and \ref{fig:f8}}. The issue will likely be resolved with higher confidence with observations of the higher sensitivity measurements made possible with {\it XRISM}/Resolve.

At this point, we address several facts related to our models and to absorption spectroscopy in general: (i) The importance of considering more than one ionic species in modeling the wind structure. More specifically, the role of $p$ in our model. (ii) They afford self-consistent inclusion of the line broadening process attributed to the wind velocity gradient by eliminating an extraneous free parameter such as the phenomenological {\tt vturb} parameter prescribed in {\tt XSTAR}.  

On the other hand, the derived best-fit slope in this analysis indicates that the  \citet{BP82} type of MHD wind and quasi-spherical radiation-driven winds of $p \sim 1.5-2$, are statistically ruled out for these BH XRB winds during \windon\ state. This is what has motivated us to exploit the power of AMD for the purpose of modeling disk winds; i.e. AMD method enables us to incorporate a multi-ion distribution in the wind that can be tested against high-quality grating data of high S/N such as those considered here. 
Spectral analyses in general often focus only on the properties of the Fe K complex  simply because this appears to be the strongest or the only feature detectable with sufficient significance in data  \citep[e.g.][]{Miller15,Shidatsu19,Tomaru19,Tomaru20,Tetarenko20,Higginbottom20,Neilsen20}. Such a single feature alone affords sufficient parameter freedom to preclude any serious constraints on the global properties of the associated outflows. 
%
%
Therefore,  in order to obtain a wind structure over a broader spatial scale,  it is paramount to model several ionic species of widely separated $\xi$ values (i.e. AMD).  An attempt along these lines is made, for example, by \citet{Trueba19} where multi-epoch {\it Chandra}/HETGS spectra of \4u1630\ are analyzed in the context of AMD method. They derived the averaged AMD by combining multi-epoch soft-state data and found $1.29 \le p \le 1.42$ in an excellent agreement with our independently derived result, although our modeling is performed over individual epoch data sets without stacking. Their analyses also support an MHD-driving origin.

As described in \S 1, thermal pressure could drive winds especially when aided by sufficient radiation force. However, these scenarios assume a somewhat arbitrary degree of freedom that determines an optimal geometric relation between the inner coronal configuration and the outer disk shape (that is under coronal illumination) in order to adjust a geometric shadowing effect \citep[e.g.][]{Shidatsu19,Tomaru19,Tomaru20,Tetarenko20}.
Also, as discussed above, these works focus mainly on the Fe K features and it is not clear if/how they can address at the same time also the soft X-ray absorbers; the importance in obtaining a broader perspective of the wind structure and dynamics.

There also seem to be conflicting suggestions among variants of thermal-radiative wind models  regarding the existence of disk-winds during {\it low/hard} state; e.g. HD simulations by \citet{Higginbottom20} have predicted powerful winds in both {\it hard} and {\it soft} states similarly to our current predictions, whereas other models  have claimed a strong suppression of winds during {\it low/hard} state, thus implying a disk dichotomy \citep[e.g.][]{Done18,Tomaru19}. Meanwhile, the MHD wind scenario provides a coherent prediction of persistent X-ray winds in {\it both} states.


The derived best-fit models can further help us estimate a number of physical quantities related to the accretion and outflows.
The accreting plasma near the BH event horizon has the Thomson optical depth on the order of the dimensionless accretion rate $\dot M_{a,o}/\dot M_E =\dot{m}_{a,o}$ \citep{K12}; since the disk is the reservoir off which the wind matter is launched we can relate this to the wind density normalization $n_{17}$ by the relation 
\begin{eqnarray}
\dot{m}_{a,o} \sim \frac{2 \sigma_T R_S n_o}{f_w} \ ,
\end{eqnarray}
where $n_o \equiv n(r_o,\theta \sim \pi/2) = n_{17} 10^{17}$ cm$^{-3}$, $f_w$ denotes the mass outflow rate relative to mass accretion rate, $\sigma_T$ is the Thomson cross section and  $R_S$ is the \sw radius (e.g. \citealt{K12}; F17; \citealt{F18,K19}). Assuming $f_w \sim 1$ as often conceived for Seyfert AGN warm absorbers \citep[e.g.][]{Turner09,Blustin05}, we find $\dot{m}_{a,o} \sim 0.4$ for the \windon\ state (obs5461) in \gro1655\ with $M=10\Msun$ in this work. Our  wind model then provides the radial scalings of the large-scale wind characteristics; The local mass outflow rate scales like $\dot{M}_{\rm out}^{\rm local} \sim n r^2 v \propto r^{3/2-p}$. The outflow  momentum rate scales as $\dot{P}_{\rm out}^{\rm local} = \dot{M}_{\rm out}^{\rm local} v \propto r^{1-p}$ and the wind kinetic power  varies like  $\dot{E}_{\rm out}^{\rm local} \sim \dot{M}_{\rm out}^{\rm local} v^2 \propto r^{1/2-p}$. That is, for $p \gsim 1$, more outflowing mass is ejected from the outer disk region, whereas a good fraction of the wind kinetic power is delivered from the interior disk region near the BH.

Considering that $n r^2 = L_X/\xi \sim 10^{37}$ erg~s$^{-1}$ $/(10^4$ erg~cm~s$^{-1}$) and $v \sim 800$ km~s$^{-1}$ (as a fiducial value) for \gro1655, one obtains $\dot{M}_{\rm out}^{\rm local} \sim 10^{17}$ g~s$^{-1}$, also implying that the {\it local} kinetic power of $\dot{E}_{\rm out}^{\rm local} \sim 4.3 \times 10^{32}$ erg~s$^{-1}$, which is broadly consistent with the previous estimates \citep[e.g.][]{Miller15,Trueba19}.
On the other hand, one can also estimate a {\it global} mass outflow rate $\dot{M}_{\rm out}^{\rm global}$ by integrating $\dot{M}_{\rm out}^{\rm local}$ over LoS radius such that
\begin{eqnarray}
\dot{M}_{\rm out}^{\rm global} &\sim&  m_p \int_{r \sim r_o}^{r \sim r_{\rm out}} n(r,70\deg) v_{\rm out}(r,70\deg) r dr \ , \\
&\sim& \frac{m_p n_{17} (3R_S)^2 c}{3/2-p} x_{\rm out}^{3/2-p}  \ , \\
&\sim& 10^{18} ~ \textmd{g~s$^{-1}$} \sim \ 10^{-8} ~ \textmd{$\Msun$/year},
\end{eqnarray}
for $M=10\Msun, \theta=70\deg$ and $p=1.3$ where $x \equiv r/r_o \sim r/(3 R_S)$ and we have used $x_{\rm out} \sim 10^8$. Accordingly, this yields a {\it global} kinetic power delivered by the disk-winds as
\begin{eqnarray}
\dot{E}_{\rm out}^{\rm global} &\sim& \frac{1}{2} \int_{r \sim r_o}^{r \sim r_{\rm out}} d\dot{M}_{\rm out}^{\rm global} v_{\rm out}^2 \sim \frac{m_p n_{17} (3R_S)^2 c}{3/2-p} x_{o}^{1/2-p} \sim 4.6 \times 10^{36} ~ \textmd{erg~s$^{-1}$} \ ,
\end{eqnarray}
where $x_o \sim 1$ and thus $\dot{E}_{\rm out}^{\rm global}/(\dot{M}_{\rm out}^{\rm global} c^2) \sim 0.005$.
Note that $\dot{M}_{\rm out}^{\rm global}$ would be even higher with less steeper value of $p$ (such as $p=1$), while $\dot{E}_{\rm out}^{\rm global}$ is practically independent of $p$ since it is dominated by $x_o \sim 1$.
%
Here, the interactive feedback provided by the X-ray wind is important not only in AGNs, but also in BH XRB systems as well since a powerful X-ray disk-wind has a potential to (1) influence the evolution of binary systems and (2) destabilize  accretion flows by extracting a significant amount of mass and energy \citep[e.g.][]{Munoz-Darias16} as estimated above.


Throughout this work, we assume that the magnetic field morphology does not change significantly such that the underlying wind geometry remains the same between \windon\ and \windoff\ state. This is an assumption to simplify and reduce the degrees of freedom in the problem. In general, it is conceivable that the wind structure can be sensitive to a number of physical properties; e.g. coronal radiation field, advective and diffusive nature of magnetic fields, mass-accretion rate and so on. Within a generic MHD framework, it is speculated that a global magnetic field of sufficiently small plasma $\beta$ may predominantly govern the structure of magnetized plasma outflows regardless of the other factors in  such a way that the field lines act like rigid wires \citep[e.g.][]{BP82,CL94}. In that case, the wind morphology may stay nearly unchanged to some extent between \windon\ and \windoff\ state, as assumed here. In a more comprehensive treatment of radiation field as in a hybrid wind scenario \citep[e.g.][]{Everett05,NeilsenHoman12}, for example, one should consider a different field geometry that leads to distinct outflow streamlines \citep[e.g.][]{F14}.


From a synergistic perspective of BH XRB winds, recent optical observations have uncovered a clear presence of disk-winds by detecting the blueshifted absorption of the P-Cygni profile with H$\alpha$ and \hei\ lines during {\it hard} state \citep[e.g.][]{Munoz-Darias19}, despite the conventional speculation of wind-jet dichotomy
\citep[e.g.][]{NeilsenLee09}.
Similarly, it is suggested, according to the simultaneous X-ray and radio observations, that jets and disk-winds are not quite mutually exclusive in neutron star LMXBs (e.g. GX~13+1) and high-mass X-ray binaries (e.g. SS~433)  as well as some BH XRBs (e.g. V404~Cyg and IGR~J17091-3624) \citep[e.g.][]{Homan16,Gatuzz20}.
Such a persistent existence of optical disk-winds in XRBs even during {\it hard} state is in fact consistent with our X-ray wind model; its spectroscopic appearance changes between the two states due to the intrinsic change in an internal physical condition of the wind such as ($p, n_{17}$) as well as photoionization balance, as found in this work. Although a physical connection between optical and X-ray disk-winds is yet to be clearly found, multi-wavelength observations will certainly help us better understand a large-scale behavior of XRB disk-winds.

%


The observed ionic velocities, $v_i$, of BH XRB absorption features in their high S/N spectra are of order of a few hundred to a thousand km/s. Therefore, one might consider these X-ray absorbers qualitatively somewhat analogous to Seyfert warm absorbers \citep[e.g.][]{McKernan07,Blustin05,Tombesi13,Laha14,Laha16} because of their similarity in velocities and column densities. Both ought to, then, be launched from similar  distances (normalized to $R_S$) such that $x \equiv r/R_S \sim c^2/v_i^2$ or $r/R_S \sim 10^5-10^6$. However, this analogy is somewhat misleading because Seyfert AGNs exhibit also UFOs of near-relativistic velocities \citep[$v_i/c \gsim 0.1$; e.g.][]{Reeves18,Reeves20,Pounds03,Chartas09,Tombesi12,Tombesi13,Gofford15,F15,F18} that must be launched at correspondingly smaller distances (e.g. $r/R_S \sim 10 - 300$), and so on with the even higher velocities in broad absorption line (BAL) quasar (QSO) \citep[e.g.][]{Chartas09}. This diversity of AGN outflow velocities is well accommodated within our model, given its density and velocity scalings. Considering that an ionic species $i$ of velocity $v_i$ is produced at a specific value of the ionization parameter $\xi_i \sim L_{\rm ion}/[n(x,\theta)x^2]$, we obtain
\begin{eqnarray}
\xi_i (x,\theta)= \frac{L_{\rm ion}}{n_o x^{2-p}} \frac{1}{f(\theta)} ~~{\rm or}~~  v_i^2 \propto \frac{1}{x} \propto \left[\frac{\xi_i(x,\theta)}{(L_{\rm ion}/n_o)}\right]^{\frac{1}{2-p}} \ ,
\end{eqnarray}
where $L_{\rm ion} \sim L_X$ is the ionizing luminosity. For $p < 2$, one sees that the velocity of an ionic species increases with $\xi_i$ with Fe K absorbers naturally having the highest velocities. The velocities also decrease with increasing $L_{\rm ion}/n_o$, the ionizing luminosity per unit mass of the wind \citep{F10b}, with $n_0$ being a proxy of the bolometric luminosity $L_{\rm bol}$ since $f_w \sim 1$. This approach then suggests that BAL QSO outflows, with the lowest $L_X/L_{\rm bol}$, should have the highest Fe K absorbers' velocities, followed by Seyferts' UFOs, while BH XRB winds of $L_X/L_{\rm bol} \sim 1$ should exhibit the lowest velocities in agreement with observed wind velocity hierarchy. These scaling laws work only if the wind, following its self-similar structure, is present from near the ISCO (but being fully ionized there) to the outer disk radius in all these source classes.

The crucial issue of $L_X/L_{\rm bol}$ depends also on the wind structure and, in particular, on the value of the parameter $p$ relative to $3/2$. This is because of the wind mass flux scaling, $\dot{M}_{\rm out}^{\rm local} \propto r^{3/2-p}$, as noted earlier. Assuming $f_w \sim {\rm constant} \sim 1$ implies \citep[see][]{K15,K19} that the mass accretion rate must decrease toward the BH; this allows for its Eddington normalized value, $\dot m(r) = \dot M_a(r)/\dot M_E$, to become smaller than $\alpha_{\rm vis}^2$ (where $\alpha_{\rm vis}$ is the disk viscosity parameter in \citealt{SS73}) interior to a transition  radius $r_{\rm tr}$ (this would depend on the outer boundary value of the mass flux provided). Thus, for $r < r_{\rm tr}$, the mass accretion rate will be sufficiently small to allow its transition into a ``hot" advection-dominated accretion flow \citep[ADAF; e.g.][]{NY94}. This segment of the flow  then plays the role of the ``corona", producing the hard X-ray ionizing radiation and gives the appearance of a disk truncated at radius $r < r_{\rm tr}$. The transition (aka. truncation) radius will decrease with increasing luminosity (i.e. total $\dot M$ at the outer disk boundary) in agreement with the disk transition described earlier and the suggestion put forward in the literature \citep[e.g.][]{Esin97}. It will also ``squeeze" the hard X-ray emission toward the ISCO reducing its fractional contribution to the bolometric luminosity as the latter approaches that of Eddington, as believed to be the case in BAL QSOs and in the XRBs during \textit{High/Soft} state.

These arguments make clear the importance of the presence of absorbers in the X-ray spectra. They are not just the radiative features of some incidental outflows  of accretion-powered sources. They are the {\it markers} that define their global accretion/outflow structure as well as their spectral states and SEDs. Should our proposal be confirmed by additional observations and modeling, it will provide a key in understanding the physics underlying both the global properties of Galactic XRBs and also AGNs.


Finally, simulated microcalorimeter spectra expected from {\it XRISM}/Resolve and {\it Athena}/X-IFU would  reveal a series of relatively weak absorption signatures in the soft X-ray band \citep[see also][]{Ratheesh21}. In addition, these simulations clearly demonstrate a
high fidelity of the current model
to resolve a quantum mechanical effect such as a spin-orbit doublet in \fexxvi\ line, which is indeed extremely challenging or only vaguely traceable even in the current third-order {\it Chandra}/HETGS spectra \citep[e.g.][]{Miller15}. These detailed spectral signatures will be within reach with the new missions as a valuable tool to perhaps help answer to more fundamental questions such as  wind launching mechanisms.

\acknowledgments
%
K.F. is grateful to Joey Neilsen for a number of illuminating questions and comments about the model and XRB disk-wind physics. We are also thankful to an anonymous referee for a number of illuminating comments and questions. This work is supported in part by NASA/ADAP (NNH18ZDA001N-2ADAP) and
{\it Chandra} Cycle 20 archival proposal grants.

\end{document}